\documentclass{aa}  %for a printer version

\usepackage{graphicx}
\usepackage{natbib}

\usepackage{booktabs}  
\usepackage[colorlinks=true,linkcolor=blue,citecolor=blue]{hyperref}
\usepackage{txfonts}

\graphicspath{{./Pictures/}} % Specifies the directory where pictures are stored

\providecommand{\abs}[1]{\lvert#1\rvert} 

\usepackage[dvipsnames]{xcolor} %color in text

\usepackage{longtable}
\usepackage{lscape}

\begin{document}

   \title{HADES RV Programme with HARPS-N at TNG}

   \subtitle{XIII. A sub-Neptune around the M dwarf GJ~720~A \thanks{Based on observations collected at the Italian Telescopio Nazionale Galileo (TNG), operated on the island of La Palma by the Fundación
Galileo Galilei of the INAF (Istituto Nazionale di Astrofisica) at the
Spanish Observatorio del Roque de los Muchachos of the Instituto
de Astrof\'isica de Canarias, in the framework of the HArps-n red Dwarf Exoplanet Survey (HADES).}}

   \author{E.~Gonz\'alez-\'Alvarez \inst{\ref{inst:CSIC-INTA1}}
   		  \and A.~Petralia \inst{\ref{inst:INAF-OAPA}}
          \and G.~Micela \inst{\ref{inst:INAF-OAPA}}
          \and J.~Maldonado \inst{\ref{inst:INAF-OAPA}}
		  \and L.~Affer \inst{\ref{inst:INAF-OAPA}}
          \and A.~Maggio \inst{\ref{inst:INAF-OAPA}}
%second group
			\and E.~Covino \inst{\ref{inst:INAF-Napoli}}
           \and M.~Damasso \inst{\ref{inst:INAF-Torino}}
           \and A.~F.~Lanza \inst{\ref{inst:INAF-Catania}}
           \and M.~Perger \inst{\ref{inst:IECC-CSIC}, \ref{inst:IECC}}
           \and M.~Pinamonti \inst{\ref{inst:INAF-Torino}}
           \and E.~Poretti \inst{\ref{inst:INAF-Merate}, \ref{inst:TNG}}
           \and G.~Scandariato \inst{\ref{inst:INAF-Catania}}
           \and A.~Sozzetti \inst{\ref{inst:INAF-Torino}}
%third group
           \and A.~Bignamini \inst{\ref{inst:INAF-Trieste}}
           \and P.~Giacobbe \inst{\ref{inst:INAF-Torino}}
           \and G.~Leto \inst{\ref{inst:INAF-Catania}}
           \and I.~Pagano \inst{\ref{inst:INAF-Catania}}
           \and R.~Zanmar~S\'anchez \inst{\ref{inst:INAF-Catania}}
%4 group
			\and J.~I.~Gonz\'alez~Hern\'andez \inst{\ref{inst:IAC}, \ref{inst:ULag}}
           \and R.~Rebolo \inst{\ref{inst:IAC}, \ref{inst:ULag}}
           \and I.~Ribas \inst{\ref{inst:IECC-CSIC}, \ref{inst:IECC}}
           \and A.~Su\'arez~Mascare{\~n}o \inst{\ref{inst:IAC}, \ref{inst:ULag}}
           \and B.~Toledo-Padr\'on \inst{\ref{inst:IAC}, \ref{inst:ULag}}
          }

			% 1 
			\institute{Centro de Astrobiolog\'ia (CSIC-INTA), Carretera de Ajalvir km 4, 28850 Torrej\'on de Ardoz, Madrid, Spain \label{inst:CSIC-INTA1}
			% 2  
            \and INAF-Osservatorio Astronomico di Palermo,
              Piazza Parlamento 1, 90134 Palermo, Italy \label{inst:INAF-OAPA}
            % 3
              \and INAF-Osservatorio Astronomico di Capodimonte, Salita Moiariello 16, 80131 Napoli, Italy \label{inst:INAF-Napoli}
             %4
              \and INAF-Osservatorio Astrofisico di Torino, via Osservatorio 20, 10025 Pino Torinese, Italy  \label{inst:INAF-Torino}
             %5
              \and INAF-Osservatorio Astrofisico di Catania, via S. Sofia 78, 95123 Catania, Italy \label{inst:INAF-Catania}
               %6
              \and Institut de Ci\`encies de l'Espai (IEEC-CSIC), Campus UAB, Carrer de Can Magrans s/n, 08193, Bellaterra, Spain  \label{inst:IECC-CSIC}
              %7
              \and Institut d'Estudis Espacials de Catalunya (IEEC), 08034 Barcelona, Spain  \label{inst:IECC}
			  %8
			  \and 
			  INAF-Osservatorio Astronomico di Brera, Via E. Bianchi 46, 23807 Merate, Italy \label{inst:INAF-Merate}
			  %9
			\and 
			%10
				Fundaci\'{o}n Galileo Galilei -- INAF, Rambla Jos\'{e} Ana Fernandez P\'{e}rez 7, 38712 -- Bre\~{n}a Baja, Spain \label{inst:TNG}
				%11
			  \and INAF-Osservatorio Astronomico di Trieste, Via Tiepolo 11, 34143 Trieste, Italy \label{inst:INAF-Trieste}
			  %12
			  \and Instituto de Astrof\'isica de Canarias, 38205 La Laguna, Tenerife, Spain \label{inst:IAC}
			  %13
		      \and Universidad de La Laguna, Dpto. Astrof\'isica, 38206 La Laguna, Tenerife, Spain \label{inst:ULag}
				}

   \offprints{E. Gonz\'alez-\'Alvarez \\ 
   \href{mailto:egonzalez@cab.inta-csic.es}{egonzalez@cab.inta-csic.es}}
   \date{Received 3 February 2021 / Accepted 17 March 2021}

% \abstract{}{}{}{}{} 
% 5 {} token are mandatory
 
  \abstract
   {The high number of super-Earth and Earth-like planets in the habitable zone detected around M-dwarf stars in the last years has revealed these stellar objects to be the key for planetary radial velocity (RV) searches.}
   {Using the HARPS-N spectrograph within The HArps-n red Dwarf Exoplanet Survey (HADES) we reach the precision needed to detect small planets with a few Earth masses using the spectroscopic radial velocity technique. HADES is mainly focused on the M-dwarf population of the northern hemisphere. }
   {We obtained 138 HARPS-N RV measurements between 2013 May and 2020 September of GJ~720~A, classified as an M0.5~V star located at a distance of 15.56\,pc. To characterize the stellar variability and to discern the periodic variation due to the Keplerian signals from those related to stellar activity, the HARPS-N spectroscopic activity indicators and the simultaneous photometric observations with the APACHE and EXORAP transit surveys were analyzed. We also took advantage of \textit{TESS}, MEarth, and SuperWASP photometric surveys. The combined analysis of HARPS-N RVs and activity indicators let us to address the nature of the periodic signals. The final model and the orbital planetary parameters were obtained by fitting simultaneously the stellar variability and the Keplerian signal using a Gaussian process regression and following a Bayesian criterion.}
   {The HARPS-N RV periodic signals around 40\,d and 100\,d have counterparts at the same frequencies in HARPS-N activity indicators and photometric light curves. Then we attribute these periodicities to stellar activity the former period being likely associated with the stellar rotation. GJ~720~A shows the most significant signal at $19.466\pm0.005$\,d with no counterparts in any stellar activity indices. We hence ascribe this RV signal, having a semiamplitude of $4.72\pm0.27$\,ms$^{-1}$, to the presence of a sub-Neptune mass planet. The planet GJ~720~Ab has a minimum mass of $13.64\pm0.79$\,$\rm M_{\oplus}$, it is in circular orbit at $0.119\pm0.002$\,AU from its parent star, and lies inside the inner boundary of the habitable zone around its parent star. }
   {}


\maketitle
%
%-------------------------------------------------------------------

\section{Introduction}
\label{Introduction}

Nowadays the high-precision spectrograph development has allowed to reach the necessary radial velocity (RV) precision to detect Neptune- and Earth-mass planets close and/or inside the habitable zone of late-type main-sequence stars. The M dwarf stars have turned out to be the ideal targets for detecting this type of planets \citep[e.g.,][]{2014MNRAS.441.1545T,2015ApJ...807...45D}. The lower mass of the parent stars results in a higher Doppler RV amplitude for a given planetary mass than those for more massive stars. However, the M dwarfs tend to be active stars \citep{1998A&A...331..581D, 2012AJ....143...93R} and it is known that the stellar activity hampers the detection of planets introducing periodic variations in the RV signals that mimic the signals with a Keplerian origin \citep{2001A&A...379..279Q, 2014Sci...345..440R}. 

Different approaches can be followed in order to disentangle the stellar activity signals from planetary induced signals. Spectroscopic activity indicators can be used to derive stellar activity variations and the stellar rotation period as well as simultaneous photometric and RV observations. The false frequencies analysis together with the coherence and the stability of the signals can also provide strong indications on the origin of the periodicity. A coherent and long-lived behaviour of the signal is expected in case the variations are caused by a Keplerian motion. The RV technique is affected by the contribution of both stellar activity and Keplerian modulations, therefore a model that considers simultaneously the stellar variability through the Gaussian Process (GP) regression together with a fit of the planetary orbital parameters can be crucial when determining the Keplerian parameters.

Here, we present the high-precision, high-resolution spectroscopic measurements of the M0.5V star GJ~720~A (HIP 91128, BD+45 2743) obtained with the HARPS-N spectrograph \citep{2012SPIE.8446E..1VC} on the Telescopio Nazionale \textit{Galileo} (TNG) as part of the HArps-n red Dwarf Exoplanet Survey (HADES). %\footnote{\tt http://www.oact.inaf.it/exoit/EXO-IT/Projects/Entries/2011/12/27_GAPS.html}.

The HADES collaboration has already produced many valuable results, regarding the statistics, activity, and characterization of M stars \citep{2017A&A...598A..26P, 2017A&A...598A..27M, 2017A&A...598A..28S, 2018A&A...612A..89S, 2019A&A...624A..27G}, and has led to the discovery of several planets \citep{2016A&A...593A.117A, 2017A&A...605A..92S, 2017A&A...598A..26P, 2018A&A...617A.104P, 2019A&A...622A.193A}.

In Section \ref{sec:GJ 720 A}, we introduce the target star (GJ~720~A) and present the stellar properties, newly derived and collected from the literature. Section \ref{sec:observations} presents the observations carried out including high-resolution spectroscopy and photometric variability monitoring. In Section \ref{sec:analysis}, we provide a detailed analysis of the HARPS-N radial velocities, spectroscopic activity indicators, and photometric light curves with the main goal of determining the presence of planet candidates. The properties of the newly discovered planet orbiting GJ~720~A are given in Section \ref{sec:GP}. A brief discussion on the implications of this finding and the conclusions of this paper appear in Section \ref{sec:summary_discussion}.

\section{GJ~720~A}
\label{sec:GJ 720 A}

GJ~720~A is an M0.5~V dwarf located at a distance of $15.557 \pm 0.006$\,pc \citep{2018AJ....156...58B} from the Sun. As published by \cite{1979lccs.book.....L} GJ~720~A (LHS~3394) has a wide companion called GJ~720~B (other names: LHS~3395 and VB 9) with relative position measured since 1960. Following the most updated classification \citep{2015A&A...577A.128A} GJ~720~B is an M2.5~V star and the projected separation between GJ~720~A and B is 112.138\,arcsec. %(1745\,AU).

In this work we focus on the primary star GJ~720~A, its basic stellar parameters (effective temperature, stellar metallicity, spectral type, mass, radius, surface gravity, and luminosity) were computed by using the same spectra used here in the RV analysis and following the procedure described in \cite{2015A&A...577A.132M} and \cite{2017A&A...598A..27M}. The most updated stellar parameters of GJ~720~A collected from the literature are compiled in Table~\ref{tab:stellar_properties_GJ720A}. During the guaranteed CARMENES exoplanet survey \citep{2018A&A...612A..49R} GJ~720~A was also observed as part of their M dwarf sample. The stellar parameters derived from the CARMENES spectra were published in \cite{2019A&A...625A..68S}. All of them agree, within the 1$\sigma$ error bars, with those derived here using the HARPS-N spectra. 

GJ~720~A is not a very active star and shows moderate chromospheric flux \citep{2017A&A...598A..27M}. This is consistent with the slow rotation of GJ~720~A, which has a projected rotational velocity $v$\,sin\,$i = 0.99\pm0.53$\,km\,s$^{-1}$ \citep{2017A&A...598A..27M}. GJ~720~A has been observed in X-rays by \textit{ROSAT} and we derived its X-ray luminosity,      
$\log L_{\rm X} = 27.26 \pm 0.15$\,erg\,s$^{-1}$ \citep{2019A&A...624A..27G}. From its X-ray luminosity, the activity level is typically found among medium active stars of its spectral type. GJ~720~A has a rotation period ($P_{\rm rot}$) of $34.5 \pm 4.7$\,days determined from Ca~{\sc ii} H \& K and $H\alpha$ spectroscopy time-series in \cite{2018A&A...612A..89S}. Also \cite{2020MNRAS.491.5216G} confirms this value in the contex of the photometric analysis of APACHE survey data. The chemical composition analysis available reveals that GJ~720~A has a slightly sub-solar metallicity.

\begin{small}
\begin{table}[!t]
\centering
\caption{Stellar parameters of GJ~720~A.}
\label{tab:stellar_properties_GJ720A}
\begin{tabular}{l c l }
\hline
\hline
\noalign{\smallskip}
Parameters  & Value & Ref.$^{a}$\\\\
\noalign{\smallskip}	
\hline	
\noalign{\smallskip}

Other name					&		HIP 91128 & \\
$\rm \alpha$ (J2000) &   18:35:19.08 &  \textit{Gaia DR3}\\
$\rm \delta$ (J2000) &   +45:44:44.4  &  \textit{Gaia DR3}\\
G (mag) & $9.1050 \pm 0.0005$ &  \textit{Gaia DR2}\\
J (mag) & $6.88 \pm 0.02$ & 2MASS\\
Spectral type				&		$\rm M0.5\,V$	& Mald17\\ 
$\pi$ (mas) & $64.236 \pm0.012$ &  \textit{Gaia DR3}\\
$d$ (pc) & $\rm 15.557 \pm 0.006$  & \textit{Bail18}	\\
$\mu _{\alpha} \cos \delta$ ($\rm mas\,yr^{-1}$) & $452.36 \pm 0.01$ & \textit{Gaia DR3}\\
$\mu _{\beta}$ ($\rm mas\,yr^{-1}$) & $363.47 \pm 0.01$ & \textit{Gaia DR3}\\

\noalign{\smallskip}
\hline
\noalign{\smallskip}

\textit{From HARPS-N spectra} & &\\
\noalign{\smallskip}

$T_{\rm eff}$ (K)  & $\rm 3837 \pm 69$ & \textit{Mald17}\\
$\log g$ (cgs) & $\rm 4.71 \pm 0.05$ & \textit{Mald17}\\
$\rm [Fe/H]$ (dex) & $\rm -0.14 \pm 0.09$ & \textit{Mald17}\\
$M$ ($\rm M_{\odot}$) & $\rm 0.57 \pm 0.06$ & \textit{Mald17}\\
$R$ ($\rm R_{\odot}$) & $\rm 0.56 \pm 0.06$ & \textit{Mald17}\\
$\log L_{\rm bol}/L_{\odot}$ & $\rm -1.217 \pm 0.0964$ & \textit{Mald17}\\

$ v \sin i$ ($\rm km\,s^{-1}$) & $\rm 0.99 \pm 0.53$ & \textit{Mald17}\\
$\log R'_{\rm HK}$	& $\rm -5.03 \pm 0.04$ & \textit{Suar18}\\
$P_{\rm rot}$ (days)$^{(*)}$ & $36.05^{+1.38}_{-1.44}$ & \textit{This work}\\
\noalign{\smallskip}

\hline
\noalign{\smallskip}
%\hline
$\log L_{\rm x}$ (erg\,s$^{-1}$) & $\rm 27.39 \pm 0.15$ & \textit{Gonz19}\\
$\log L_{\rm x}/L_{\rm bol}$ & $\rm -5.11 \pm 0.18$ & \textit{Gonz19}\\
\noalign{\smallskip}
\hline

\end{tabular}
\tablefoot{$^{(a)}$\textit{Gaia} DR3: \cite{2020arXiv201202061G}; \textit{Gaia} DR2: \cite{2018A&A...616A...1G}; 2MASS: \cite{2003yCat.2246....0C}; Mald17 : \cite{2017A&A...598A..27M}; Bail18: \cite{2018AJ....156...58B}; Gonz19: \cite{2019A&A...624A..27G}; Suar18: \cite{2018A&A...612A..89S}. $^{(*)}$ $P_{\rm rot}$ value derived from the S-index activity indicator. 

}
\end{table}
\end{small}

\section{Observations}
\label{sec:observations}

\subsection{HARPS-N radial velocities}

\begin{figure}[!tb]
\centering
\includegraphics[width=0.5\textwidth]{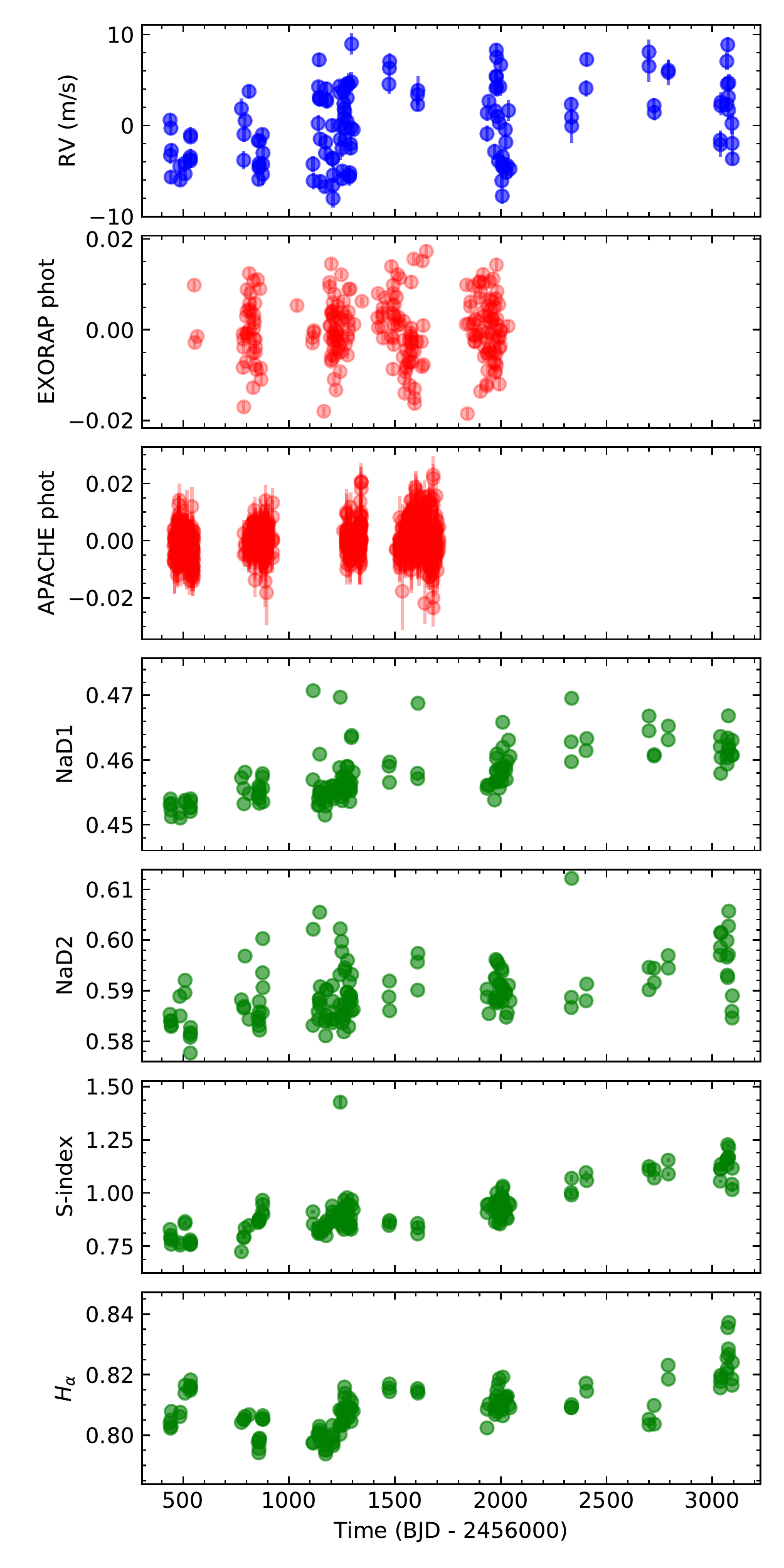}
\caption{Radial velocity (blue dots), EXORAP ($B$-band) and APACHE ($V$-band) photometric (red dots), and spectroscopic (green dots) activity indicators time-series for GJ~720~A. }
\label{M80_rv_time_TERRA}
\end{figure}

GJ~720~A has been monitored from the 26th of May 2013 to the 1st of September 2020 for a total of 138 data points. Of the 138 HARPS-N epochs, 75 were obtained within the GAPS observing program and 63 within the Spanish observing program. The spectra were obtained with the high resolution (resolving power $R \sim 115,000$) optical echelle spectrograph HARPS-N. The exposure time was set to 15 minutes yielding an average signal-to-noise ratio (S/N) of 76 at 5500\,\AA. Data were reduced using the latest version of the Data Reduction Software \citep[DRS V3.7,][]{2007A&A...468.1115L}. For GJ~720~A the M2 mask was used. The RVs were computed by matching the spectra with a high S/N template obtained by co-adding the spectra of the target, as implemented in the Java-based Template-Enhanced Radial velocity Re-analysis Application \citep[TERRA,][]{2012ApJS..200...15A}. TERRA provides more accurate RVs when it is applied to M dwarfs, considering orders redder than the 22nd. The GJ~720~A TERRA RVs show a root mean square ($rms$) dispersion of 4.19\,m\,s$^{-1}$ and a mean internal error of 0.9\,m\,s$^{-1}$. The HARPS-N RV time-serie is shown in the top panel of Figure \ref{M80_rv_time_TERRA} while the RV data are provided in Table \ref{tab:gj720a_rv_act_data}. 

The TERRA pipeline also provides measurements for a number of spectral features and other diagnostics of stellar activity (e.g., Ca~{\sc ii} H \& K (S-index), the Na I D line, and H$\alpha$). The derived values are given in Table \ref{tab:gj720a_rv_act_data} and the corresponding time-series are shown in Figure \ref{M80_rv_time_TERRA}.

%----------------------------------

%phot Mearth and SuperWASP
\begin{figure}[!tb]
\centering
\includegraphics[width=0.5\textwidth]{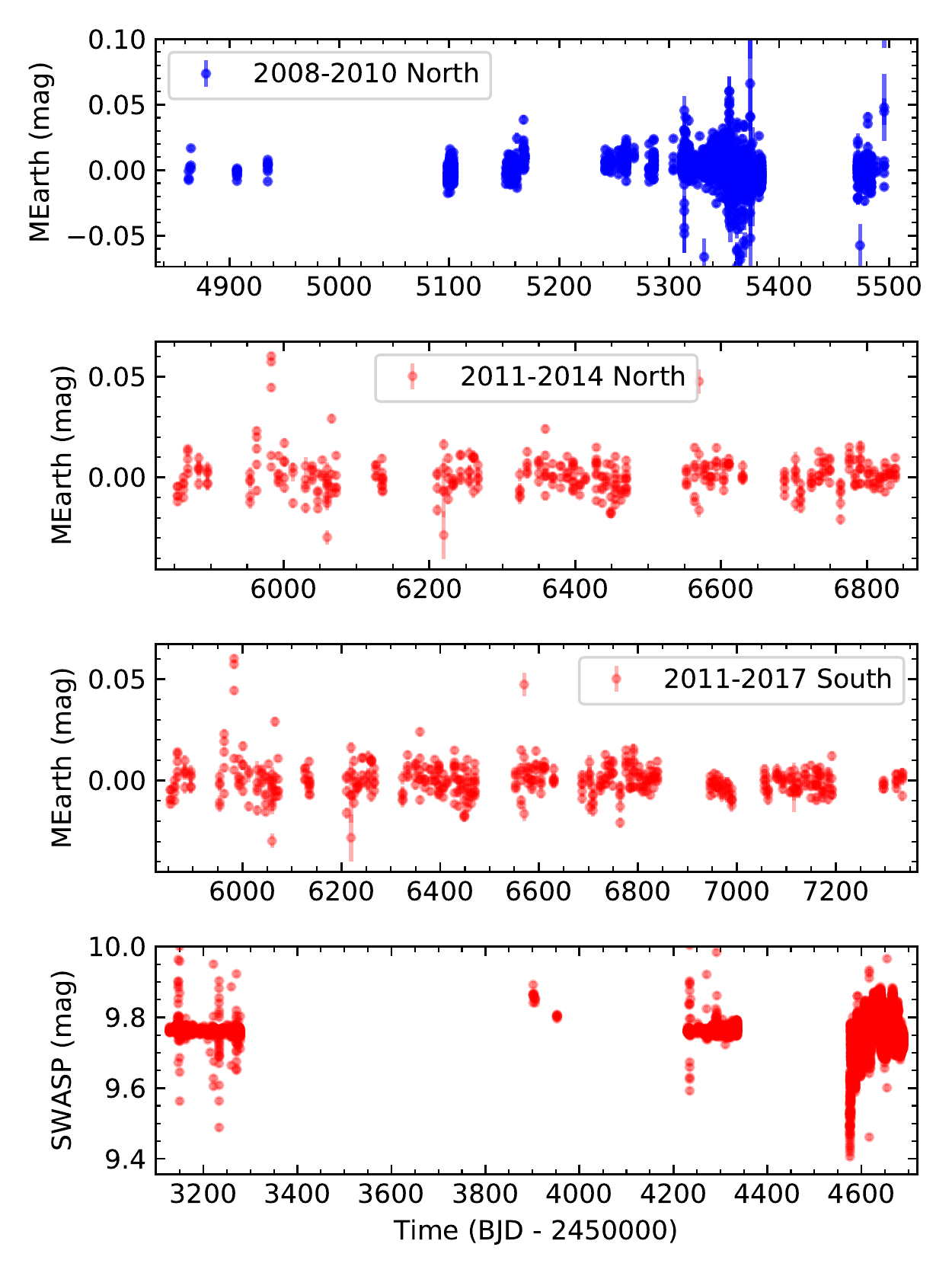}
\caption{MEarth and SuperWASP photometric time series for GJ~720~A after 2.5$\sigma$ clipping applied.}
\label{fig:M80_phot_mearth_superwasp}
\end{figure}

\subsection{Photometric time series}
\label{subsec:photometric time serie}

\begin{figure}[!tb]
\centering
\includegraphics[width=0.24\textwidth]{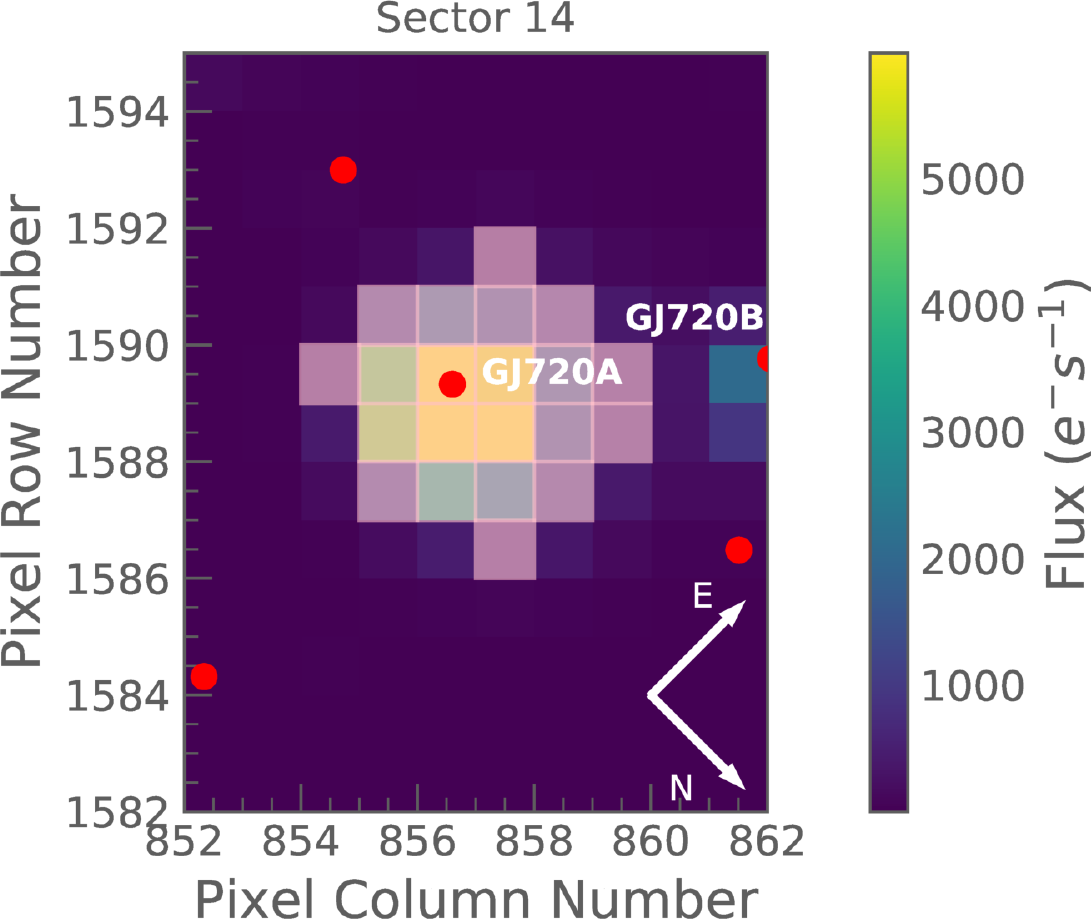}
\includegraphics[width=0.24\textwidth]{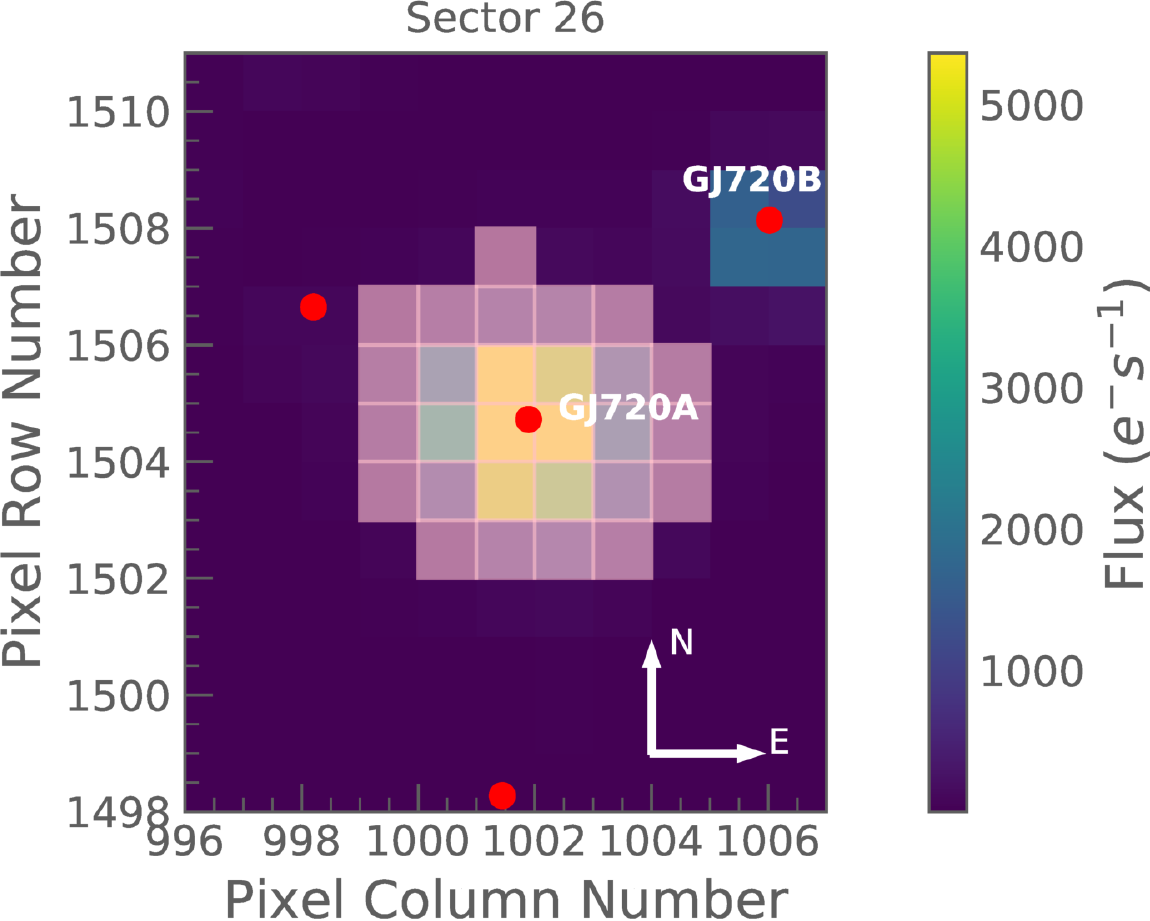}

\caption{Target pixel files (TPF) of GJ~720~A (TIC122958010) in \textit{TESS} Sectors 14 and 26. The electron counts are color-coded. The shadowed pixels correspond to the \textit{TESS} optimal photometric aperture used to obtain the simple aperture photometry (SAP) fluxes. The red dots correspond with the bright nearby stars with \textit{TESS} magnitude less than 16. The positions of GJ~720~A and GJ~720~B are indicated. }
\label{fig:apertures}
\end{figure}

\paragraph{SuperWASP and MEarth.} GJ~720~A was photometrically observed by the Wide Angle Search for Planets (SuperWASP) exoplanet transit survey \citep{2014CoSka..43..500S} and the MEarth \citep{2012AJ....144..145B} survey. There were three photometric campaigns using the MEarth telescopes between 2008 and 2010, from 2011 to 2014, and between 2011 and 2017. The first two campaigns were conducted in the Northern Hemisphere with a total number of datapoits of 632 ($rms$ of 8.8\,mmag) and 984 ($rms$ of 6.9\,mmag), respectively. While the last one was carried out in the South with 1,480 measurements ($rms$ of 6.2\,mmag). The original photometric data presented several outliers and we applied a 2.5$\sigma$-clipping algorithm to remove them. The outliers were also removed from the SuperWASP photometric data taken between 2004 and 2008. The different photometric time-series are presented if Fig. \ref{fig:M80_phot_mearth_superwasp}.

\paragraph{EXORAP.} We monitored GJ~720~A  also in the framework of the EXORAP project at the INAF-Catania Astrophysical Observatory with an 80\,cm f/8 Ritchey-Chretien robotic telescope (APT2) located at Serra la Nave on Mt. Etna. We collected $\sim$5\,yr of B, V, R, and I-band photometry in order to obtain simultaneously photometric and spectroscopic data. We performed data reduction by applying overscan, bias, dark subtraction, and flat fielding with IRAF\footnote{IRAF is distributed by the National Optical Astronomy Observatories, which are operated by the Association of Universities for Research in Astronomy, Inc., under cooperative agreement with the National Science Foundation.} procedures and visual inspection to check the image quality (see \citealt{2016A&A...593A.117A} for details). Errors in the individual photometric points include the intrinsic noise (photon noise and sky noise) and the $rms$ of the ensemble stars used for computing the differential photometry. The final dataset contains $\sim$240 photometric points for each of the $B$, $V$, $R$ and $I$ bands distributed over 5 consecutive seasons, between MJD=56555 and MJD=58034 ($B$ filter shown in Fig.~\ref{M80_rv_time_TERRA}).

\paragraph{APACHE.} 44 of the HADES targets (including GJ~720~A) were also monitored photometrically by APACHE (A PAthway towards the CHaracterization of Habitable Earths) photometric transit search project \citep{2013EPJWC..4703006S}. Our target was very intensively observed by APACHE, having 163 nights over a timespan of ~1250\,days, for a total of 5900 points in the $V$-band. The APACHE photometric observing epochs (binned data are shown in Fig.~\ref{M80_rv_time_TERRA}) partially overlap with the spectroscopic observations carried out within HADES, therefore the photometric and spectroscopic activity data analyzed here are partially simultaneous.

\paragraph{\textit{TESS}.} GJ~720~A (TIC122958010) star was observed by \textit{TESS} in sector 14 between 2019 July 18 and August 15 and in sector 26 between 2020 July 18 and August 15. The light curves and the target pixel (TPFs) files for the different sectors were downloaded from the Mikulski Archive for Space Telescopes (MAST) which is a NASA founded project. We verified the pixels in the aperture FITS extension and marked those that were used in the optimal photometric aperture in order to check that there is no other source contamination that could affect the transit search. The TPFs files of GJ~720~A with the standard pipeline apertures are shown in Fig. \ref{fig:apertures}. We also included in the figure the bright nearby stars with \textit{TESS} magnitude less than 16. The visual binary companion GJ~720~B is located outside the standard pipeline apertures and therefore we can discard some kind of contaminating flux from it. The light curve files provide simple aperture photometry (SAP) fluxes and photometry corrected for systematics effects \cite[PDC,][]{2012PASP..124.1000S, 2014PASP..126..100S}, being the last ones optimized for \textit{TESS} transit searches. 

The information regarding the different photometric surveys, the used observing filters/sectors, the number of days covered per observing season, the number of photometric measurements, and the standard deviation of the differential photometry are summarized in Table \ref{tab:phot:propert}.

\begin{table}[]
\begin{small}
\caption{Photometric seasons available for GJ~720~A.}
\label{tab:phot:propert}
\centering  

\setlength{\tabcolsep}{1pt}
\begin{tabular}{l c c c c r }
\hline
\hline
\noalign{\smallskip}
Obs.$^a$  & Filter/sector & Season & $\Delta$T & $\rm N_{obs}$ & $\sigma$\\
 		&		& & (d) & &  \\
\noalign{\smallskip}
\hline
\noalign{\smallskip}

MEarth-North & $RG715$ & 2008--2010 & 3,532 & 632 & 8.8\,mmag\\
MEarth-North & $RG715$ & 2011--2014  & 397 	 &  984 & 6.9\,mmag\\
MEarth-South & $RG715$ & 2011--2017 &  536  & 1,480 & 6.2\,mmag\\

SuperWASP & 			...         & 2004--2008 &  1,559   & 12,922  &  36.7\,mmag\\
EXORAP      & 	$B$				  & 2013--2017 &  1,479   & 248  &  8.6\,mmag\\
APACHE     & 	$V$  					& 2013--2016 &  1,250   & 5,900  &  5.5\,mmag\\
\textit{TESS} & Sector 14			& 	July 2019 & 27 & 18,522 &  4.8 $\rm \cdot 10^{-4}\, e^{-}/s$ \\
\textit{TESS} & Sector 26		   & 	July 2020 & 27 & 16,941 & 4.8 $\rm \cdot 10^{-4}\, e^{-}/s$ \\

\hline
\end{tabular}
\end{small}
\end{table}

%----------------------------------
\section{Analysis of GJ~720~A}
\label{sec:analysis}

\begin{figure}[]
\centering
\includegraphics[width=0.9\columnwidth]{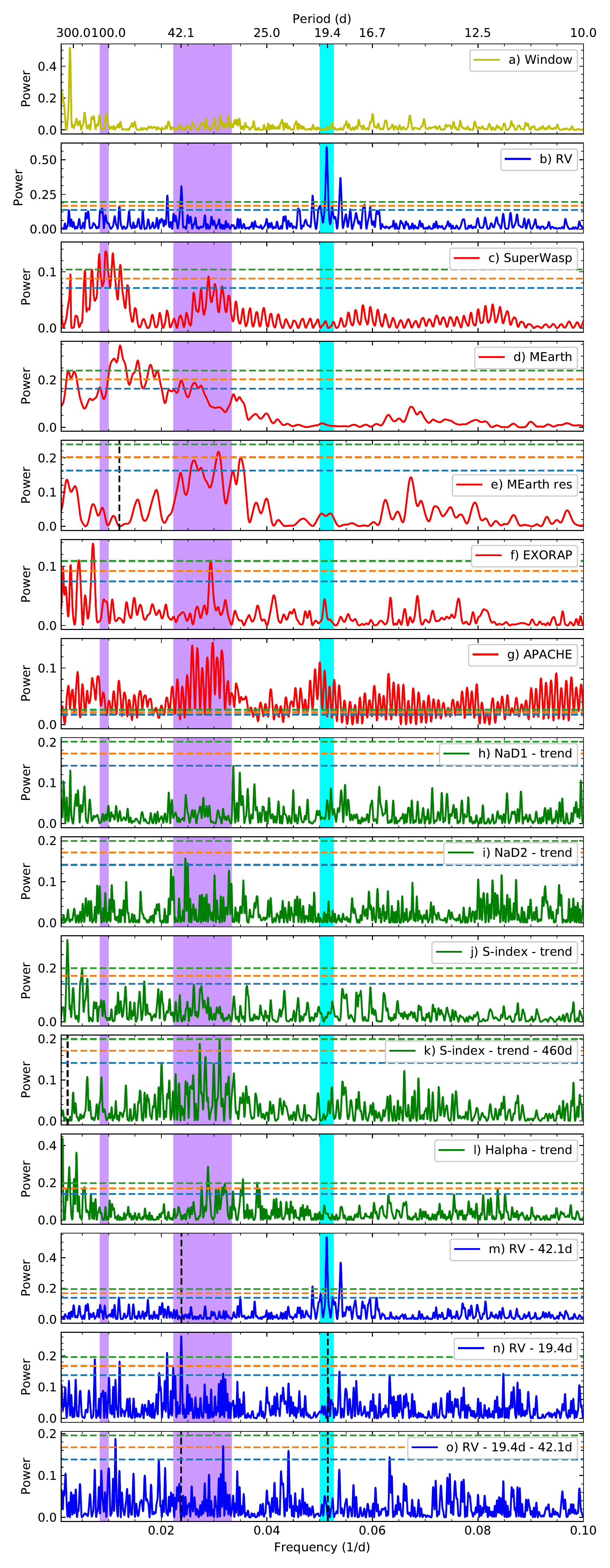}
\caption{\tiny{GLS periodograms for GJ~720~A RV data (blue solid lines), photospheric stellar activity (red solid lines), and spectroscopic stellar activity (green solid lines) in the frequency range 0.001--0.1\,$\rm d^{-1}$ (1000--10\,d in time range). In all the panels, the horizontal dashed lines indicate FAP levels of 10\%~(blue), 1\%~(orange), and 0.1\%~(green). The shadowed areas indicate the region where the RV highest peaks are found. {\sl a) and b) panels:} Spectral window (yellow line) and HARPS-N RVs (blue line), respectively. {\sl c) panel:} SuperWASP binned photometric data. {\sl d) and e) panels:} MEarth binned photometric data and the residuals after removing the highest peak found at $\sim$100\,d (black vertical dashed line). {\sl f) panel:} EXORAP $B$ filter. {\sl g) panel:} APACHE $V$ filter binned photometric data. {\sl h) through l) panels:} HARPS-N NaD1, NaD2, Ca~{\sc ii} H \& K (S-index), and H$\alpha$ HARPS-N spectroscopic activity indicators with a linear trend removed. {\sl m) panel:} HARPS-N RV residuals after removing the 42.1\,d signal, marked with a black vertical dashed line. {\sl n) panel:} HARPS-N RV residuals after removing the planet candidate signal at 19.5\,d. {\sl o) panel:} HARPS-N RV residuals after removing the 19.5 and 42.1\,d signals. All the activity indicators and the RV data show a significant, broad peak between 35--45\,d. This is likely associated with the rotation period of GJ~720~A.}}
\label{fig:GLS_activity_multiplot}
\end{figure}

\begin{figure*}[]
\centering
\includegraphics[width=0.67\columnwidth]{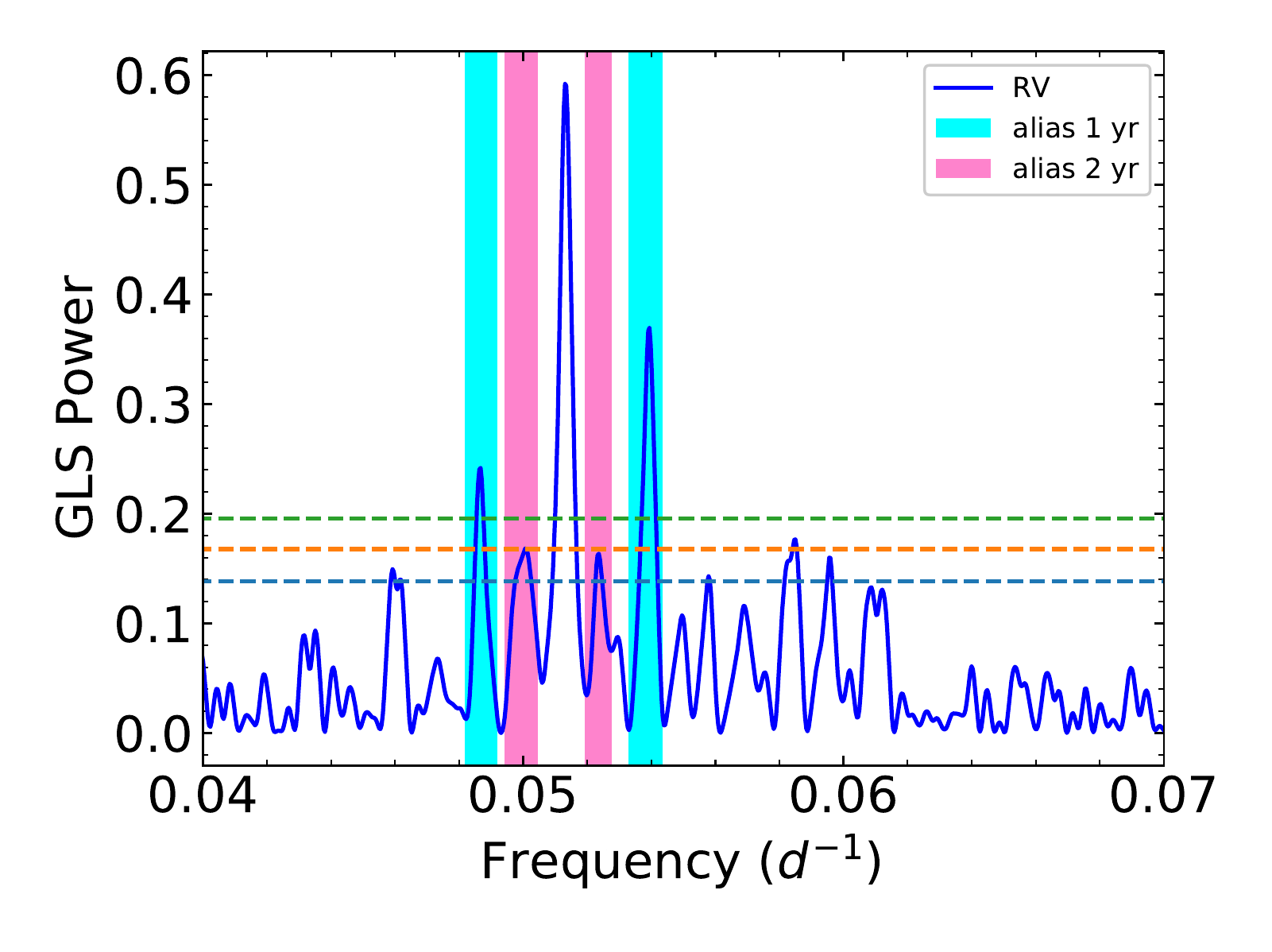}
\includegraphics[width=0.67\columnwidth]{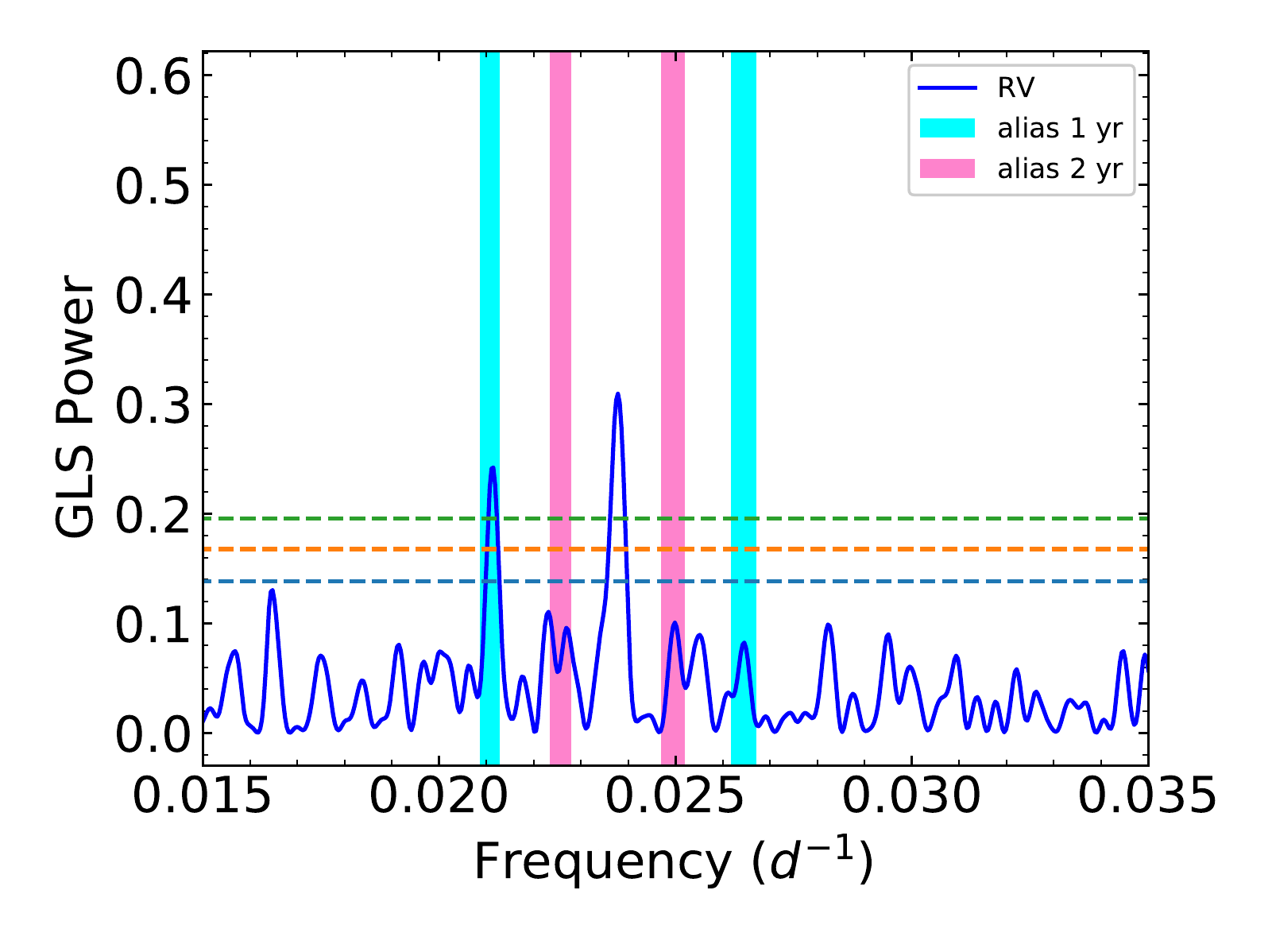}
\includegraphics[width=0.67\columnwidth]{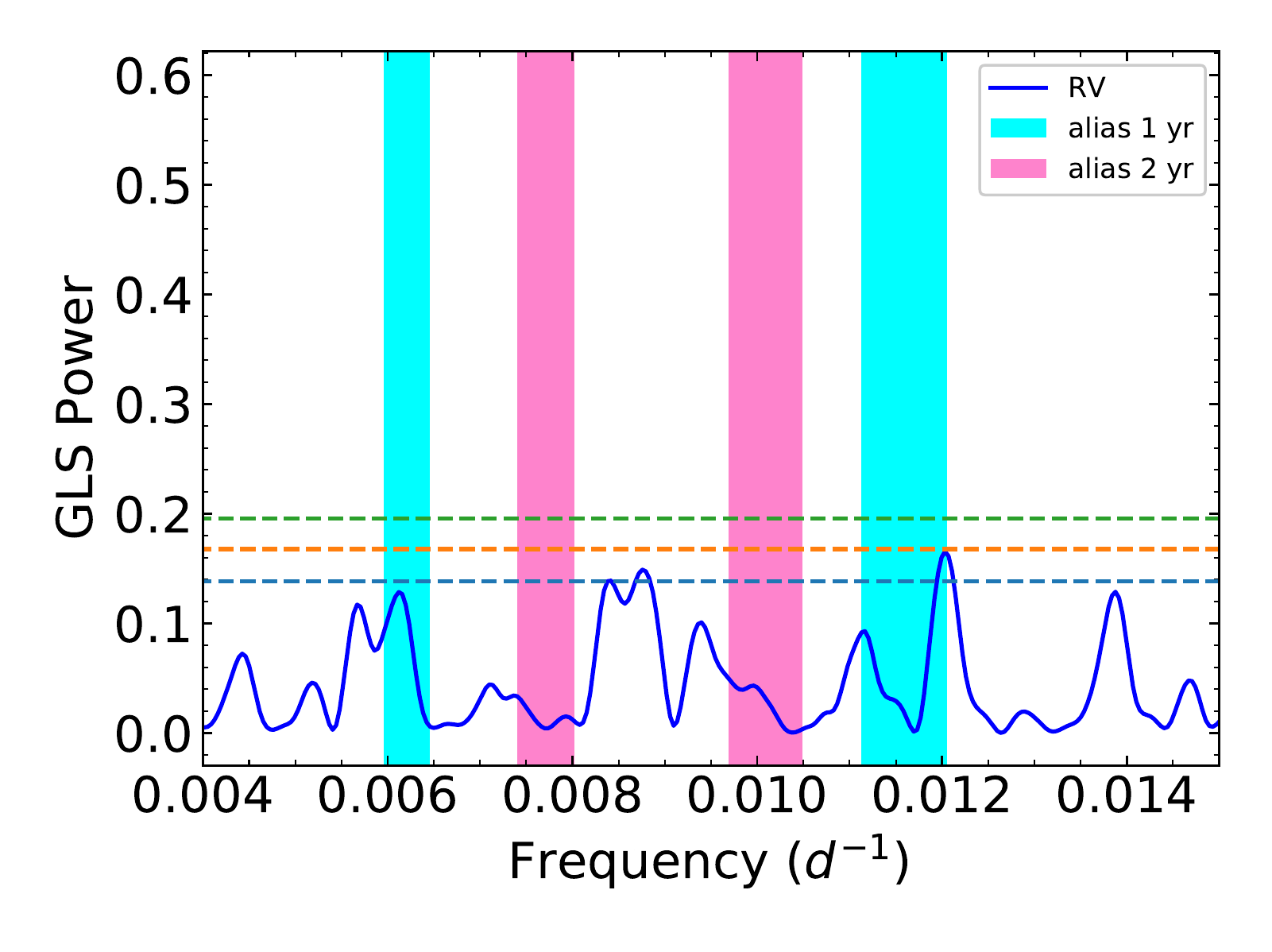}
\caption{Zoom-in of the GLS periodogram of the HARPS-N RV data of GJ~720~A (blue solid line) around the strongest signals at 19.5 (\textit{left panel}), 42.1 (\textit{center panel}), and 112.4 \,days (\textit{right panel}). The corresponding values in the frequency domain are 0.05128, 0.02375, and 0.0890\,$\rm d^{-1}$, respectively. Horizontal dashed lines indicate the different FAPs: 0.1\% (green), 1\% (orange) and 10\% (blue). The 1 and 2-sidereal-year aliases around each of the strongest signals are indicated with cian and purple vertical solid lines, respectively.}
\label{fig:gj720a_window}
\end{figure*}

\begin{figure*}[]
\centering
\begin{minipage}{0.48\linewidth}
\includegraphics[angle=0,scale=0.37]{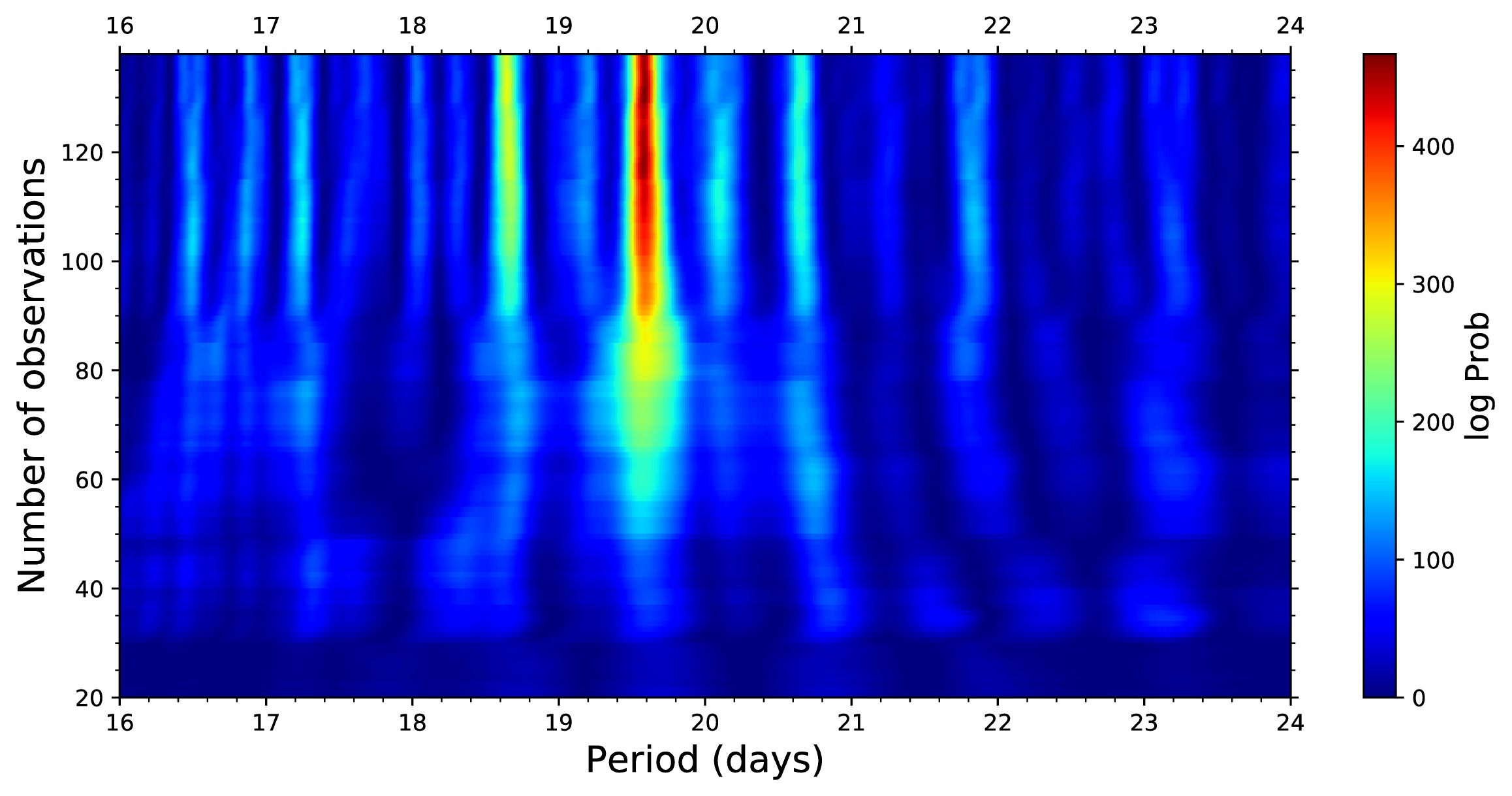}
\end{minipage}
\begin{minipage}{0.48\linewidth}
\includegraphics[angle=0,scale=0.37]{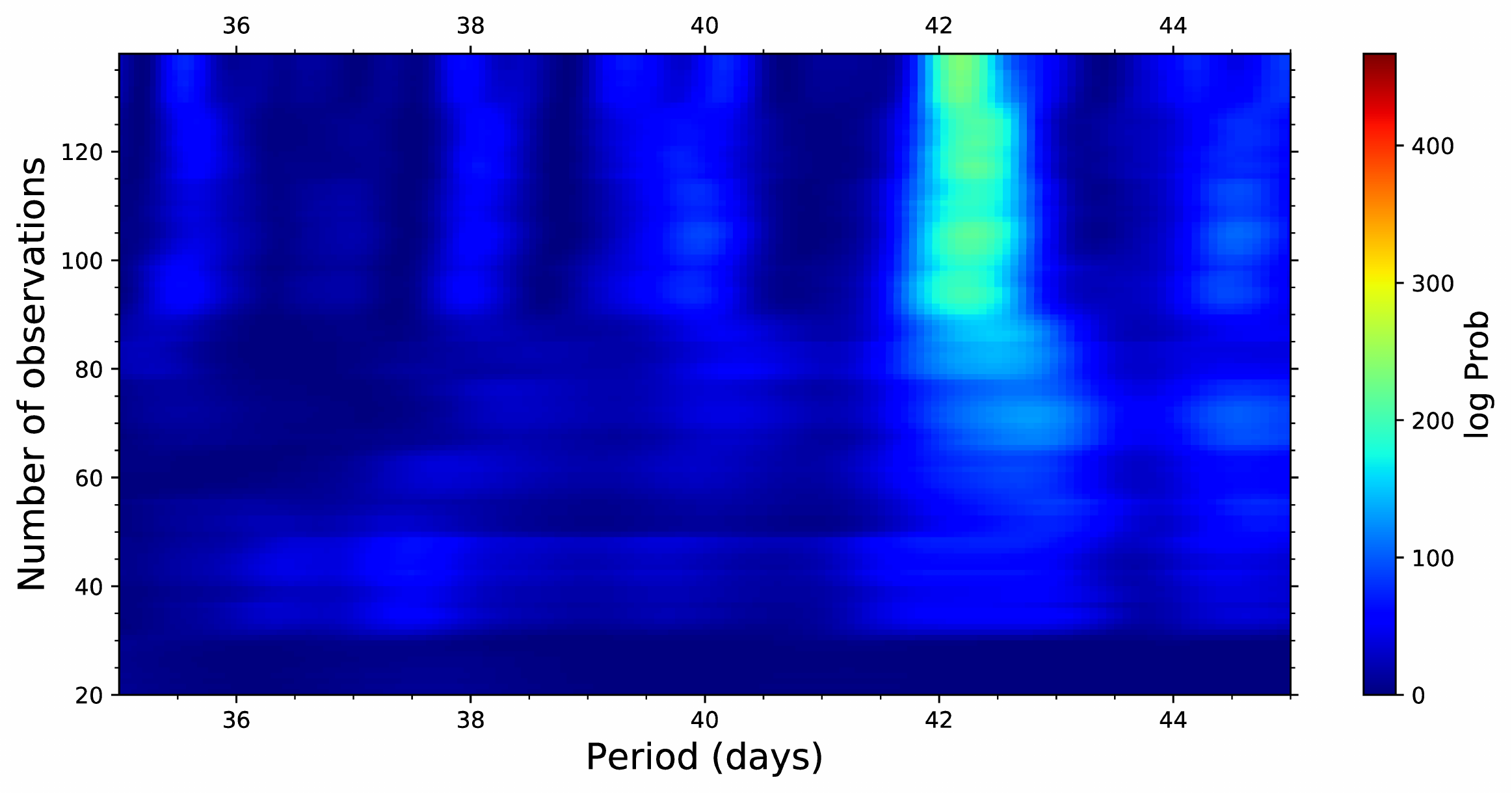}
\end{minipage}
\caption{Evolution of the s-BGLS periodogram of the HARPS-N RV data of GJ~720~A.
\textit{Left panel:}The most probable period (the reddest one) at 19.5\,days is clearly visible after around $\rm N_{obs}$ > 80. \textit{Right panel:} s-BGLS periodogram around 40\,d signal produced by the stellar variability.}
\label{M80_color_s-BGLS}
\end{figure*}
%----------------------------------

\subsection{HARPS-N radial velocities, pre-whitening}

The first step of the RV data analysis is the identification of significant periodic signals in the time series. The procedure was applied to the full RV data using the Generalized Lomb-Scargle (GLS) periodogram algorithm \citep{2009A&A...496..577Z}. We consider significant periods if the power is higher than a chosen false alarm probability \citep[FAP,][]{2009A&A...496..577Z} level of 10\% (notable), 1\% (prominent), and 0.1\%(significant). In Fig. \ref{fig:GLS_activity_multiplot} we included the GLS periodograms for the RVs (blue), the spectroscopic activity indicators (green) and the photometry (red) in the frequency range 0.001--0.1\,$\rm d^{-1}$ (1000--10\,d in time range). The window function of the HARPS-N RV data is depicted in the top panel of Fig. \ref{fig:GLS_activity_multiplot} (yellow line). The second panel of Fig. \ref{fig:GLS_activity_multiplot} reports the GLS periodogram for the HARPS-N RV dataset. There are several significant peaks higher than the 0.1\% FAP located at 19.5\,d (light blue shadowed area), $\sim$40\,d and $\sim$100\,d (purple shadowed areas). The most prominent peak is the one at 19.5\,d, that we will consider our most interesting signal from now on.

In what follows, we demonstrate that these three signals are not related to each other by an aliasing effect, which is typically caused by the gaps in the time coverage of the observations \citep[e.g.,][]{2010ApJ...722..937D}. To identify the presence of possible aliasing phenomena, the spectral window has to be considered. If peaks are seen in the window function, their corresponding aliases will be present in the RV periodograms as $f_{\rm alias}$ = $f_{\rm true} \pm m f_{\rm window}$ , where $m$ is an integer, $f_{\rm true}$ is the frequency identified in the RV periodogram and $f_{\rm window}$ the frequency from the window function \citep{1975Ap&SS..36..137D}. Typical aliases are those associated with the sidereal year, synodic month, sidereal day, and solar day. There are two significant peaks in the window function at the sidereal 1 and 2 years. Searching for these modulations around the principal peaks of the GLS periodogram of the RV data (19.5, $\sim$40, $\sim$100\,d), we found that secondary (lower) peaks around these three values are due to the aliases at 1 and 2 years of the spectral windows (Fig. \ref{fig:gj720a_window}). The three main signals in the RV time series are not related to each other by the observation sampling.

In order to verify whether the 19.5\,d signal was coherent over the whole observational time baseline, we produced the stacked Bayesian generalized Lomb-Scargle periodogram \citep[s-BGLS,][]{2015A&A...573A.101M} which computes the relative probability between peaks. Figure \ref{M80_color_s-BGLS} shows the s-BGLS periodogram of the HARPS-N RV data around 19.5\,d and also around the 40\,d activity signal. The s-BGLS showed a continuous increasing of the probability at 19.5\,d (left panel of Fig. \ref{M80_color_s-BGLS}) after around 90 observations and thereafter also the signal became narrower, as expected for a Keplerian signal. On the contrary the behaviour of the signal at $\sim$40\,d did not show such high probability and narrow structure. The first maximum of the probability for the $\sim$40\,d signal is produced after around 90 observations and thereafter decreased and increased again for some time. Also exact value of the $\sim$40\,d signal is erratically changing in time, following the increase of the number of observations. This behaviour is typical of an incoherent (in amplitude and phase) signal, as that due to the rotational modulation of a star. The coherence of the 19.5\,d period established above does not support its identification as the first harmonic of the $\sim$40\,d period, despite its close value, since it should vary accordingly.

The $\sim$40\,d signal that we could attribute to stellar activity, following the pre-whitening method, can be modeled with a sinusoidal curve of $42.1\pm0.1$\,d with a semi-amplitude of $3.02\pm0.44$\,m/s in the RV data. After its removal from the HARPS-N RV data (see $m$ panel of Figure \ref{fig:GLS_activity_multiplot}), we found an $rms$ of the residuals of 3.36\,m/s while the signal at $\sim$100\,d, that we could also relate to stellar activity disappears. The signal at 19.5\,d remains in the RV GLS periodogram of the residuals and its significance is still far above the FAP level 0.1\%. Now we subtract first the 19.5\,d signal (see $n$ panel of Figure \ref{fig:GLS_activity_multiplot}) to observe the behaviour of the $\sim$40\,d signal. The corresponding RV residuals still present the $\sim$40\,d signal with the same GLS power and therefore the same significance. When removing both contributions (19.4 and $\sim$40\,d peaks, see $o$ panel of Fig. \ref{fig:GLS_activity_multiplot}) no additional peaks above a 0.1\% FAP are present. Those remaining peaks with 1\% and 10\% are considered as result of the stellar noise.

%----------------------------------
\subsection{Spectroscopic stellar activity}
\label{subsec:Spectroscopic stellar activity}

From the previous section we identified three principal periodic signals (19.5, $\sim$40, $\sim$100\,d) in the GLS periodogram of the RV data being mandatory the knowledge of their origin. The M dwarfs are on average more active than solar-like stars \citep{1997A&A...327.1114L,2005ApJ...621..398O} and therefore the effects of the stellar activity (chromospheric or photospheric) can be confused with planetary signals or even hide them. In order to disentangle the effects of activity from true RV variations we analysed two commonly used chromospheric activity indicators based on measurements of the H$\alpha$ 6562.82\,\AA\,and Ca~{\sc ii} H \& K 3933.7, 3968.5\,\AA\,lines (S-index) and also the sodium doublet (NaD1 and NaD2) provided by TERRA pipeline. 

The associated GLS periodograms of all analyzed activity indicators are shown in Fig. \ref{fig:GLS_activity_multiplot} (green panels). We also include the FAP levels and the shadowed colored areas that correspond with the principal peaks (19.5, $\sim$40, $\sim$100\,d) identified from the GLS of the RVs. In particular, the NaD1, NaD2, H$\alpha$, and S-index indices present a clear trend in their time-series (green panels of Fig. \ref{M80_rv_time_TERRA}). Analyzing the GLS periodogram we observed this trend as a long-term variability of period $>$\,350\,d. For this reason we detrended the NaD1, NaD2, S-index, and H$\alpha$ time series subtracting a straight line before the analysis of the activity indicators. With the activity indices detrended, all of them present some kind of activity centered around 40\,d. 

In the H$\alpha$ case, the $\sim$40\,d signal is also present (FAP level $<$0.1\%) but it is not the highest one. While for the S-index case, after data detrending, the GLS periodogram shows another long-term variation at 450\,d that we also removed (black vertical line in the $k$ panel of Fig. \ref{fig:GLS_activity_multiplot}). After that we obtained a clear, unique, and significant period at $32.21\pm0.05$\,days. The same technique was used to obtain the rotation period of the star published by \cite{2018A&A...612A..89S}. The authors derived the stellar rotation at $P_{\rm rot}=34.5 \pm 4.7$\,days with the HARPS-N spectra available to date.

Taking advantage of the available 138 HARPS-N RV data points, we modeled the stellar variability of the S-index (original data that include the trend) using a Gaussian process (GP) regression, which is a more sophisticated method than the pre-whitening one. The fit was performed using {\tt{juliet}} \citep{2019MNRAS.490.2262E} that used {\tt radvel} \citep{2018PASP..130d4504F} to model Keplerian RV signals, and {\tt george} \citep{2015ITPAM..38..252A} to model the stellar variability with GP. We used an exp-sin-squared kernel multiplied by a squared-exponential kernel, which is included as a default kernel within {\tt juliet}. This kernel has the form:

\begin{equation}
\label{eq:exp-sin-sqr_kernel}
k(\tau) = \sigma_{GP}^2 \exp \left( -\alpha_{GP} \tau^2 - \Gamma \sin^2 \left( \frac{\pi \tau}{P_{rot}} \right)  \right)
\end{equation}

where $\sigma_{GP}$ is the amplitude of the GP component given in the same units of the data, $\Gamma$ is the amplitude of the GP sine-squared component and is dimensionless, $\alpha$ is the inverse squared length-scale of the GP exponential component given in $\rm d^{-2}$, $\rm P_{rot}$ the period of the GP quasi-periodic component given in $\rm d$, and $\tau$ is the time lag. 

\begin{figure}[t]
\includegraphics[width=\columnwidth]{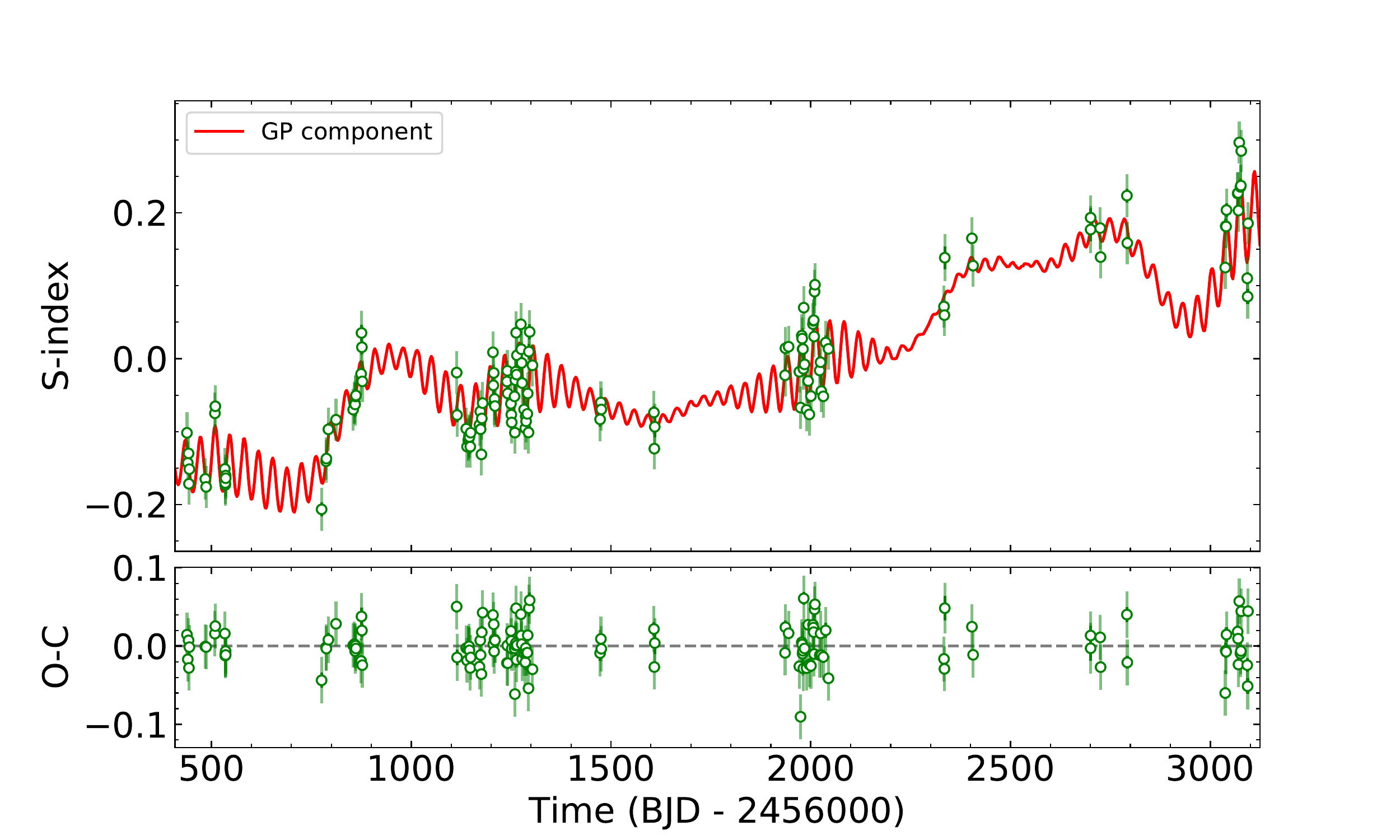}
\caption{GJ~720~A original S-index data together with the best model and the residuals. The fitted GP model (red line) corresponds to an exp-sin-squared kernel multiplied with a squared-exponential kernel modeling the stellar variability. The error bars (color green) include the jitter (light green color) taken into account.}
\label{fig:sindex_GP_model}
\end{figure}

\begin{figure}[t]
\includegraphics[width=\columnwidth]{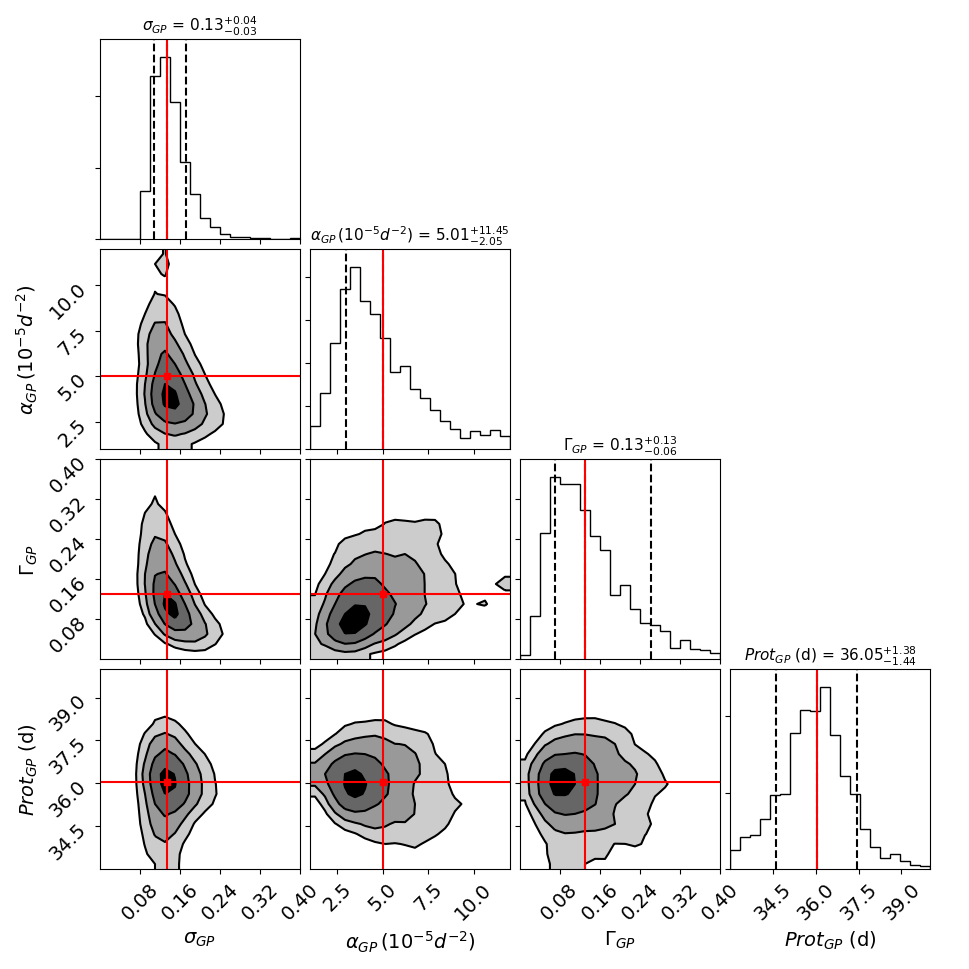}
\caption{Posterior distributions for the parameters that model the stellar variability using the S-index activity indicator. The vertical dashed lines indicate the 16, 50, and 84\% quantiles of the fitted parameters; this corresponds to 1$\sigma$ uncertainty. The red line shows the median value of each fitted parameter. }
\label{fig:corner_gp_sindex}
\end{figure}

All parameters were set with wide priors and in particular the $P_{\rm rot}$ was set free to vary in the range 1--500\,d (see Table \ref{tab:gj720a_priors_details_sindex}). The found stellar rotation period, using the GP technique, corresponds with $P_{\rm rot}=36.05^{+1.39}_{-1.44}$\,d with a length-scale median value of 141.28\,d ($\alpha_{GP}$=5.01$\times 10^{-5}$\,$\rm d^{-2}$). Figure~\ref{fig:sindex_GP_model} shows the GP model that best fits the S-index data while in Fig.~\ref{fig:corner_gp_sindex} we show the posterior distributions for the GP parameters of the model.

Finally we note that in all the studied chromospheric activity indicators, no significant signals were identified around 19.5\,d (light blue colored area in Fig.~\ref{fig:GLS_activity_multiplot}) and therefore the hypothesis of a planet candidate at 19.5\,d is now more reliable. All the chromospheric activity indicators show significant peaks in the range 35--45\,d (purple colored area). In particular the S-index GP regression analysis established the stellar rotation period at $P_{\rm rot}=36.05^{+1.39}_{-1.44}$\,d. We conclude that all the possible signals identified in the RV GLS periodogram around this value could be related to stellar activity effects or to the stellar rotation period and therefore their Keplerian nature can be ruled out. The different activity indicators used here track different features in the stellar atmosphere and considering the differential rotation of the star it is plausible that they do not yield exactly the same periods found in RV data. In fact the closest period that we could identify in the RV GLS periodogram is 42.1\,d.

%----------------------------------
\subsection{Photometric stellar activity}

The M dwarfs have inhomogeneities on their surface that rotate with the star. These inhomogeneities cause RV variations due to the distortion of the spectral line profile and can be misinterpreted as signals of Keplerian nature. Those inhomogeneities affect also the photometric measurements that is why a photospheric analysis is crucial in order to avoid unwanted signals as planetary candidates.

\paragraph{Super-WASP and MEarth.} 
We analyzed the GLS periodograms of Super-WASP light curve and MEarth differential light curves ($c$ and $d$ panels of Fig. \ref{fig:GLS_activity_multiplot}). There are several observing seasons for GJ~720~A within the MEarth survey, thus we studied the differential light curves separately per observing campaign and all the seasons together. Due to the huge number of data points obtained with MEarth we also analyzed the binned differential light curve data. The different analysis for the MEarth available seasons yielded that the variability of the star is better seen in the binned 2008--2010 season, the rest of them are not shown here for clarity. No obvious, significant peak can be extracted from the GLS periodograms of the Super-WASP and MEarth light curves. In both cases, there is no clear and narrow peak which can be attributed to the stellar rotation period. The Super-WASP GLS periodogram of the binned data shows the two highest peaks around 40 and 100\,d (panel $c$ of Fig. \ref{fig:GLS_activity_multiplot}). The MEarth GLS periodogram (using the binned data of the season 2008-2010) shows the highest peak around 90\,d ($d$ panel of Fig. \ref{fig:GLS_activity_multiplot}) and, after subtracting its contribution (black vertical line of $e$ panel of Fig. \ref{fig:GLS_activity_multiplot}), the highest period moves towards $\sim$40\,d. 
The periodicities showed in the GLS light curves periodograms agree with those found in the GLS analysis of both RV and spectroscopic activity indicators (light blue and purple colored areas in Fig. \ref{fig:GLS_activity_multiplot}). No photospheric or spectroscopic activity signal is detected in the region of interest around 19\,d.

\paragraph{EXORAP.} We first analyze the four differential light curves (one for each band) using the GLS periodogram. As the observed photometry shows long term trends, we pre-whiten the light curves by subtracting from each data series the corresponding third-order polynomial best-fit. 
The periodogram of the pre-whitened $B$ light curve (see panel $f$ of Fig.~\ref{fig:GLS_activity_multiplot}) shows two peaks with FAP$<$1\% at $\sim$34 and $\sim$140\,days. In the $V$, $R$ and $I$ bands we do not detect any signal more significant than 5\% and therefore we do not show them here for clarity. These results suggest a scenario where the photometric variability is due to the effects of an irregularly spotted stellar surface coupled with stellar rotation. This is consistent with the fact that the activity signal is stronger at bluer wavelengths, where the contrast between photosphere and cool spots is larger.

\paragraph{APACHE.} \cite{2020MNRAS.491.5216G} has recently published the GJ~720~A rotation period at 33.6\,days using the APACHE differential photometric observations. Using the same APACHE binned photometric data we computed the GLS periodogram here (panel $g$ of Fig.~\ref{fig:GLS_activity_multiplot}) corroborating that the highest peak value corresponds to that published. But we consider that such value can be regarded as an approximate rotation period of the star because of the presence of other nearby peaks (e.g., $\sim$37\,days) with a comparable significance. Folding in phase the APACHE light curve at 33.6\,d, the amplitude is relatively small (2.5$\pm$0.2\,mmag) compared with the $rms$ of the data (5.5\,mmag) and the mean weighted internal errors (3.9\,mmag), which explains why this signal can not be clearly identified in the GLS periodogram. Analyzing the different photometric epochs separately we also found the $\sim$33 and $\sim$37-d signals as the highest ones in the GLS periodogram of the first and fourth APACHE epochs.

\begin{figure*}[!tb]
\centering
\includegraphics[width=2.1\columnwidth]{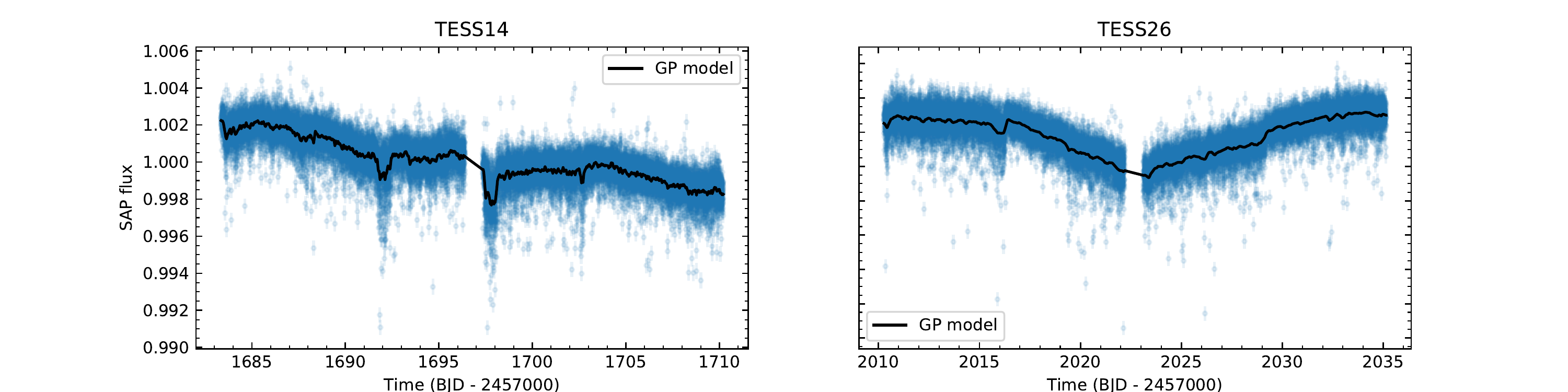}
\includegraphics[width=2.1\columnwidth]{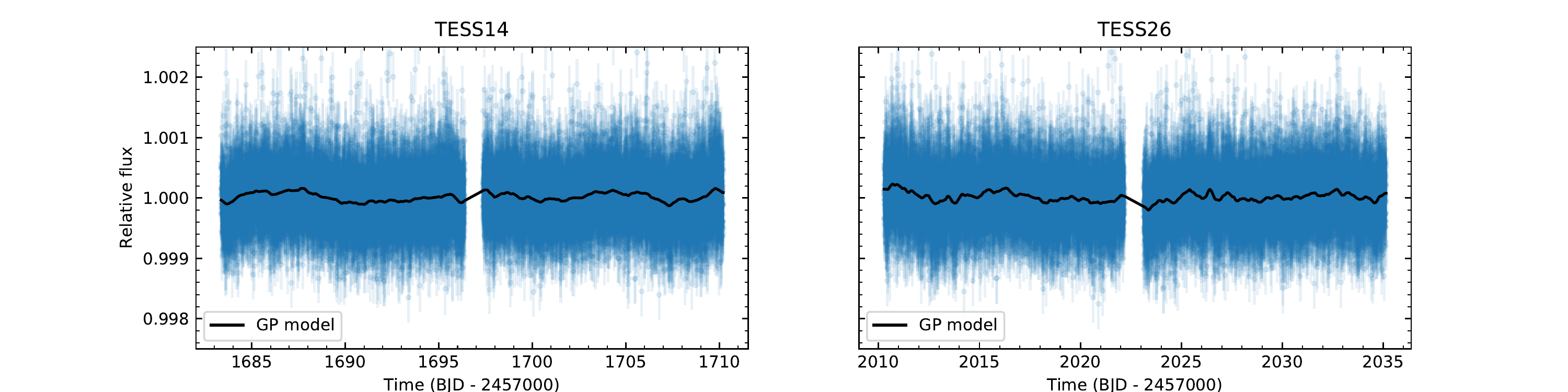}
\caption{\textit{Top panel:} GJ~720~A \textit{TESS} SAP fluxes (blue points) for the two sectors with the best stellar activity model (black line). \textit{Bottom panel:} \textit{TESS} PDC fluxes (blue points) with the best GP model fitted (black line).}
\label{fig:tess_detrend}
\end{figure*}

\paragraph{\textit{TESS}.} We looked at the \textit{TESS} light curve using the SAP fluxes to model the stellar activity signatures in order to find a possible stellar rotational period and we used the PDC fluxes for transit searches. The two sectors \textit{TESS} light curves were analyzed at the same time using a quasi-periodic kernel (QPK) introduced by \cite{2017AJ....154..220F} of the form

\begin{equation}
\label{eq:GP}
k_{i,j}(\tau) = \frac{B}{2+C} e^{-\tau/L} \left[ \cos \left( \frac{2\pi \tau}{P_{rot}} \right) + (1 + C) \right],
\end{equation} 

where $\tau= \abs{t_{i}-t_{j}}$ is the time-lag, $B$ and $C$ define the amplitude of the GP, $L$ is a timescale for the amplitude-modulation of the GP, and $P_{\rm rot}$ is the period of the quasi-periodic modulations. 
The {\tt juliet} lightcurve models include a dilution factor ($D_i$) which allows to account for possible contaminating sources in the aperture that might produce a smaller transit depth than the real one. Also the model takes into account the relative out-of-transit target flux ($M_i$) which is a multiplicative term and not an additive offset. For the transit modeling, {\tt juliet} uses the {\tt batman} package \citep{2015PASP..127.1161K}. The limb-darkening effect was taken into account with $q_1$ and $q_2$ coefficients, as defined by \cite{2013MNRAS.435.2152K}, and a quadratic law. Figure \ref{fig:tess_detrend} shows the SAP (top panels) and PDC data (bottom panels) for the two \textit{TESS} sectors with the best GP model found. In our case the median value of the posterior distribution for the rotational period for the two \textit{TESS} different sectors using the SAP fluxes is $49.8^{+7.3}_{-11.5}$ and $28.9^{+16.0}_{-7.1}$\,d, respectively. While using the PDC fluxes the median values of the rotational period for the different sectors is $33.7^{+11.0}_{-15.9}$ and $34.8^{+10.1}_{-13.8}$\,d, respectively. The corresponding error bars in both cases (SAP and PDC fluxes) are slightly high which suggests that there is not a precise determination of the rotation period for GJ~720~A star when using the \textit{TESS} photometric data. \textit{TESS} time series are shorter than the period we are looking for. Therefore, searching for the rotation period with \textit{TESS} data is not providing precise results. But the obtained values agree with those obtained before when analyzing the rest of spectroscopic and photometric activity indicators. 

A GP model can be also used in order to detrend the \textit{TESS} light curves before the search of possible transit features. Therefore using the optimized PDC fluxes and the corresponding previous GP model fit to detrend the light curve, we proceed to search transits. In a first approach searching possible transits we set a wide uniform prior, 1--25\,d, for the planetary signal. In a second approach we took advantage from the times of the inferior conjuctions as derived from the RV curve, in order to estimate the expected times of the transits and to look specifically at those times in the \textit{TESS} light curves. In both approaches no transiting planets for GJ~720~A were found.

We can conclude after the chromospheric and photospheric analysis that all activity indicators show a significant but broad peak, always in the range of 35--45\,d. Therefore, this range can be associated with stellar active regions probably at different latitudes on a differentially rotating star. In fact the GP analysis with the S-index revealed the stellar rotation period to be $P_{\rm rot}=36.05^{+1.39}_{-1.44}$\,d. While the other identified signal by the activity indicators (also seen in RV data) at around 100\,d is more likely related to the life cycle of the active regions, as explained by \cite{2017A&A...598A..28S} where was established that the active regions could persist some stellar rotations. Due to the complex mechanism that the differential rotation can exhibit on M dwarf stars due to their convective layers, we could not find a narrow signal that determines a precise value for the rotation period in each one of the analyzed activity indicators.

%----------------------------------
\section{Gaussian process regression}
\label{sec:GP}

The impact that the stellar activity effects can induce in the RVs could be different as a function of the stellar magnetic phenomena (e.g., evolving spot configurations among others). Each target star can have a specfic behaviour for accounting the effects of its stellar variability (e.g., rotating spots, faculae) and the flexibility of the GP algorithms makes their use essential in order to reproduce the stellar phenomena \citep{2020arXiv201201862P}. However the diversity of the mathematical GP kernels associated with true physical phenomena has not been well evaluated to date. Therefore it could be possible that the stellar activity of a specific target could be explicitly better reproduced by one kernel than by another. We decide in this section to test two of the commonly used kernels (exp-sin-squared and QP) to reproduce the stellar variability in order to obtain a robust result of the planet parameters and to know the goodness of the kernels for this specific target.

\subsection{Exp-sin-squared kernel, {\tt juliet}}

\begin{table*}[!t]
\begin{small}
\centering
\caption{Comparison of different solutions for GJ~720~A using {\tt juliet}.}
\label{tab:gj720a_model_comparison}
\begin{tabular}{l c c c c}

\hline
\hline
\noalign{\smallskip}
Model$^{(1)}$ & Params. & Description & $\ln \mathcal{L}$ & BIC$^{(2)}$\\
\noalign{\smallskip}	
\hline	
\noalign{\smallskip}
BM &  $\gamma_0$  & RV offset & -394 & 798\\
								 & $\sigma$ &  RV jitter & &\\
\noalign{\smallskip}	
\hline	
\noalign{\smallskip}
								 
BM+GP				    & $\sigma_{\rm GP}$   & Amplitude of GP & -315 & 660\\
							    & $\alpha_{\rm GP}$ & Inverse (squared) length-scale &  &\\
							    & $\Gamma_{\rm GP}$ & Amplitude of the sine-part & &\\
							    & $P_{\rm {rot, GP}}$ & Period of GP & &\\
							    
BM+GP+LT     		& slope & Slope RV data & -312 & 665 \\
								& inter. & Intercept coeff. & & \\
\noalign{\smallskip}	
\hline	
\noalign{\smallskip}
							    
BM+1pl      			    & $P$, $T_0$, $e$, $\omega$, $K$ & Planet params  & -346 & 718 \\

\noalign{\smallskip}	
\hline	
\noalign{\smallskip}

BM+GP+1pl			& ...  & ... & -298 & 650 \\

BM+GP+1pl+LT  & ...  & ... & -295 & 655 \\

\noalign{\smallskip}	
\hline	
\noalign{\smallskip}

BM+2pl      			 &   ... & ...  & -319  & 698 \\
BM+GP+2pl			& ...  & ... & -299 & 667\\

\noalign{\smallskip}	
\hline	
\noalign{\smallskip}
\end{tabular}
\tablefoot{$^{(1)}$ BM stands for the base model containing RV offsets and jitter. GP corresponds to a quasi-periodic GP kernel and 1pl means one planet model.$^{(2)}$ BIC corresponds to the Bayesian Information Criterion.  }
\end{small}
\end{table*}

We used a different approach to analyze the HARPS-N RV data in search for planet candidates using {\tt juliet}. This technique foresees a simultaneous fit of the stellar activity and the planetary signals. The stellar activity contamination has a significant effect on the derived planetary parameters and to model both signals (stellar and Keplerian) at the same time through the GP regression is essential. The GP kernel implemented here was the exp-sin-squared kernel, previously used and described by the equation \ref{eq:exp-sin-sqr_kernel}. The different implemented models will be judged on the basis of the Bayesian Information Criterion \citep[BIC,][]{2007MNRAS.377L..74L}. It is based on the log-evidences (ln\,$\mathcal{L}$) introducing a penalty term for the parameters used in the model avoiding an overfitting of the data. The BIC value can be described by the equation:

\begin{equation}
{\rm BIC} = k\, \ln(n) - 2\,\ln(\mathcal{L}),
\end{equation}

where $n$ is the number of data points, $k$ the number of free parameters to be estimated, and $\mathcal{L}$ the maximized value of the likelihood function of the model \citep[see][for details]{2019MNRAS.490.2262E}. The models are better when the BIC value is lower. The $\Delta$BIC threshold for considering one model more probable than another with the BIC criterion correspond to: \textit{i)} $\Delta$BIC = 0--2 not worth more than a bare mention, \textit{ii)} $\Delta$BIC = 2--6 is positive, \textit{iii)} $\Delta$BIC = 6--10 is a strong evidence of preferred, and \textit{vi)} $\Delta$BIC > 10 the model is very strongly preferred.

\begin{figure}[!tb]
\centering
\includegraphics[width=0.5\textwidth]{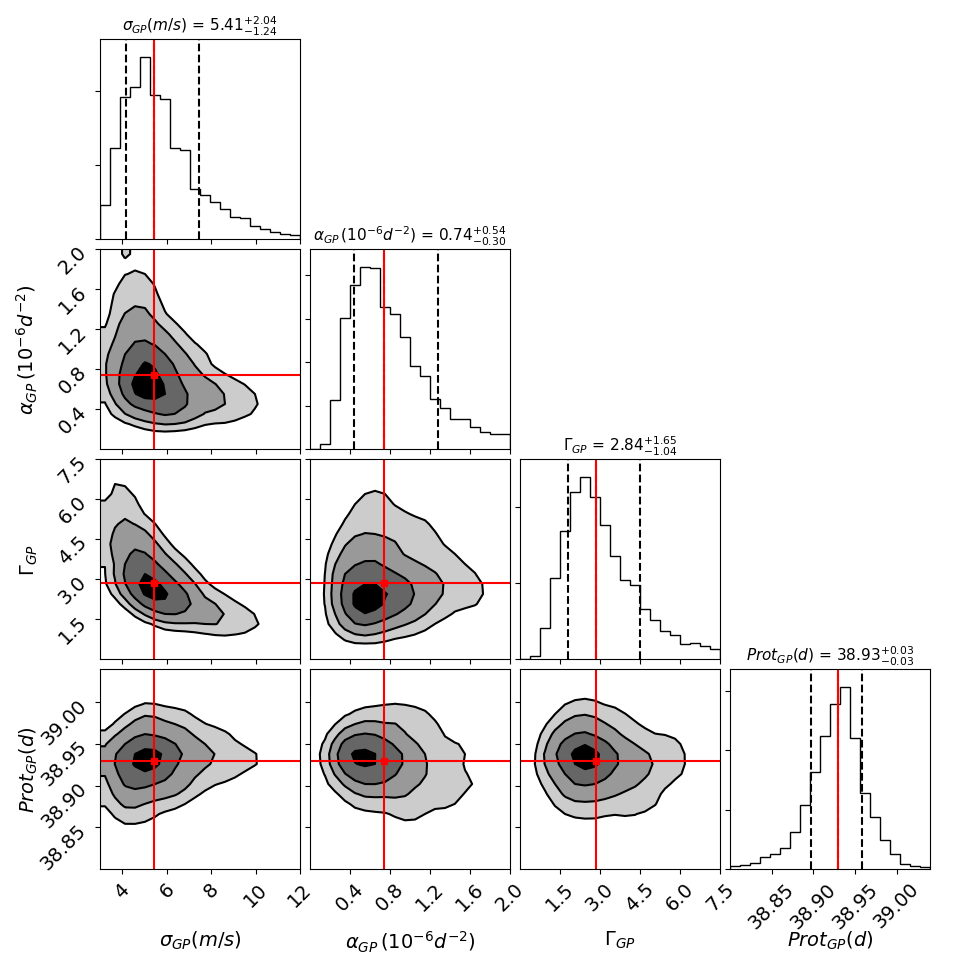}
\caption{Posterior distributions for the parameters of the GJ~720~A HARPS-N RV data fitting a base model plus an exp-sin-squared GP kernel setting wide prior values for the GP parameters (e.g., $P_{\rm rot} = \mathcal{U}$(1, 1000)\,d).
}
\label{fig:gj720a_GP_BM+GP}
\end{figure}

In a first approach we used a base model (BM, only includes individual offset and RV jitter), plus a GP kernel that we used to model the stellar variations observed in the RV data and when analyzing the stellar activity indicators, already discussed in previous sections. We consider uniform distributions for the $\sigma_{GP}$ and $P_{\rm rot}$ parameters of the exp-sin-squared GP kernel with the priors set to $\mathcal{U}$(0, 15)\,m/s and $\mathcal{U}$(1, 1000)\,d, respectively. For $\alpha_{GP}$ and $\Gamma_{GP}$ GP parameters we followed a Jeffrey's distributions \citep{1946RSPSA.186..453J} setting the prior values as $\mathcal{J}$($10^{-20}$,\,$10^{4}$)\,$\rm d^{-2}$ and $\mathcal{J}$(0.01,\,100), respectively. We set wide prior values for all these four GP parameters in order to test the stellar variability present in the RV data. The expected result with these wide priors, especially with the $P_{\rm rot}$ prior that includes both highest signals (19.5 and $\sim$40\,d) of the RV data, was that the BM+GP model would identify some of them as the $P_{\rm rot}$ value. Surprisingly, the value found for the GP $P_{\rm rot}$ parameter using the HARPS-N RV data corresponds to $38.9\pm0.05$\,d which is close to the $P_{\rm rot}$ value found from the S-index ($36.05^{+1.38}_{-1.44}$\,d), and it is placed in the same activity region (30--50\,d) determined from the other activity indicators. The final posterior distributions obtained with a base model plus a GP kernel are shown in Fig. \ref{fig:gj720a_GP_BM+GP}. The length-scale of the GP signal corresponds to a value of $\sim$1000\,d ($\alpha_{GP}$=0.74$\times10^{-6}$\,$\rm d^{-2}$) while the obtained amplitude of the GP model is close to the $rms$ of the RV data. In order to test if including a linear trend (LT) helps the improvement of the final model we also fitted the RV data considering the BM plus a GP kernel plus a LT.

In the second approach we modeled the BM plus one planet. In order to explore a blind search of the planet period, without taking into account the recovered information from the GLS periodogram of the RV data, we set a wide uniform prior value for the period of the planet, $\mathcal{U}_{P1}$(1, 50)\,d. This prior includes the highest signals of the RV GLS periodogram (19.5 and $\sim$40\,d). The rest of the priors were set as follows: uniform distributions for the eccentricity $\mathcal{U}_{ecc}$(0, 0.8), the argument of periastron $\mathcal{U}_{\omega}$(0, 360)\,deg, the semi-amplitude $\mathcal{U}_K$(0, 10)\,m/s, and the time of periastron passage $\mathcal{U}_{t_0}$(0, 50)\,d with respect to the time reference 2,456,000. The period found for the planet candidate was $\rm P1 = 19.484^{+0.007}_{-0.006}$\,d with a time of periastron passage at $t_0 = 6.05^{+0.53}_{-0.43}$ (BJD - 2,456,400).

In a third approach, our model was composed of the BM, the GP kernel modeling the activity with a uniform distribution of $\mathcal{U}$(30, 50)\,d for the $P_{\rm rot}$ GP parameter plus one planet with a normal distribution for the planet orbital period of $\mathcal{N}$(19.5, 0.5)\,d. The same model as the previous one but considering the $P_{\rm rot}$ GP parameter as an open uniform distribution, $\mathcal{U}$(1, 1000)\,d, was also considered. The planet and GP parameters obtained for these two models were compatible within 1$\sigma$ error bars. Another test was also considered including a linear trend in the model. We note that the narrow prior adopted here for the planet, $\mathcal{N}_{P1}$(19.5, 0.5)\,d, is larger than its final posterior distribution and it is also larger than the final posterior distribution obtained following the second approach where we set a wide uniform prior for the planet ($\mathcal{U}_{P1}$(1, 50)\,d). Therefore the 19.5\,d signal is well characterized and in what follows the assumption of this narrow prior ($\mathcal{N}_{P1}$(19.5, 0.5)\,d) is justified.

In the fourth and last approach we considered the two highest RV signals (19.5 and $\sim$40\,d) of Keplerian origin. The first model took into account the BM plus two planets. The orbital planetary periods were considered with normal distributions $\mathcal{N}_{P1}$(19.5, 0.5)\,d and $\mathcal{N}_{P2}$(42.1, 0.5)\,d. The second model took into account the BM plus the same two Keplerian signals (19.5 and $\sim$40\,d) plus a GP kernel with a wide uniform prior for the $P_{\rm rot}$ in the range 1--1000\,d.

The comparison between the different models together with their log-likelihood and BIC values are summarized in Table \ref{tab:gj720a_model_comparison}. After that and following the BIC criterion, the “best” model corresponds to the third approach where together with the base model, an exp-sin-squared GP kernel simulating the stellar contribution and a Keplerian orbit for the planet candidate at 19.5\,d were employed. The $\Delta$BIC between the model BM+GP and the model BM+GP+1 planet has a value of 10 for the latter to be strongly preferred. The BIC criterion retains more likely the BM+GP+1 planet model than the star-only model and therefore supports the planetary hypothesis at 19.5\,d. The distributions and the corresponding priors used to fit the “best” model are listed in detail in Table \ref{tab:gj720a_priors_details}. The $P_{\rm rot}$ value derived from the “best” RV+GP model (35.23$\pm$0.11\,d) is compatible, within 1\,$\sigma$, with the $P_{\rm rot}$ obtained from the S-index GP analysis ($36.05^{+1.38}_{-1.44}$\,d). The amplitude and the length-scale of the GP are $4.44^{+2.32}_{-1.36}$\,m/s and 842\,d ($\alpha_{GP}$=1.41$\times10^{-6}$\,$\rm d^{-2}$), respectively.

\begin{figure}[!tb]
\centering
\includegraphics[width=0.5\textwidth]{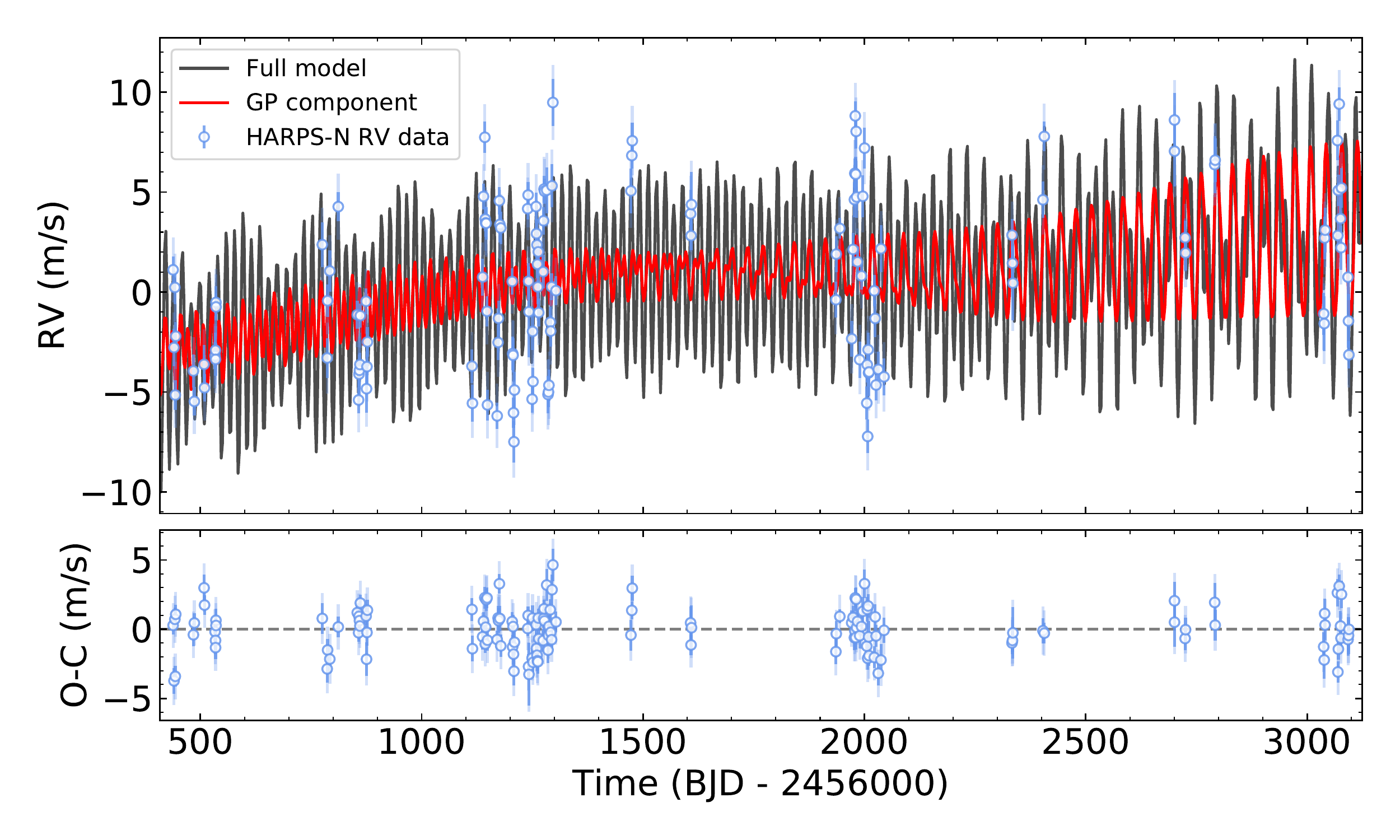}
\includegraphics[width=0.5\textwidth]{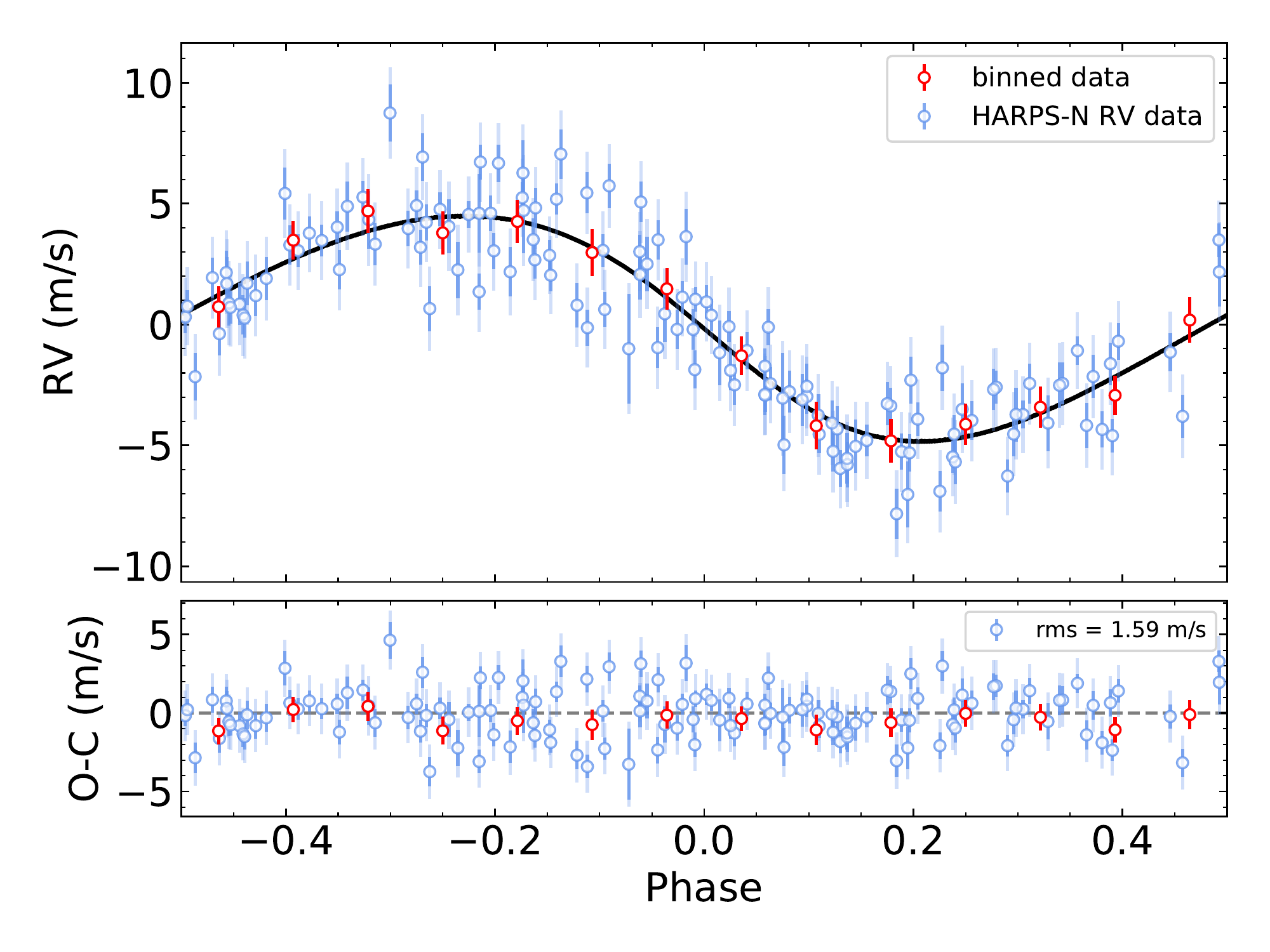}
\caption{ \textit{Top panel:} RV time series (blue dots) together with the “best” model and the residuals. The fitted model (black line) corresponds to the base model plus an exp-sin-squared GP kernel that models the stellar activity at $\rm P_{rot} = 35.23\pm0.11$\,d, and the planetary signal at $19.466\pm 0.005$\,d. The GP contribution is shown with red colored line. The error bars (color blue) include the RV jitter (light blue color) taken into account. \textit{Bottom panel:} RVs (blue dots) folded in phase (the base model and the stellar activity were removed) with the orbital period of the planet and its residuals. The best Keplerian solution (black line) has an RV amplitude of $4.72 \pm 0.27$\,m$\rm s^{-1}$. The red dots correspond to the binned data. The $rms$ of the residuals is 1.59\,m$\rm s^{-1}$. 
}
\label{fig:gj720a_GP_plots}
\end{figure}

\begin{figure}[!tb]
\centering
\includegraphics[width=0.5\textwidth]{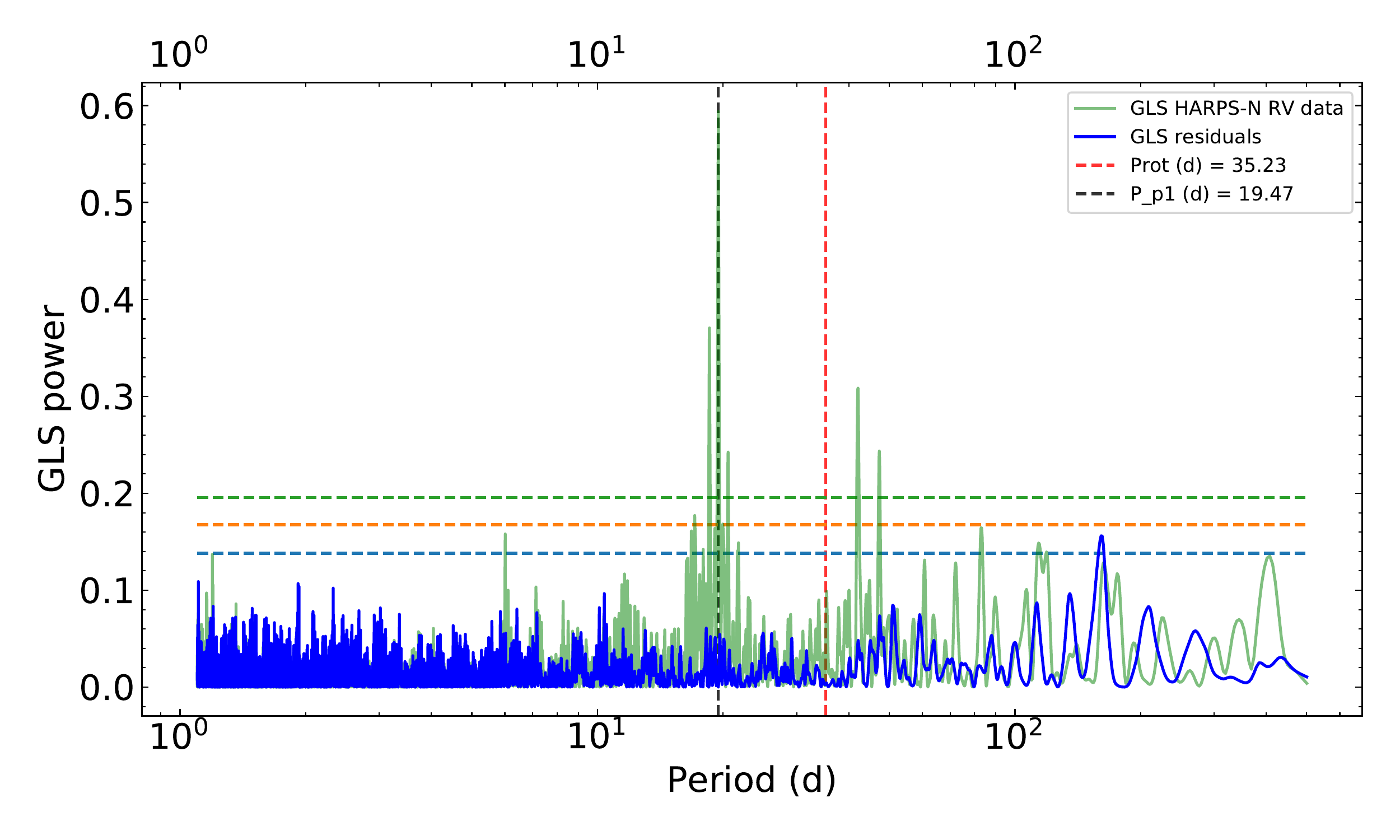}
\caption{GLS periodograms of the RV data (green line) and the GLS of the residuals (blue line) after substracting the “best” model that models the RV stellar activity ($\rm P_{rot} = 35.23\pm0.11$\,d) and the Keplerian signal ($19.466\pm 0.005$\,d) at the same time. The horizontal dashed lines indicate FAP levels of 10\%~(blue), 1\%~(orange), and 0.1\%~(green). The black and red vertical dashed lines indicate the orbital rotational period of the planet and the rotation period of the star, respectively. 
}
\label{fig:gj720a_GLS_res}
\end{figure}

Figure \ref{fig:gj720a_GP_plots} shows the simultaneous fit RV+GP with the “best” model as a function of the time and the planetary signal of GJ~720~A folded in phase with the orbital period. The final orbital parameters of the planet are listed in Table \ref{tab:gj720a_from_GP_and_RV}. Figure \ref{fig:gj720a_GLS_res} shows the GLS periodogram of the original RVs (green line) and the corresponding RV GLS periodogram of the residuals after substracting the “best” model (blue line). This figure shows how the final model produced an optimal removal of all the signals present in the RV data. The $rms$ of the residuals is 1.59\,m$\rm s^{-1}$, around three times smaller than the $rms$ (4.19\,m$\rm s^{-1}$) of the original RV data.

\begin{figure*}[]
\centering
\includegraphics[width=\textwidth]{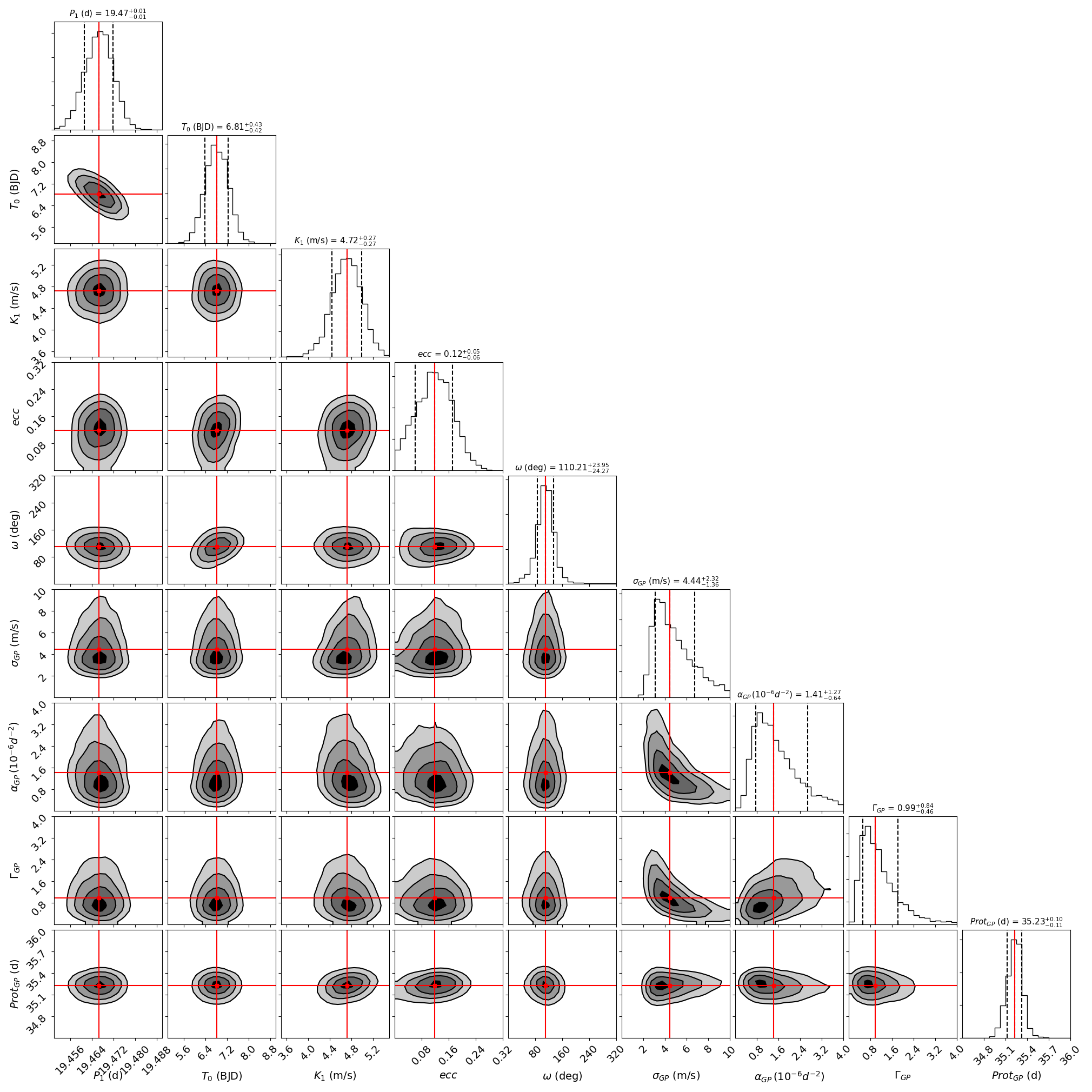}
\caption{Posterior distributions for the parameters of the “best” fitted model (BM+GP+1pl) that describes the planet orbiting GJ~720~A and the stellar variability using {\tt juliet}. The vertical dashed lines indicate the 16, 50, and 84\% quantiles of the fitted parameters; this corresponds to 1$\sigma$ uncertainty. The red line shows the median value of each parameter. 
}
\label{fig:gj720a_cornerplot}
\end{figure*}

In Figure \ref{fig:gj720a_cornerplot} we show the posterior distributions of the fitted parameters, this is one planet plus the stellar activity. The planet, GJ~720~Ab, has a minimum mass of $13.64\pm0.79$\,$M_{\oplus}$ located at a distance of $0.119\pm 0.002$\,AU from the host star with an orbital period of $19.466\pm0.005$\,d. Due to the low eccentricity value obtained and its error bars we can conclude that the eccentricity is compatible with zero and therefore our planetary orbit is circular. 

A detailed summary of the hyperparameters and priors values used for all the different models followed with {\tt juliet} are listed in Table \ref{tab:summary_prios}. While the final parameter values obtained for each model are summarized in Table \ref{tab:gj720a_final_values_diff_models_juliet}.

%BM+GP corner

\begin{table}[!tb]
\centering
\caption{Keplerian orbital parameters of GJ~720~Ab from Gaussian process regression method for the two different approaches that we followed. The first one with {\tt juliet} where the explored parameters were $e$ and $\omega$ and the second one using {\tt emcee} where the parameters $e$ and $\omega$ are derived from the explored parameters $\sqrt{e} sin(\omega)$ and $\sqrt{e} cos(\omega)$.}
\label{tab:gj720a_from_GP_and_RV}
\begin{tabular}{l c c c}

\hline
\hline
\noalign{\smallskip}

Parameter & 		GJ~720~A b & GJ~720~A b \\
 					& {\tt juliet} &  {\tt emcee}\\
\noalign{\smallskip}	
\hline	
\noalign{\smallskip}
$P$  (d) 								&$19.466^{+0.005}_{-0.005}$ &  ${19.47}_{-0.01}^{+0.01}$\\\\
$T_0$ (BJD-2,456,400)$^{(1)}$ &$6.81^{+0.43}_{-0.42} $ & ${7.02}_{-1.82}^{+1.63}$ \\
$e$ 									&  $0.12^{+0.05}_{-0.06} $ & ${0.10}_{-0.06}^{+0.06}$\\
$\omega$ (deg) 				&	 $110.22^{+23.97}_{-24.28}$ & ${102.16}_{-31.48}^{+29.77}$\\
$K$ (m/s)							&   $4.72^{+0.27}_{-0.27}$ & ${4.60}_{-0.29}^{+0.28}$\\
$\gamma_0$ (m/s)$^{(2)}$ 		&  $-0.53^{+2.29}_{-2.54}$ & ${0.04}_{-1.04}^{+0.97}$\\

\noalign{\smallskip}	
\hline	
\noalign{\smallskip}
\textit{Derived physical parameters} &  \\
\noalign{\smallskip}

$m_{\rm p}\sin i$ ($\rm M_{ \oplus}$)   & \multicolumn{2}{c}{$13.64^{+0.78}_{-0.79}$ } \\

$a$ (au) 															  & \multicolumn{2}{c}{$0.119^{+0.002}_{-0.001}$}\\
$T_{\rm eq}$ (K)$^{(3)}$								 &  \multicolumn{2}{c}{309$\pm$24 -- 401$\pm$32}\\

\noalign{\smallskip}
\hline
\end{tabular}
\tablefoot{$^{(1)}$ $T_0$ corresponds to the periastron passage. $^{(2)}$ Arbitrary zero point applied to HARPS-N RVs. $^{(3)}$\,For Bond albedo in the interval 0.65--0.0. }
\end{table}

%------------------------
\subsection{Quasi-periodic kernel, {\tt emcee}}

For completeness, we also have performed another GP analysis on the BM+GP and BM+GP+1pl models that differ in the adopted covariance function, the chain sampler and in priors definition, which are uninformative (see Tab.~\ref{tab:nino_t1}). This analysis employs the celerite quasi-periodic kernel, described in equation \ref{eq:GP}, and the {\tt emcee} \citep{2013PASP..125..306F} package based on the affine-invariant ensemble sampler for Markov chain Monte Carlo (MCMC) \citep{2010CAMCS...5...65G}.

The parameter space is covered by 32 walkers, whose initial positions are randomly selected within the priors boundaries. This choice on the initial position of the walkers requires a burn-in phase to free the chain from very low probability values. Therefore, we have set a chain of 50K steps as burn-in, at the end of which a blob, centered at the maximum probability position, is initialized to feed a second chain. We run the second chain until convergence occurs, i.e. the autocorrelation time of each parameter \citep[see][]{Sokal1996MonteCM}, evaluated every 10K steps, varies less than 1\% and the chain is 100 times longer than the estimated autocorrelation time. Based on this criterion, the chains converged after 160K and 440K steps, respectively, for the BM+GP and BM+GP+1pl models.

Posterion distributions of the latter model is presented in Fig.~\ref{fig:nino_f1}. BIC parameters for the two models are $766.3$ and $649.5$, with the lowest being that of the BM+GP+1pl model. As in the previous analysis, the BIC comparison supports the model in which RV data are described with a Keplerian signal whose planetary parameters are in agreement with the ones obtained in the previous analysis and reported in Tab.~\ref{tab:gj720a_from_GP_and_RV}.

\subsection{Observing seasons analysis}

Looking at the time cadence of the observations we can identify five different observing seasons for our HARPS-N RV data. Only the third and fourth seasons (SIII $\sim$[588, 888] and SIV $\sim$[1238, 1638] days (BJD-2,456,400)) have a significant number of observations in order to investigate the stability of the planetary signal by applying a GP analysis. Priors of BM+GP+1pl model using the {\tt emcee} analysis have been adopted, except for the parameter $T_{0}$ whose prior is (BJD-2456400) [0,30]. The estimated planetary parameters for the two subsets are presented in Tab.~\ref{tab:nino_t3}. Obtained values are generally in agreement with respect to the estimates from the former analysis within 1$\sigma$ confidence. This does not apply to the semi-amplitude ($K_1$) of the planetary signal which is systematically larger than the one obtained previously and, in the case of SIV, it is compatible with the previously estimated value within 1.3$\sigma$. This effect is related to the inability of the quasi-periodic kernel to model the stellar activity in these two seasons, therefore, being the kernel parameters not well constrained, a fraction of the stellar signal is absorbed by the semi-amplitude producing larger values of this parameter.

\begin{figure*}[]
\centering
\includegraphics[width=\textwidth]{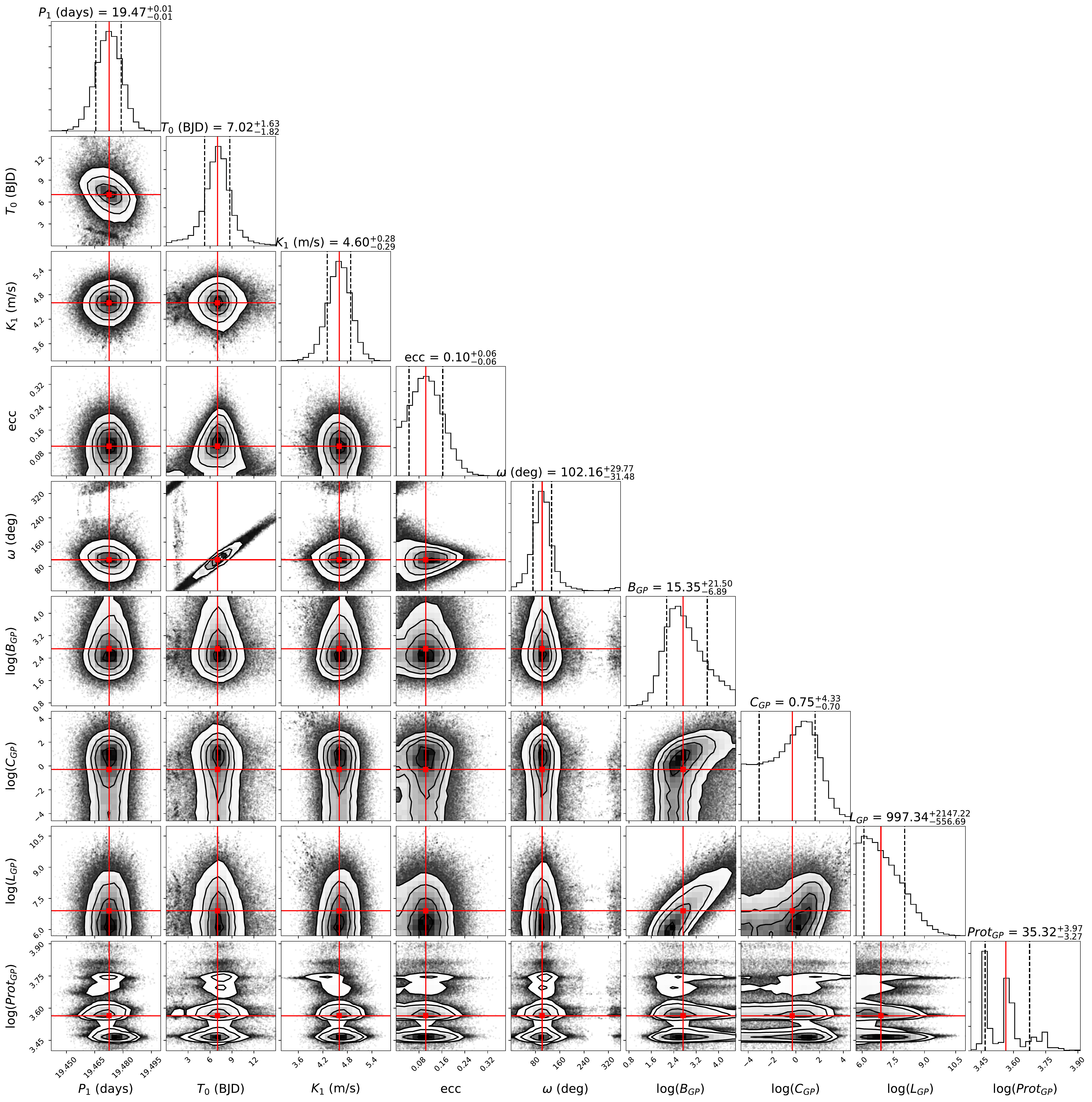}
\caption{Posterior distribution of the BM+GP+1pl model in which one sample from the chain over 100 have been displayed. To lighten the visualization, the offset and jitter parameters have not been displayed. The median value of each parameter is plotted with the red line. The results were obtained using {\tt emcee} code. The $\log$ describing the GP parameters stands for neperian logarithm}.
\label{fig:nino_f1}
\end{figure*}

Looking at the two different posterior distributions obtained with each kernel (exp-sin-squared kernel in Fig. \ref{fig:gj720a_cornerplot} and QP kenel in Fig. \ref{fig:nino_f1}), we conclude that the exp-sin-squared kernel is the optimal kernel for our target in order to model the stellar variability. The exp-sin-squared kernel is able to identify the same stellar rotational period in the RV data as that obtained from the activity indicators even setting a wide range prior for the GP $P_{\rm rot}$ parameter.

\begin{table}[!tb]
\centering
\caption{Comparison of the planetary parameters for the two different observed seasons analyzed and the full dataset using {\tt emcee}. SIII corresponds to the RV data in the range $\sim$[588, 888] days (BJD-2,456,000) and SIV stands for RV data $\sim$[1238, 1638] days (BJD-2,456,000).}
\label{tab:nino_t3}

\begin{tabular}{l c c c c r}

\hline
\hline
\noalign{\smallskip}
Parameter & SIII & SIV & Full Dataset \\
\noalign{\smallskip}	
\hline	

$P$ (d) &  ${19.52}_{-0.07}^{+0.07}$ & ${19.35}_{-0.17}^{+0.16}$& ${19.47}_{-0.01}^{+0.01}$\\
$K$ (m/s) & ${5.26}_{-0.41}^{+0.40}$& ${5.65}_{-0.54}^{+0.52}$& ${4.60}_{-0.29}^{+0.28}$\\
$e$	 & ${0.11}_{-0.07}^{+0.08}$& ${0.13}_{-0.09}^{+0.10}$& ${0.10}_{-0.06}^{+0.06}$\\
$\omega$ (deg)	& ${165.01}_{-36.97}^{+43.11}$ & ${77.34}_{-34.58}^{+64.58}$& ${102.16}_{-31.48}^{+29.77}$\\
\noalign{\smallskip}	
\hline	
\noalign{\smallskip}
\end{tabular}
\end{table}

%----------------------------------
\section{Summary and conclusions}
\label{sec:summary_discussion}

The monitoring of GJ~720~A M dwarf with the HARPS-N spectrograph during our observing campaign within the HADES program resulted in a sub-Neptune mass detection of a minimum mass of $13.64^{+0.78}_{-0.79}$\,M$_{\oplus}$ with a semi-major axis of $0.119^{+0.002}_{-0.001}$\,AU in a circular orbit ($e=0.12^{+0.05}_{-0.06} $) that revolves with a period of $19.466^{+0.005}_{-0.005}$\,d. All spectroscopic activity indicators (H$\alpha$, Ca~{\sc ii} H \& K, NaD lines) together with the available photometry from the MEarth, SuperWASP, EXORAP, APACHE, and \textit{TESS} surveys indicate that the other two detected periodicities ($\sim$40 and $\sim$100\,d) in the HARPS-N RV data are related to stellar activity phenomena. In fact the stellar rotation period derived here from the S-index activity indicator using a GP regression determined its value at $P_{\rm rot}=36.05^{+1.38}_{-1.44}$\,d (compatible, within the error bars, with the $P_{\rm rot}$ value ($35.23\pm0.11$\,d) obatined from the best fit RV data analysis). The RV signal around 100\,d is more likely related to a life cycle of the active regions that persists for some stellar rotations. The different approaches we followed here provided strong arguments in favor of the GJ~720~Ab planet detection. No counterparts in any stellar activity indices were found at the planet orbital period, the stability and the coherence of the planetary signal indicates a long-lived behaviour, and the activity and planetary signals are not related with each other by a possible alias phenomena. Also we modeled the stellar variability and the possible Keplearian signals in a simultaneous way using a GP regression and employing two independent analyses ({\tt juliet} and {\tt emcee}) for completeness. We analyzed different possible models (e.g., only GP, GP+1pl, GP+2pl, only Keplerian signals) that could reproduced the RV signals. The one planet model at 19.5\,d plus a GP kernel modeling the stellar activity was the fit statistically preferred among the different implemented models and was the fit that determined the final planetary orbital parameters.

We looked at the \cite{2019A&A...623A..72K} catalog of proper motion anomalies in order to find possible evidence of outer, massive companions in the Hipparcos-Gaia absolute astrometry. There is no statistically significant proper motion variation reported for GJ~720~A. Based on the analytical formulation of \cite{2019A&A...623A..72K}, the sensitivity curve for the star implies that companions with masses of 0.27\,M$_{\rm J}$ at 1\,AU are ruled out, and massive planets with a few Jupiter masses, or larger, would produce detectable effects out to a few tens AU. At the exact separation of GJ~720~Ab the detectable mass from proper motion anomaly is found to be around 15\,M$_{\rm J}$.

\begin{figure}[!htb]
\centering
\includegraphics[width=0.5\textwidth]{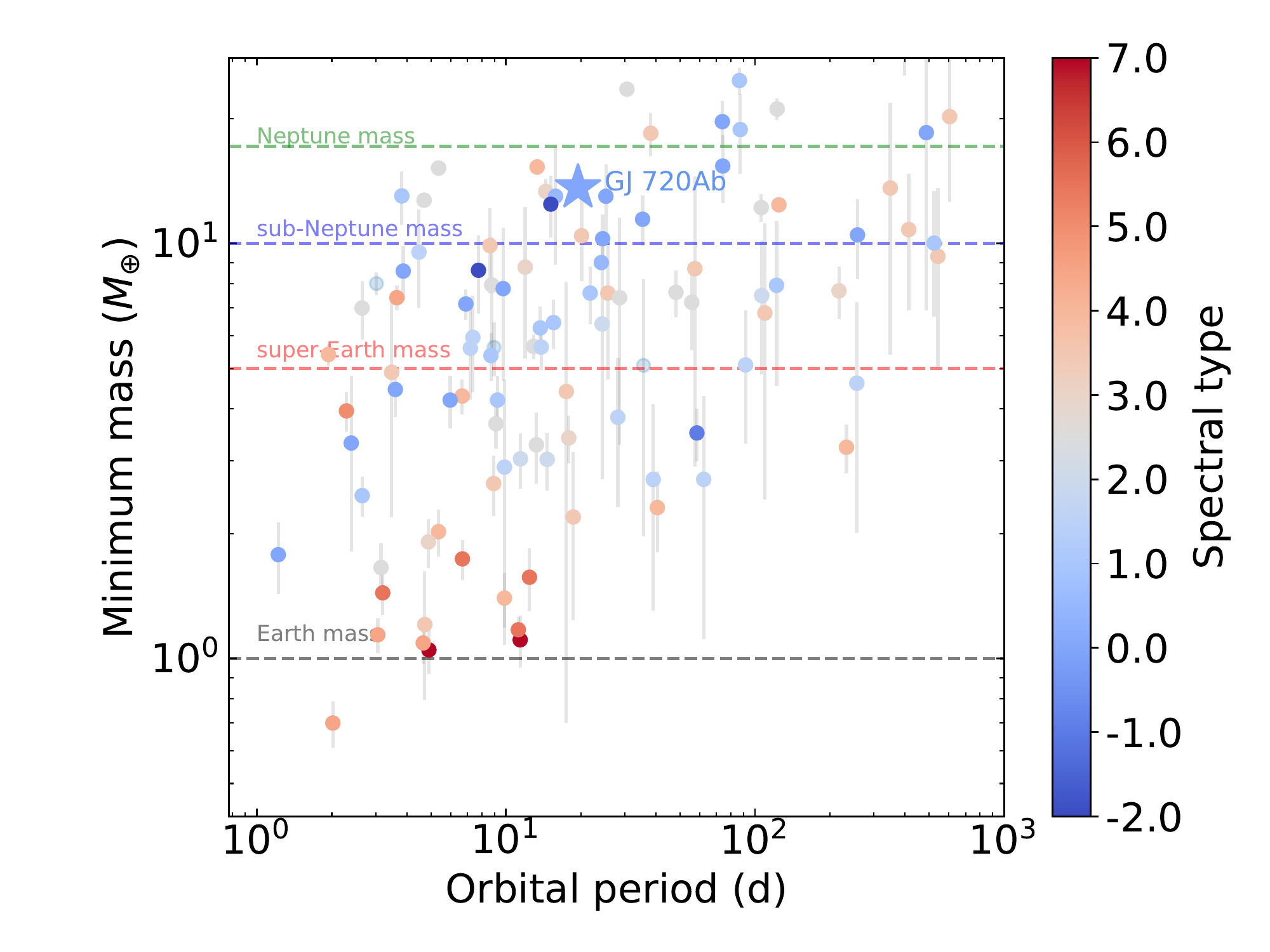}
\caption{Minimum mass vs. orbital period diagram for known Neptune-type and super-Earth planets (colored dots) around M dwarfs only detected with RV method (available data at \url{http://www.exoplanet.eu}). The star symbol indicates the location of the planet GJ~720~Ab. The colorcode divides the sample by the spectral type of the host star with -2.0, 0.0 and 7.0 corresponding to K5.0V, M0.0V, and M7.0V, respectively.
}
\label{fig:M80_planets_around_dM}
\end{figure}

In Figure \ref{fig:M80_planets_around_dM} we represent the position occupied by GJ~720~Ab in the diagram of known Neptune-type and super-Earth planets discovered only with the RV method around M dwarfs. From this diagram we can observe that more massive planets are located around early-M dwarfs (blue dots) while the Earth-like planets are found around mid/late-M spectral type (red/orange dots). We note that  could be a selection effect because smaller planets can be detected more easily around smaller stars. Our target populates the more massive sub-Neptune part of the diagram at intermediate orbital periods. Thanks to unbiased fore-casting model presented by \cite{2017ApJ...834...17C}, we predicted the planetary radius as 3.84$^{+1.53}_{-1.44}$\,R$_{\oplus}$. Once obtained an estimation of the planet radius we can derive the probability that GJ~720~Ab transits in front of the disk of its host star and the depth that the planet would infer. The corresponding values correspond to 2.2$^{+2.5}_{-1.9}$\% of transit probability and $4575^{+7105}_{-3016}$\,(ppm) for the transit depth. 

Following the models by \cite{2013ApJ...765..131K, 2014ApJ...787L..29K} we estimated the conservative habitable zone limits following the Runaway Greenhouse for 5\,$\rm M_{\oplus}$ coefficients. The received effective stellar flux compared to the Sun corresponds with $S_{\rm eff}$ = 1.01\,$\rm S_{\odot}$ and the inner edge of the GJ~720~A habitable zone is placed at 0.24\,AU. The habitable zone determination, following \cite{2014ApJ...787L..29K}, implies that GJ~720~Ab lies inside the inner boundary of the habitable zone where the insolation flux has a value of $4.28^{+1.25}_{-0.97}$\,$\rm S_{\oplus}$. The theoretical equilibrium temperature ($T_{\rm eq}$) of GJ~720~Ab was derived by using the Stefan–Boltzmann equation, the stellar parameters of Table \ref{tab:stellar_properties_GJ720A}, and two extreme values of the albedo. In the two extreme cases, $A = 0.0$ and $A = 0.65$, the $T_{\rm eq}$ for GJ~720~Ab is 401$\pm$32\,K for a non-reflecting planet and 309$\pm$24\,K for the high-reflectance planet.

%-----------------------------------------------------------------

\begin{acknowledgements}

This research was supported by the Italian Ministry of Education, University, and Research through the PREMIALE WOW 2013 research project under grant \textit{Ricerca di pianeti intorno a stelle di piccola massa}. This paper makes use of data from the MEarth Project, which is a collaboration between Harvard University and the Smithsonian Astrophysical Observatory. The MEarth Project acknowledges funding from the David and Lucile Packard Fellowship for Science and Engineering and the National Science Foundation under grants AST-0807690, AST-1109468, AST-1616624 and AST-1004488 (Alan T. Waterman Award), and a grant from the John Templeton Foundation. This work has made use of data from the European Space Agency (ESA) mission {\it Gaia} (\url{https://www.cosmos.esa.int/gaia}), processed by the {\it Gaia} Data Processing and Analysis Consortium (DPAC, \url{https://www.cosmos.esa.int/web/gaia/dpac/consortium}). Funding for the DPAC has been provided by national institutions, in particular the institutions participating in the {\it Gaia} Multilateral Agreement. E.G.A. acknowledge support from the Spanish Ministery for Science, Innovation, and Universities through projects AYA-2016-79425-C3-1/2/3-P, AYA2015-69350-C3-2-P, ESP2017-87676-C5-2- R, ESP2017-87143-R. The Centro de Astrobiolog\'ia (CAB, CSIC-INTA) is a Center of Excellence “Maria de Maeztu“. A.P. acknowledges support from ASI-INAF agreement 2018-22-HH.0 \emph{Partecipazione alla fase B1 della missione ARIEL}. We acknowledge support from the Accordo Attuativo ASI-INAF n. 2018.22.HH.O, Partecipazione alla fase B1 della missione Ariel (ref. G. Micela). A.S., M.Pi. acknowledge the financial contribution from the agreement ASI-INAF n.2018-16-HH.0. M.P., I.R. acknowledge support from the Spanish Ministry of Science and Innovation and the European Regional Development Fund through grant PGC2018-098153-B-C33, as well as the support of the Generalitat de Catalunya/CERCA programme. J.I.G.H. acknowledges financial support from Spanish MICINN under the 2013 Ram\'on y Cajal program RYC-2013-14875. A.S.M. acknowledges financial support from the Spanish Ministry of Science and Innovation (MICINN) under the 2019 Juan de la Cierva Programme. B.T.P. acknowledges Fundaci\'on La Caixa for the financial support received in the form of a Ph.D. contract. J.I.G.H., R.R., A.S.M., B.T.P. acknowledge financial support from the Spanish MICINN AYA2017-86389-P.

\end{acknowledgements}


\bibliographystyle{aa} % style aa.bst
\bibliography{bibliography.bib} % your references Yourfile.bib

\begin{thebibliography}{55}
\expandafter\ifx\csname natexlab\endcsname\relax\def\natexlab#1{#1}\fi

\bibitem[{{Affer} {et~al.}(2019){Affer}, {Damasso}, {Micela}, {Poretti},
  {Scandariato}, {Maldonado}, {Lanza}, {Covino}, {Garrido Rubio}, {Gonz{\'a}lez
  Hern{\'a}ndez}, {Gratton}, {Leto}, {Maggio}, {Perger}, {Sozzetti},
  {Su{\'a}rez Mascare{\~n}o}, {Bonomo}, {Borsa}, {Claudi}, {Cosentino},
  {Desidera}, {Giacobbe}, {Molinari}, {Pedani}, {Pinamonti}, {Rebolo}, {Ribas},
  \& {Toledo-Padr{\'o}n}}]{2019A&A...622A.193A}
{Affer}, L., {Damasso}, M., {Micela}, G., {et~al.} 2019, \aap, 622, A193

\bibitem[{{Affer} {et~al.}(2016){Affer}, {Micela}, {Damasso}, {Perger},
  {Ribas}, {Su{\'a}rez Mascare{\~n}o}, {Gonz{\'a}lez Hern{\'a}ndez}, {Rebolo},
  {Poretti}, {Maldonado}, {Leto}, {Pagano}, {Scandariato}, {Zanmar Sanchez},
  {Sozzetti}, {Bonomo}, {Malavolta}, {Morales}, {Rosich}, {Bignamini},
  {Gratton}, {Velasco}, {Cenadelli}, {Claudi}, {Cosentino}, {Desidera},
  {Giacobbe}, {Herrero}, {Lafarga}, {Lanza}, {Molinari}, \&
  {Piotto}}]{2016A&A...593A.117A}
{Affer}, L., {Micela}, G., {Damasso}, M., {et~al.} 2016, \aap, 593, A117

\bibitem[{{Alonso-Floriano} {et~al.}(2015){Alonso-Floriano}, {Morales},
  {Caballero}, {Montes}, {Klutsch}, {Mundt}, {Cort{\'e}s-Contreras}, {Ribas},
  {Reiners}, {Amado}, {Quirrenbach}, \& {Jeffers}}]{2015A&A...577A.128A}
{Alonso-Floriano}, F.~J., {Morales}, J.~C., {Caballero}, J.~A., {et~al.} 2015,
  \aap, 577, A128

\bibitem[{{Ambikasaran} {et~al.}(2015){Ambikasaran}, {Foreman-Mackey},
  {Greengard}, {Hogg}, \& {O'Neil}}]{2015ITPAM..38..252A}
{Ambikasaran}, S., {Foreman-Mackey}, D., {Greengard}, L., {Hogg}, D.~W., \&
  {O'Neil}, M. 2015, IEEE Transactions on Pattern Analysis and Machine
  Intelligence, 38, 252

\bibitem[{{Anglada-Escud{\'e}} \& {Butler}(2012)}]{2012ApJS..200...15A}
{Anglada-Escud{\'e}}, G. \& {Butler}, R.~P. 2012, \apjs, 200, 15

\bibitem[{{Bailer-Jones} {et~al.}(2018){Bailer-Jones}, {Rybizki}, {Fouesneau},
  {Mantelet}, \& {Andrae}}]{2018AJ....156...58B}
{Bailer-Jones}, C.~A.~L., {Rybizki}, J., {Fouesneau}, M., {Mantelet}, G., \&
  {Andrae}, R. 2018, \aj, 156, 58

\bibitem[{{Berta} {et~al.}(2012){Berta}, {Irwin}, {Charbonneau}, {Burke}, \&
  {Falco}}]{2012AJ....144..145B}
{Berta}, Z.~K., {Irwin}, J., {Charbonneau}, D., {Burke}, C.~J., \& {Falco},
  E.~E. 2012, \aj, 144, 145

\bibitem[{{Chen} \& {Kipping}(2017)}]{2017ApJ...834...17C}
{Chen}, J. \& {Kipping}, D. 2017, \apj, 834, 17

\bibitem[{{Cosentino} {et~al.}(2012){Cosentino}, {Lovis}, {Pepe}, {Collier
  Cameron}, {Latham}, {Molinari}, {Udry}, {Bezawada}, {Black}, {Born},
  {Buchschacher}, {Charbonneau}, {Figueira}, {Fleury}, {Galli}, {Gallie},
  {Gao}, {Ghedina}, {Gonzalez}, {Gonzalez}, {Guerra}, {Henry}, {Horne},
  {Hughes}, {Kelly}, {Lodi}, {Lunney}, {Maire}, {Mayor}, {Micela}, {Ordway},
  {Peacock}, {Phillips}, {Piotto}, {Pollacco}, {Queloz}, {Rice}, {Riverol},
  {Riverol}, {San Juan}, {Sasselov}, {Segransan}, {Sozzetti}, {Sosnowska},
  {Stobie}, {Szentgyorgyi}, {Vick}, \& {Weber}}]{2012SPIE.8446E..1VC}
{Cosentino}, R., {Lovis}, C., {Pepe}, F., {et~al.} 2012, in \procspie, Vol.
  8446, Ground-based and Airborne Instrumentation for Astronomy IV, 84461V

\bibitem[{{Cutri} {et~al.}(2003){Cutri}, {Skrutskie}, {van Dyk}, {Beichman},
  {Carpenter}, {Chester}, {Cambresy}, {Evans}, {Fowler}, {Gizis}, {Howard},
  {Huchra}, {Jarrett}, {Kopan}, {Kirkpatrick}, {Light}, {Marsh}, {McCallon},
  {Schneider}, {Stiening}, {Sykes}, {Weinberg}, {Wheaton}, {Wheelock}, \&
  {Zacarias}}]{2003yCat.2246....0C}
{Cutri}, R.~M., {Skrutskie}, M.~F., {van Dyk}, S., {et~al.} 2003, VizieR Online
  Data Catalog, 2246

\bibitem[{{Dawson} \& {Fabrycky}(2010)}]{2010ApJ...722..937D}
{Dawson}, R.~I. \& {Fabrycky}, D.~C. 2010, \apj, 722, 937

\bibitem[{{Deeming}(1975)}]{1975Ap&SS..36..137D}
{Deeming}, T.~J. 1975, \apss, 36, 137

\bibitem[{{Delfosse} {et~al.}(1998){Delfosse}, {Forveille}, {Perrier}, \&
  {Mayor}}]{1998A&A...331..581D}
{Delfosse}, X., {Forveille}, T., {Perrier}, C., \& {Mayor}, M. 1998, \aap, 331,
  581

\bibitem[{{Dressing} \& {Charbonneau}(2015)}]{2015ApJ...807...45D}
{Dressing}, C.~D. \& {Charbonneau}, D. 2015, \apj, 807, 45

\bibitem[{{Espinoza} {et~al.}(2019){Espinoza}, {Kossakowski}, \&
  {Brahm}}]{2019MNRAS.490.2262E}
{Espinoza}, N., {Kossakowski}, D., \& {Brahm}, R. 2019, \mnras, 490, 2262

\bibitem[{{Foreman-Mackey} {et~al.}(2017){Foreman-Mackey}, {Agol},
  {Ambikasaran}, \& {Angus}}]{2017AJ....154..220F}
{Foreman-Mackey}, D., {Agol}, E., {Ambikasaran}, S., \& {Angus}, R. 2017, \aj,
  154, 220

\bibitem[{{Foreman-Mackey} {et~al.}(2013){Foreman-Mackey}, {Hogg}, {Lang}, \&
  {Goodman}}]{2013PASP..125..306F}
{Foreman-Mackey}, D., {Hogg}, D.~W., {Lang}, D., \& {Goodman}, J. 2013, \pasp,
  125, 306

\bibitem[{{Fulton} {et~al.}(2018){Fulton}, {Petigura}, {Blunt}, \&
  {Sinukoff}}]{2018PASP..130d4504F}
{Fulton}, B.~J., {Petigura}, E.~A., {Blunt}, S., \& {Sinukoff}, E. 2018, \pasp,
  130, 044504

\bibitem[{{Gaia Collaboration} {et~al.}(2018){Gaia Collaboration}, {Brown},
  {Vallenari}, {Prusti}, {de Bruijne}, {Babusiaux}, {Bailer-Jones}, {Biermann},
  {Evans}, {Eyer}, {Jansen}, {Jordi}, {Klioner}, {Lammers}, {Lindegren},
  {Luri}, {Mignard}, {Panem}, {Pourbaix}, {Randich}, {Sartoretti}, {Siddiqui},
  {Soubiran}, {van Leeuwen}, {Walton}, {Arenou}, {Bastian}, {Cropper},
  {Drimmel}, {Katz}, {Lattanzi}, {Bakker}, {Cacciari}, {Casta{\~n}eda},
  {Chaoul}, {Cheek}, {De Angeli}, {Fabricius}, {Guerra}, {Holl}, {Masana},
  {Messineo}, {Mowlavi}, {Nienartowicz}, {Panuzzo}, {Portell}, {Riello},
  {Seabroke}, {Tanga}, {Th{\'e}venin}, {Gracia-Abril}, {Comoretto},
  {Garcia-Reinaldos}, {Teyssier}, {Altmann}, {Andrae}, {Audard},
  {Bellas-Velidis}, {Benson}, {Berthier}, {Blomme}, {Burgess}, {Busso},
  {Carry}, {Cellino}, {Clementini}, {Clotet}, {Creevey}, {Davidson}, {De
  Ridder}, {Delchambre}, {Dell'Oro}, {Ducourant}, {Fern{\'a}ndez-
  Hern{\'a}ndez}, {Fouesneau}, {Fr{\'e}mat}, {Galluccio}, {Garc{\'\i}a-Torres},
  {Gonz{\'a}lez-N{\'u}{\~n}ez}, {Gonz{\'a}lez-Vidal}, {Gosset}, {Guy},
  {Halbwachs}, {Hambly}, {Harrison}, {Hern{\'a}ndez}, {Hestroffer}, {Hodgkin},
  {Hutton}, {Jasniewicz}, {Jean-Antoine-Piccolo}, {Jordan}, {Korn},
  {Krone-Martins}, {Lanzafame}, {Lebzelter}, {L{\"o}ffler}, {Manteiga},
  {Marrese}, {Mart{\'\i}n-Fleitas}, {Moitinho}, {Mora}, {Muinonen}, {Osinde},
  {Pancino}, {Pauwels}, {Petit}, {Recio-Blanco}, {Richards}, {Rimoldini},
  {Robin}, {Sarro}, {Siopis}, {Smith}, {Sozzetti}, {S{\"u}veges}, {Torra}, {van
  Reeven}, {Abbas}, {Abreu Aramburu}, {Accart}, {Aerts}, {Altavilla},
  {{\'A}lvarez}, {Alvarez}, {Alves}, {Anderson}, {Andrei}, {Anglada Varela},
  {Antiche}, {Antoja}, {Arcay}, {Astraatmadja}, {Bach}, {Baker},
  {Balaguer-N{\'u}{\~n}ez}, {Balm}, {Barache}, {Barata}, {Barbato}, {Barblan},
  {Barklem}, {Barrado}, {Barros}, {Barstow}, {Bartholom{\'e} Mu{\~n}oz},
  {Bassilana}, {Becciani}, {Bellazzini}, {Berihuete}, {Bertone}, {Bianchi},
  {Bienaym{\'e}}, {Blanco-Cuaresma}, {Boch}, {Boeche}, {Bombrun}, {Borrachero},
  {Bossini}, {Bouquillon}, {Bourda}, {Bragaglia}, {Bramante}, {Breddels},
  {Bressan}, {Brouillet}, {Br{\"u}semeister}, {Brugaletta}, {Bucciarelli},
  {Burlacu}, {Busonero}, {Butkevich}, {Buzzi}, {Caffau}, {Cancelliere},
  {Cannizzaro}, {Cantat-Gaudin}, {Carballo}, {Carlucci}, {Carrasco},
  {Casamiquela}, {Castellani}, {Castro-Ginard}, {Charlot}, {Chemin},
  {Chiavassa}, {Cocozza}, {Costigan}, {Cowell}, {Crifo}, {Crosta}, {Crowley},
  {Cuypers}, {Dafonte}, {Damerdji}, {Dapergolas}, {David}, {David}, {de
  Laverny}, {De Luise}, {De March}, {de Martino}, {de Souza}, {de Torres},
  {Debosscher}, {del Pozo}, {Delbo}, {Delgado}, {Delgado}, {Di Matteo},
  {Diakite}, {Diener}, {Distefano}, {Dolding}, {Drazinos}, {Dur{\'a}n},
  {Edvardsson}, {Enke}, {Eriksson}, {Esquej}, {Eynard Bontemps}, {Fabre},
  {Fabrizio}, {Faigler}, {Falc{\~a}o}, {Farr{\`a}s Casas}, {Federici},
  {Fedorets}, {Fernique}, {Figueras}, {Filippi}, {Findeisen}, {Fonti},
  {Fraile}, {Fraser}, {Fr{\'e}zouls}, {Gai}, {Galleti}, {Garabato},
  {Garc{\'\i}a-Sedano}, {Garofalo}, {Garralda}, {Gavel}, {Gavras}, {Gerssen},
  {Geyer}, {Giacobbe}, {Gilmore}, {Girona}, {Giuffrida}, {Glass}, {Gomes},
  {Granvik}, {Gueguen}, {Guerrier}, {Guiraud}, {Guti{\'e}rrez-S{\'a}nchez},
  {Haigron}, {Hatzidimitriou}, {Hauser}, {Haywood}, {Heiter}, {Helmi}, {Heu},
  {Hilger}, {Hobbs}, {Hofmann}, {Holland}, {Huckle}, {Hypki}, {Icardi},
  {Jan{\ss}en}, {Jevardat de Fombelle}, {Jonker}, {Juh{\'a}sz}, {Julbe},
  {Karampelas}, {Kewley}, {Klar}, {Kochoska}, {Kohley}, {Kolenberg},
  {Kontizas}, {Kontizas}, {Koposov}, {Kordopatis}, {Kostrzewa-Rutkowska},
  {Koubsky}, {Lambert}, {Lanza}, {Lasne}, {Lavigne}, {Le Fustec}, {Le
  Poncin-Lafitte}, {Lebreton}, {Leccia}, {Leclerc}, {Lecoeur-Taibi},
  {Lenhardt}, {Leroux}, {Liao}, {Licata}, {Lindstr{\o}m}, {Lister}, {Livanou},
  {Lobel}, {L{\'o}pez}, {Managau}, {Mann}, {Mantelet}, {Marchal}, {Marchant},
  {Marconi}, {Marinoni}, {Marschalk{\'o}}, {Marshall}, {Martino}, {Marton},
  {Mary}, {Massari}, {Matijevi{\v{c}}}, {Mazeh}, {McMillan}, {Messina},
  {Michalik}, {Millar}, {Molina}, {Molinaro}, {Moln{\'a}r}, {Montegriffo},
  {Mor}, {Morbidelli}, {Morel}, {Morris}, {Mulone}, {Muraveva}, {Musella},
  {Nelemans}, {Nicastro}, {Noval}, {O'Mullane}, {Ord{\'e}novic},
  {Ord{\'o}{\~n}ez-Blanco}, {Osborne}, {Pagani}, {Pagano}, {Pailler},
  {Palacin}, {Palaversa}, {Panahi}, {Pawlak}, {Piersimoni}, {Pineau}, {Plachy},
  {Plum}, {Poggio}, {Poujoulet}, {Pr{\v{s}}a}, {Pulone}, {Racero}, {Ragaini},
  {Rambaux}, {Ramos-Lerate}, {Regibo}, {Reyl{\'e}}, {Riclet}, {Ripepi}, {Riva},
  {Rivard}, {Rixon}, {Roegiers}, {Roelens}, {Romero-G{\'o}mez}, {Rowell},
  {Royer}, {Ruiz-Dern}, {Sadowski}, {Sagrist{\`a} Sell{\'e}s}, {Sahlmann},
  {Salgado}, {Salguero}, {Sanna}, {Santana- Ros}, {Sarasso}, {Savietto},
  {Schultheis}, {Sciacca}, {Segol}, {Segovia}, {S{\'e}gransan}, {Shih},
  {Siltala}, {Silva}, {Smart}, {Smith}, {Solano}, {Solitro}, {Sordo}, {Soria
  Nieto}, {Souchay}, {Spagna}, {Spoto}, {Stampa}, {Steele},
  {Steidelm{\"u}ller}, {Stephenson}, {Stoev}, {Suess}, {Surdej}, {Szabados},
  {Szegedi-Elek}, {Tapiador}, {Taris}, {Tauran}, {Taylor}, {Teixeira},
  {Terrett}, {Teyssandier}, {Thuillot}, {Titarenko}, {Torra Clotet}, {Turon},
  {Ulla}, {Utrilla}, {Uzzi}, {Vaillant}, {Valentini}, {Valette}, {van Elteren},
  {Van Hemelryck}, {van Leeuwen}, {Vaschetto}, {Vecchiato}, {Veljanoski},
  {Viala}, {Vicente}, {Vogt}, {von Essen}, {Voss}, {Votruba}, {Voutsinas},
  {Walmsley}, {Weiler}, {Wertz}, {Wevers}, {Wyrzykowski}, {Yoldas},
  {{\v{Z}}erjal}, {Ziaeepour}, {Zorec}, {Zschocke}, {Zucker}, {Zurbach}, \&
  {Zwitter}}]{2018A&A...616A...1G}
{Gaia Collaboration}, {Brown}, A.~G.~A., {Vallenari}, A., {et~al.} 2018, \aap,
  616, A1

\bibitem[{{Gaia Collaboration} {et~al.}(2020){Gaia Collaboration}, {Smart},
  {Sarro}, {Rybizki}, {Reyl{\'e}}, {Robin}, {Hambly}, {Abbas}, {Barstow}, {de
  Bruijne}, {Bucciarelli}, {Carrasco}, {Cooper}, {Hodgkin}, {Masana},
  {Michalik}, {Sahlmann}, {Sozzetti}, {Brown}, {Vallenari}, {Prusti},
  {Babusiaux}, {Biermann}, {Creevey}, {Evans}, {Eyer}, {Hutton}, {Jansen},
  {Jordi}, {Klioner}, {Lammers}, {Lindegren}, {Luri}, {Mignard}, {Panem},
  {Pourbaix}, {Randich}, {Sartoretti}, {Soubiran}, {Walton}, {Arenou},
  {Bailer-Jones}, {Bastian}, {Cropper}, {Drimmel}, {Katz}, {Lattanzi}, {van
  Leeuwen}, {Bakker}, {Casta{\~n}eda}, {De Angeli}, {Ducourant}, {Fabricius},
  {Fouesneau}, {Fr{\'e}mat}, {Guerra}, {Guerrier}, {Guiraud}, {Jean-Antoine
  Piccolo}, {Messineo}, {Mowlavi}, {Nicolas}, {Nienartowicz}, {Pailler},
  {Panuzzo}, {Riclet}, {Roux}, {Seabroke}, {Sordo}, {Tanga}, {Th{\'e}venin},
  {Gracia-Abril}, {Portell}, {Teyssier}, {Altmann}, {Andrae}, {Bellas-Velidis},
  {Benson}, {Berthier}, {Blomme}, {Brugaletta}, {Burgess}, {Busso}, {Carry},
  {Cellino}, {Cheek}, {Clementini}, {Damerdji}, {Davidson}, {Delchambre},
  {Dell'Oro}, {Fern{\'a}ndez-Hern{\'a}ndez}, {Galluccio}, {Garc{\'\i}a-Lario},
  {Garcia-Reinaldos}, {Gonz{\'a}lez-N{\'u}{\~n}ez}, {Gosset}, {Haigron},
  {Halbwachs}, {Harrison}, {Hatzidimitriou}, {Heiter}, {Hern{\'a}ndez},
  {Hestroffer}, {Holl}, {Jan{\ss}en}, {Jevardat de Fombelle}, {Jordan},
  {Krone-Martins}, {Lanzafame}, {L{\"o}ffler}, {Lorca}, {Manteiga}, {Marchal},
  {Marrese}, {Moitinho}, {Mora}, {Muinonen}, {Osborne}, {Pancino}, {Pauwels},
  {Recio-Blanco}, {Richards}, {Riello}, {Rimoldini}, {Roegiers}, {Siopis},
  {Smith}, {Ulla}, {Utrilla}, {van Leeuwen}, {van Reeven}, {Abreu Aramburu},
  {Accart}, {Aerts}, {Aguado}, {Ajaj}, {Altavilla}, {{\'A}lvarez}, {{\'A}lvarez
  Cid-Fuentes}, {Alves}, {Anderson}, {Anglada Varela}, {Antoja}, {Audard},
  {Baines}, {Baker}, {Balaguer-N{\'u}{\~n}ez}, {Balbinot}, {Balog}, {Barache},
  {Barbato}, {Barros}, {Bartolom{\'e}}, {Bassilana}, {Bauchet},
  {Baudesson-Stella}, {Becciani}, {Bellazzini}, {Bernet}, {Bertone}, {Bianchi},
  {Blanco-Cuaresma}, {Boch}, {Bombrun}, {Bossini}, {Bouquillon}, {Bragaglia},
  {Bramante}, {Breedt}, {Bressan}, {Brouillet}, {Burlacu}, {Busonero},
  {Butkevich}, {Buzzi}, {Caffau}, {Cancelliere}, {C{\'a}novas},
  {Cantat-Gaudin}, {Carballo}, {Carlucci}, {Carnerero}, {Casamiquela},
  {Castellani}, {Castro-Ginard}, {Castro Sampol}, {Chaoul}, {Charlot},
  {Chemin}, {Chiavassa}, {Cioni}, {Comoretto}, {Cornez}, {Cowell}, {Crifo},
  {Crosta}, {Crowley}, {Dafonte}, {Dapergolas}, {David}, {David}, {de Laverny},
  {De Luise}, {De March}, {De Ridder}, {de Souza}, {de Teodoro}, {de Torres},
  {del Peloso}, {del Pozo}, {Delgado}, {Delgado}, {Delisle}, {Di Matteo},
  {Diakite}, {Diener}, {Distefano}, {Dolding}, {Eappachen}, {Edvardsson},
  {Enke}, {Esquej}, {Fabre}, {Fabrizio}, {Faigler}, {Fedorets}, {Fernique},
  {Fienga}, {Figueras}, {Fouron}, {Fragkoudi}, {Fraile}, {Franke}, {Gai},
  {Garabato}, {Garcia-Gutierrez}, {Garc{\'\i}a-Torres}, {Garofalo}, {Gavras},
  {Gerlach}, {Geyer}, {Giacobbe}, {Gilmore}, {Girona}, {Giuffrida}, {Gomel},
  {Gomez}, {Gonzalez-Santamaria}, {Gonz{\'a}lez-Vidal}, {Granvik},
  {Guti{\'e}rrez-S{\'a}nchez}, {Guy}, {Hauser}, {Haywood}, {Helmi}, {Hidalgo},
  {Hilger}, {H{\l}adczuk}, {Hobbs}, {Holland}, {Huckle}, {Jasniewicz},
  {Jonker}, {Juaristi Campillo}, {Julbe}, {Karbevska}, {Kervella}, {Khanna},
  {Kochoska}, {Kontizas}, {Kordopatis}, {Korn}, {Kostrzewa-Rutkowska},
  {Kruszy{\'n}ska}, {Lambert}, {Lanza}, {Lasne}, {Le Campion}, {Le Fustec},
  {Lebreton}, {Lebzelter}, {Leccia}, {Leclerc}, {Lecoeur-Taibi}, {Liao},
  {Licata}, {Lindstr{\o}m}, {Lister}, {Livanou}, {Lobel}, {Madrero Pardo},
  {Managau}, {Mann}, {Marchant}, {Marconi}, {Marcos Santos}, {Marinoni},
  {Marocco}, {Marshall}, {Polo}, {Mart{\'\i}n-Fleitas}, {Masip}, {Massari},
  {Mastrobuono-Battisti}, {Mazeh}, {McMillan}, {Messina}, {Millar}, {Mints},
  {Molina}, {Molinaro}, {Moln{\'a}r}, {Montegriffo}, {Mor}, {Morbidelli},
  {Morel}, {Morris}, {Mulone}, {Munoz}, {Muraveva}, {Murphy}, {Musella},
  {Noval}, {Ord{\'e}novic}, {Orr{\`u}}, {Osinde}, {Pagani}, {Pagano},
  {Palaversa}, {Palicio}, {Panahi}, {Pawlak}, {Pe{\~n}alosa Esteller},
  {Penttil{\"a}}, {Piersimoni}, {Pineau}, {Plachy}, {Plum}, {Poggio},
  {Poretti}, {Poujoulet}, {Pr{\v{s}}a}, {Pulone}, {Racero}, {Ragaini},
  {Rainer}, {Raiteri}, {Rambaux}, {Ramos}, {Ramos-Lerate}, {Re Fiorentin},
  {Regibo}, {Ripepi}, {Riva}, {Rixon}, {Robichon}, {Robin}, {Roelens},
  {Rohrbasser}, {Romero-G{\'o}mez}, {Rowell}, {Royer}, {Rybicki}, {Sadowski},
  {Sagrist{\`a} Sell{\'e}s}, {Salgado}, {Salguero}, {Samaras}, {Sanchez
  Gimenez}, {Sanna}, {Santove{\~n}a}, {Sarasso}, {Schultheis}, {Sciacca},
  {Segol}, {Segovia}, {S{\'e}gransan}, {Semeux}, {Shahaf}, {Siddiqui},
  {Siebert}, {Siltala}, {Slezak}, {Solano}, {Solitro}, {Souami}, {Souchay},
  {Spagna}, {Spoto}, {Steele}, {Steidelm{\"u}ller}, {Stephenson},
  {S{\"u}veges}, {Szabados}, {Szegedi-Elek}, {Taris}, {Tauran}, {Taylor},
  {Teixeira}, {Thuillot}, {Tonello}, {Torra}, {Torra}, {Turon}, {Unger},
  {Vaillant}, {van Dillen}, {Vanel}, {Vecchiato}, {Viala}, {Vicente},
  {Voutsinas}, {Weiler}, {Wevers}, {Wyrzykowski}, {Yoldas}, {Yvard}, {Zhao},
  {Zorec}, {Zucker}, {Zurbach}, \& {Zwitter}}]{2020arXiv201202061G}
{Gaia Collaboration}, {Smart}, R.~L., {Sarro}, L.~M., {et~al.} 2020, arXiv
  e-prints, arXiv:2012.02061

\bibitem[{{Giacobbe} {et~al.}(2020){Giacobbe}, {Benedetto}, {Damasso},
  {Sozzetti}, {Christille}, {Lattanzi}, {Calcidese}, {Carbognani}, {Barbato},
  {Pinamonti}, {Poggio}, {Lanza}, {Bernagozzi}, {Cenadelli}, {Lanteri}, \&
  {Bertolini}}]{2020MNRAS.491.5216G}
{Giacobbe}, P., {Benedetto}, M., {Damasso}, M., {et~al.} 2020, \mnras, 491,
  5216

\bibitem[{{Gonz{\'a}lez-{\'A}lvarez} {et~al.}(2019){Gonz{\'a}lez-{\'A}lvarez},
  {Micela}, {Maldonado}, {Affer}, {Maggio}, {Lanza}, {Covino}, {Benatti},
  {Bignamini}, {Cosentino}, {Damasso}, {Desidera}, {Gonz{\'a}lez
  Hern{\'a}ndez}, {Mart{\'\i}nez-Fiorenzano}, {Pagano}, {Perger}, {Piotto},
  {Pinamonti}, {Rainer}, {Rebolo}, {Ribas}, {Scandariato}, {Sozzetti},
  {Su{\'a}rez Mascare{\~n}o}, \& {Toledo-Padr{\'o}n}}]{2019A&A...624A..27G}
{Gonz{\'a}lez-{\'A}lvarez}, E., {Micela}, G., {Maldonado}, J., {et~al.} 2019,
  \aap, 624, A27

\bibitem[{{Goodman} \& {Weare}(2010)}]{2010CAMCS...5...65G}
{Goodman}, J. \& {Weare}, J. 2010, Communications in Applied Mathematics and
  Computational Science, 5, 65

\bibitem[{{Jeffreys}(1946)}]{1946RSPSA.186..453J}
{Jeffreys}, H. 1946, Proceedings of the Royal Society of London Series A, 186,
  453

\bibitem[{{Kervella} {et~al.}(2019){Kervella}, {Arenou}, {Mignard}, \&
  {Th{\'e}venin}}]{2019A&A...623A..72K}
{Kervella}, P., {Arenou}, F., {Mignard}, F., \& {Th{\'e}venin}, F. 2019, \aap,
  623, A72

\bibitem[{{Kipping}(2013)}]{2013MNRAS.435.2152K}
{Kipping}, D.~M. 2013, \mnras, 435, 2152

\bibitem[{{Kopparapu} {et~al.}(2013){Kopparapu}, {Ramirez}, {Kasting}, {Eymet},
  {Robinson}, {Mahadevan}, {Terrien}, {Domagal-Goldman}, {Meadows}, \&
  {Deshpande}}]{2013ApJ...765..131K}
{Kopparapu}, R.~K., {Ramirez}, R., {Kasting}, J.~F., {et~al.} 2013, \apj, 765,
  131

\bibitem[{{Kopparapu} {et~al.}(2014){Kopparapu}, {Ramirez}, {SchottelKotte},
  {Kasting}, {Domagal-Goldman}, \& {Eymet}}]{2014ApJ...787L..29K}
{Kopparapu}, R.~K., {Ramirez}, R.~M., {SchottelKotte}, J., {et~al.} 2014,
  \apjl, 787, L29

\bibitem[{{Kreidberg}(2015)}]{2015PASP..127.1161K}
{Kreidberg}, L. 2015, \pasp, 127, 1161

\bibitem[{{Leto} {et~al.}(1997){Leto}, {Pagano}, {Buemi}, \&
  {Rodono}}]{1997A&A...327.1114L}
{Leto}, G., {Pagano}, I., {Buemi}, C.~S., \& {Rodono}, M. 1997, \aap, 327, 1114

\bibitem[{{Liddle}(2007)}]{2007MNRAS.377L..74L}
{Liddle}, A.~R. 2007, \mnras, 377, L74

\bibitem[{{Lovis} \& {Pepe}(2007)}]{2007A&A...468.1115L}
{Lovis}, C. \& {Pepe}, F. 2007, \aap, 468, 1115

\bibitem[{{Luyten}(1979)}]{1979lccs.book.....L}
{Luyten}, W.~J. 1979, {LHS catalogue. A catalogue of stars with proper motions
  exceeding 0''5 annually}

\bibitem[{{Maldonado} {et~al.}(2015){Maldonado}, {Affer}, {Micela},
  {Scandariato}, {Damasso}, {Stelzer}, {Barbieri}, {Bedin}, {Biazzo},
  {Bignamini}, {Borsa}, {Claudi}, {Covino}, {Desidera}, {Esposito}, {Gratton},
  {Gonz{\'a}lez Hern{\'a}ndez}, {Lanza}, {Maggio}, {Molinari}, {Pagano},
  {Perger}, {Pillitteri}, {Piotto}, {Poretti}, {Prisinzano}, {Rebolo}, {Ribas},
  {Shkolnik}, {Southworth}, {Sozzetti}, \& {Su{\'a}rez
  Mascare{\~n}o}}]{2015A&A...577A.132M}
{Maldonado}, J., {Affer}, L., {Micela}, G., {et~al.} 2015, \aap, 577, A132

\bibitem[{{Maldonado} {et~al.}(2017){Maldonado}, {Scandariato}, {Stelzer},
  {Biazzo}, {Lanza}, {Maggio}, {Micela}, {Gonz{\'a}lez-{\'A}lvarez}, {Affer},
  {Claudi}, {Cosentino}, {Damasso}, {Desidera}, {Gonz{\'a}lez Hern{\'a}ndez},
  {Gratton}, {Leto}, {Messina}, {Molinari}, {Pagano}, {Perger}, {Piotto},
  {Rebolo}, {Ribas}, {Sozzetti}, {Su{\'a}rez Mascare{\~n}o}, \& {Zanmar
  Sanchez}}]{2017A&A...598A..27M}
{Maldonado}, J., {Scandariato}, G., {Stelzer}, B., {et~al.} 2017, \aap, 598,
  A27

\bibitem[{{Mortier} {et~al.}(2015){Mortier}, {Faria}, {Correia}, {Santerne}, \&
  {Santos}}]{2015A&A...573A.101M}
{Mortier}, A., {Faria}, J.~P., {Correia}, C.~M., {Santerne}, A., \& {Santos},
  N.~C. 2015, \aap, 573, A101

\bibitem[{{Osten} {et~al.}(2005){Osten}, {Hawley}, {Allred}, {Johns-Krull}, \&
  {Roark}}]{2005ApJ...621..398O}
{Osten}, R.~A., {Hawley}, S.~L., {Allred}, J.~C., {Johns-Krull}, C.~M., \&
  {Roark}, C. 2005, \apj, 621, 398

\bibitem[{{Perger} {et~al.}(2020){Perger}, {Anglada-Escud{\'e}}, {Ribas},
  {Rosich}, {Herrero}, \& {Morales}}]{2020arXiv201201862P}
{Perger}, M., {Anglada-Escud{\'e}}, G., {Ribas}, I., {et~al.} 2020, arXiv
  e-prints, arXiv:2012.01862

\bibitem[{{Perger} {et~al.}(2017){Perger}, {Garc{\'{\i}}a-Piquer}, {Ribas},
  {Morales}, {Affer}, {Micela}, {Damasso}, {Su{\'a}rez-Mascare{\~n}o},
  {Gonz{\'a}lez-Hern{\'a}ndez}, {Rebolo}, {Herrero}, {Rosich}, {Lafarga},
  {Bignamini}, {Sozzetti}, {Claudi}, {Cosentino}, {Molinari}, {Maldonado},
  {Maggio}, {Lanza}, {Poretti}, {Pagano}, {Desidera}, {Gratton}, {Piotto},
  {Bonomo}, {Martinez Fiorenzano}, {Giacobbe}, {Malavolta}, {Nascimbeni},
  {Rainer}, \& {Scandariato}}]{2017A&A...598A..26P}
{Perger}, M., {Garc{\'{\i}}a-Piquer}, A., {Ribas}, I., {et~al.} 2017, \aap,
  598, A26

\bibitem[{{Pinamonti} {et~al.}(2018){Pinamonti}, {Damasso}, {Marzari},
  {Sozzetti}, {Desidera}, {Maldonado}, {Scandariato}, {Affer}, {Lanza},
  {Bignamini}, {Bonomo}, {Borsa}, {Claudi}, {Cosentino}, {Giacobbe},
  {Gonz{\'a}lez-{\'A}lvarez}, {Gonz{\'a}lez Hern{\'a}ndez}, {Gratton}, {Leto},
  {Malavolta}, {Martinez Fiorenzano}, {Micela}, {Molinari}, {Pagano}, {Pedani},
  {Perger}, {Piotto}, {Rebolo}, {Ribas}, {Su{\'a}rez Mascare{\~n}o}, \&
  {Toledo-Padr{\'o}n}}]{2018A&A...617A.104P}
{Pinamonti}, M., {Damasso}, M., {Marzari}, F., {et~al.} 2018, \aap, 617, A104

\bibitem[{{Queloz} {et~al.}(2001){Queloz}, {Henry}, {Sivan}, {Baliunas},
  {Beuzit}, {Donahue}, {Mayor}, {Naef}, {Perrier}, \&
  {Udry}}]{2001A&A...379..279Q}
{Queloz}, D., {Henry}, G.~W., {Sivan}, J.~P., {et~al.} 2001, \aap, 379, 279

\bibitem[{{Reiners} {et~al.}(2012){Reiners}, {Joshi}, \&
  {Goldman}}]{2012AJ....143...93R}
{Reiners}, A., {Joshi}, N., \& {Goldman}, B. 2012, \aj, 143, 93

\bibitem[{{Reiners} {et~al.}(2018){Reiners}, {Zechmeister}, {Caballero},
  {Ribas}, {Morales}, {Jeffers}, {Sch{\"o}fer}, {Tal-Or}, {Quirrenbach},
  {Amado}, {Kaminski}, {Seifert}, {Abril}, {Aceituno}, {Alonso-Floriano},
  {Ammler-von Eiff}, {Antona}, {Anglada-Escud{\'e}}, {Anwand-Heerwart},
  {Arroyo-Torres}, {Azzaro}, {Baroch}, {Barrado}, {Bauer}, {Becerril},
  {B{\'e}jar}, {Ben{\'{\i}}tez}, {Berdi{\~n}as}, {Bergond}, {Bl{\"u}mcke},
  {Brinkm{\"o}ller}, {del Burgo}, {Cano}, {C{\'a}rdenas V{\'a}zquez}, {Casal},
  {Cifuentes}, {Claret}, {Colom{\'e}}, {Cort{\'e}s-Contreras}, {Czesla},
  {D{\'{\i}}ez-Alonso}, {Dreizler}, {Feiz}, {Fern{\'a}ndez}, {Ferro},
  {Fuhrmeister}, {Galad{\'{\i}}-Enr{\'{\i}}quez}, {Garcia-Piquer},
  {Garc{\'{\i}}a Vargas}, {Gesa}, {G{\'o}mez Galera}, {Gonz{\'a}lez
  Hern{\'a}ndez}, {Gonz{\'a}lez-Peinado}, {Gr{\"o}zinger}, {Grohnert},
  {Gu{\`a}rdia}, {Guenther}, {Guijarro}, {de Guindos}, {Guti{\'e}rrez-Soto},
  {Hagen}, {Hatzes}, {Hauschildt}, {Hedrosa}, {Helmling}, {Henning}, {Hermelo},
  {Hern{\'a}ndez Arab{\'{\i}}}, {Hern{\'a}ndez Casta{\~n}o}, {Hern{\'a}ndez
  Hernando}, {Herrero}, {Huber}, {Huke}, {Johnson}, {de Juan}, {Kim}, {Klein},
  {Kl{\"u}ter}, {Klutsch}, {K{\"u}rster}, {Lafarga}, {Lamert}, {Lamp{\'o}n},
  {Lara}, {Laun}, {Lemke}, {Lenzen}, {Launhardt}, {L{\'o}pez del Fresno},
  {L{\'o}pez-Gonz{\'a}lez}, {L{\'o}pez-Puertas}, {L{\'o}pez Salas},
  {L{\'o}pez-Santiago}, {Luque}, {Mag{\'a}n Madinabeitia}, {Mall}, {Mancini},
  {Mandel}, {Marfil}, {Mar{\'{\i}}n Molina}, {Maroto Fern{\'a}ndez},
  {Mart{\'{\i}}n}, {Mart{\'{\i}}n-Ruiz}, {Marvin}, {Mathar}, {Mirabet},
  {Montes}, {Moreno-Raya}, {Moya}, {Mundt}, {Nagel}, {Naranjo}, {Nortmann},
  {Nowak}, {Ofir}, {Oreiro}, {Pall{\'e}}, {Panduro}, {Pascual}, {Passegger},
  {Pavlov}, {Pedraz}, {P{\'e}rez-Calpena}, {P{\'e}rez Medialdea}, {Perger},
  {Perryman}, {Pluto}, {Rabaza}, {Ram{\'o}n}, {Rebolo}, {Redondo}, {Reffert},
  {Reinhart}, {Rhode}, {Rix}, {Rodler}, {Rodr{\'{\i}}guez},
  {Rodr{\'{\i}}guez-L{\'o}pez}, {Rodr{\'{\i}}guez Trinidad}, {Rohloff},
  {Rosich}, {Sadegi}, {S{\'a}nchez-Blanco}, {S{\'a}nchez Carrasco},
  {S{\'a}nchez-L{\'o}pez}, {Sanz-Forcada}, {Sarkis}, {Sarmiento},
  {Sch{\"a}fer}, {Schmitt}, {Schiller}, {Schweitzer}, {Solano}, {Stahl},
  {Strachan}, {St{\"u}rmer}, {Su{\'a}rez}, {Tabernero}, {Tala}, {Trifonov},
  {Tulloch}, {Ulbrich}, {Veredas}, {Vico Linares}, {Vilardell}, {Wagner},
  {Winkler}, {Wolthoff}, {Xu}, {Yan}, \& {Zapatero
  Osorio}}]{2018A&A...612A..49R}
{Reiners}, A., {Zechmeister}, M., {Caballero}, J.~A., {et~al.} 2018, \aap, 612,
  A49

\bibitem[{{Robertson} {et~al.}(2014){Robertson}, {Mahadevan}, {Endl}, \&
  {Roy}}]{2014Sci...345..440R}
{Robertson}, P., {Mahadevan}, S., {Endl}, M., \& {Roy}, A. 2014, Science, 345,
  440

\bibitem[{{Scandariato} {et~al.}(2017){Scandariato}, {Maldonado}, {Affer},
  {Biazzo}, {Leto}, {Stelzer}, {Zanmar Sanchez}, {Claudi}, {Cosentino},
  {Damasso}, {Desidera}, {Gonz{\'a}lez {\'A}lvarez}, {Gonz{\'a}lez
  Hern{\'a}ndez}, {Gratton}, {Lanza}, {Maggio}, {Messina}, {Micela}, {Pagano},
  {Perger}, {Piotto}, {Rebolo}, {Ribas}, {Rosich}, {Sozzetti}, \& {Su{\'a}rez
  Mascare{\~n}o}}]{2017A&A...598A..28S}
{Scandariato}, G., {Maldonado}, J., {Affer}, L., {et~al.} 2017, \aap, 598, A28

\bibitem[{{Schweitzer} {et~al.}(2019){Schweitzer}, {Passegger}, {Cifuentes},
  {B{\'e}jar}, {Cort{\'e}s-Contreras}, {Caballero}, {del Burgo}, {Czesla},
  {K{\"u}rster}, {Montes}, {Zapatero Osorio}, {Ribas}, {Reiners},
  {Quirrenbach}, {Amado}, {Aceituno}, {Anglada-Escud{\'e}}, {Bauer},
  {Dreizler}, {Jeffers}, {Guenther}, {Henning}, {Kaminski}, {Lafarga},
  {Marfil}, {Morales}, {Schmitt}, {Seifert}, {Solano}, {Tabernero}, \&
  {Zechmeister}}]{2019A&A...625A..68S}
{Schweitzer}, A., {Passegger}, V.~M., {Cifuentes}, C., {et~al.} 2019, \aap,
  625, A68

\bibitem[{{Smith} \& {WASP Consortium}(2014)}]{2014CoSka..43..500S}
{Smith}, A.~M.~S. \& {WASP Consortium}. 2014, Contributions of the Astronomical
  Observatory Skalnate Pleso, 43, 500

\bibitem[{{Smith} {et~al.}(2012){Smith}, {Stumpe}, {Van Cleve}, {Jenkins},
  {Barclay}, {Fanelli}, {Girouard}, {Kolodziejczak}, {McCauliff}, {Morris}, \&
  {Twicken}}]{2012PASP..124.1000S}
{Smith}, J.~C., {Stumpe}, M.~C., {Van Cleve}, J.~E., {et~al.} 2012, \pasp, 124,
  1000

\bibitem[{Sokal(1996)}]{Sokal1996MonteCM}
Sokal, A. 1996, in Monte Carlo Methods in Statistical Mechanics: Foundations
  and New Algorithms Note to the Reader

\bibitem[{{Sozzetti} {et~al.}(2013){Sozzetti}, {Bernagozzi}, {Bertolini},
  {Calcidese}, {Carbognani}, {Cenadelli}, {Christille}, {Damasso}, {Giacobbe},
  {Lanteri}, {Lattanzi}, \& {Smart}}]{2013EPJWC..4703006S}
{Sozzetti}, A., {Bernagozzi}, A., {Bertolini}, E., {et~al.} 2013, in European
  Physical Journal Web of Conferences, Vol.~47, European Physical Journal Web
  of Conferences, 03006

\bibitem[{{Stumpe} {et~al.}(2014){Stumpe}, {Smith}, {Catanzarite}, {Van Cleve},
  {Jenkins}, {Twicken}, \& {Girouard}}]{2014PASP..126..100S}
{Stumpe}, M.~C., {Smith}, J.~C., {Catanzarite}, J.~H., {et~al.} 2014, \pasp,
  126, 100

\bibitem[{{Su{\'a}rez Mascare{\~n}o} {et~al.}(2017){Su{\'a}rez Mascare{\~n}o},
  {Gonz{\'a}lez Hern{\'a}ndez}, {Rebolo}, {Velasco}, {Toledo-Padr{\'o}n},
  {Affer}, {Perger}, {Micela}, {Ribas}, {Maldonado}, {Leto}, {Zanmar Sanchez},
  {Scandariato}, {Damasso}, {Sozzetti}, {Esposito}, {Covino}, {Maggio},
  {Lanza}, {Desidera}, {Rosich}, {Bignamini}, {Claudi}, {Benatti}, {Borsa},
  {Pedani}, {Molinari}, {Morales}, {Herrero}, \&
  {Lafarga}}]{2017A&A...605A..92S}
{Su{\'a}rez Mascare{\~n}o}, A., {Gonz{\'a}lez Hern{\'a}ndez}, J.~I., {Rebolo},
  R., {et~al.} 2017, \aap, 605, A92

\bibitem[{{Su{\'a}rez Mascare{\~n}o} {et~al.}(2018){Su{\'a}rez Mascare{\~n}o},
  {Rebolo}, {Gonz{\'a}lez Hern{\'a}ndez}, {Toledo-Padr{\'o}n}, {Perger},
  {Ribas}, {Affer}, {Micela}, {Damasso}, {Maldonado}, {Gonz{\'a}lez-Alvarez},
  {Leto}, {Pagano}, {Scandariato}, {Sozzetti}, {Lanza}, {Malavolta}, {Claudi},
  {Cosentino}, {Desidera}, {Giacobbe}, {Maggio}, {Rainer}, {Esposito},
  {Benatti}, {Pedani}, {Morales}, {Herrero}, {Lafarga}, {Rosich}, \&
  {Pinamonti}}]{2018A&A...612A..89S}
{Su{\'a}rez Mascare{\~n}o}, A., {Rebolo}, R., {Gonz{\'a}lez Hern{\'a}ndez},
  J.~I., {et~al.} 2018, \aap, 612, A89

\bibitem[{{Tuomi} {et~al.}(2014){Tuomi}, {Jones}, {Barnes},
  {Anglada-Escud{\'e}}, \& {Jenkins}}]{2014MNRAS.441.1545T}
{Tuomi}, M., {Jones}, H. R.~A., {Barnes}, J.~R., {Anglada-Escud{\'e}}, G., \&
  {Jenkins}, J.~S. 2014, \mnras, 441, 1545

\bibitem[{{Zechmeister} \& {K{\"u}rster}(2009)}]{2009A&A...496..577Z}
{Zechmeister}, M. \& {K{\"u}rster}, M. 2009, \aap, 496, 577

\end{thebibliography}

%\newpage
%----------------------------------------------------------------------------------------
%	APPENDIX
%----------------------------------------------------------------------------------------

\begin{appendix} %First appendix
\section{Tables}

%GJ 720 A rv and activity data

\longtab[1]{

\begin{scriptsize}
\begin{longtable}{lccccccr}
\caption{\label{tab:gj720a_rv_act_data} GJ 720 A data of the HARPS-N observations.}\\
\hline\hline
\noalign{\smallskip}
BJD-2,400,000 & RV & eRV & NaD1 & NaD2  & S-index & eS-index &  H$\alpha$ \\
(d)  & (m\,s$^{-1}$) &  (m\,s$^{-1}$) &  &  &  & \\
 \noalign{\smallskip}
\hline
\noalign{\smallskip}
\endfirsthead
\caption{continued.}\\
\hline\hline
\noalign{\smallskip}
BJD-2,400,000 & RV & eRV & NaD1 & NaD2  & S-index & eS-index &  H$\alpha$ \\
(d)  & (m\,s$^{-1}$) &  (m\,s$^{-1}$) &  &  &  & \\
\noalign{\smallskip}
\hline
\endhead
\hline
\endfoot
56438.6283 & 0.8556 & 0.6549 & 0.4531 & 0.5853 & 0.8289 & 0.0040 & 0.8039 \\ 
56440.6376 & -3.0450 & 0.9396 & 0.4540 & 0.5840 & 0.7878 & 0.0057 & 0.8023 \\ 
56442.6065 & -0.0730 & 0.8321 & 0.4523 & 0.5840 & 0.8007 & 0.0046 & 0.8051 \\ 
56443.5709 & -5.4419 & 0.7409 & 0.4533 & 0.5829 & 0.7591 & 0.0062 & 0.8028 \\ 
56444.5464 & -2.4062 & 0.6942 & 0.4512 & 0.5831 & 0.7791 & 0.0046 & 0.8079 \\ 
56484.4726 & -4.1991 & 0.8349 & 0.4519 & 0.5888 & 0.7657 & 0.0053 & 0.8061 \\ 
56486.5896 & -5.7892 & 0.7041 & 0.4510 & 0.5850 & 0.7548 & 0.0049 & 0.8076 \\ 
56508.5780 & -3.9796 & 0.9532 & 0.4535 & 0.5896 & 0.8557 & 0.0057 & 0.8139 \\ 
56509.5798 & -5.1460 & 0.6858 & 0.4538 & 0.5920 & 0.8653 & 0.0056 & 0.8165 \\ 
56533.4210 & -3.4373 & 0.6672 & 0.4526 & 0.5808 & 0.7794 & 0.0043 & 0.8148 \\ 
56534.4324 & -3.1517 & 0.8646 & 0.4521 & 0.5777 & 0.7576 & 0.0057 & 0.8160 \\ 
56534.4956 & -3.6094 & 0.8079 & 0.4527 & 0.5813 & 0.7601 & 0.0055 & 0.8154 \\ 
56535.3663 & -0.7889 & 0.8062 & 0.4536 & 0.5817 & 0.7703 & 0.0058 & 0.8163 \\ 
56535.5215 & -1.0090 & 0.6777 & 0.4540 & 0.5827 & 0.7667 & 0.0044 & 0.8183 \\ 
56775.6149 & 2.1136 & 1.1157 & 0.4573 & 0.5882 & 0.7240 & 0.0097 & 0.8042 \\ 
56786.6700 & -3.5345 & 0.9845 & 0.4556 & 0.5865 & 0.7900 & 0.0082 & 0.8053 \\ 
56787.5918 & -0.7065 & 0.8225 & 0.4532 & 0.5868 & 0.7936 & 0.0065 & 0.8051 \\ 
56792.5339 & 0.7418 & 1.0174 & 0.4581 & 0.5968 & 0.8337 & 0.0089 & 0.8063 \\ 
56811.6383 & 4.0310 & 0.7325 & 0.4548 & 0.5843 & 0.8468 & 0.0056 & 0.8068 \\ 
56854.5890 & -1.4405 & 0.6970 & 0.4545 & 0.5847 & 0.8607 & 0.0049 & 0.7978 \\ 
56857.5711 & -5.7692 & 0.6496 & 0.4560 & 0.5840 & 0.8697 & 0.0051 & 0.7942 \\ 
56858.5186 & -4.3603 & 0.6858 & 0.4541 & 0.5831 & 0.8706 & 0.0052 & 0.7956 \\ 
56859.5271 & -4.2436 & 0.8009 & 0.4556 & 0.5878 & 0.8678 & 0.0061 & 0.7987 \\ 
56860.4789 & -3.9977 & 0.5824 & 0.4533 & 0.5858 & 0.8787 & 0.0045 & 0.7977 \\ 
56861.4923 & -1.4096 & 0.6056 & 0.4551 & 0.5822 & 0.8802 & 0.0048 & 0.7990 \\ 
56874.4739 & -0.7580 & 0.6499 & 0.4574 & 0.5857 & 0.9098 & 0.0054 & 0.8053 \\ 
56875.4972 & -5.1943 & 1.2135 & 0.4579 & 0.5935 & 0.9659 & 0.0111 & 0.8055 \\ 
56876.4863 & -4.0069 & 0.9428 & 0.4556 & 0.6003 & 0.9463 & 0.0088 & 0.8065 \\ 
56877.4838 & -2.8328 & 0.7407 & 0.4535 & 0.5906 & 0.8995 & 0.0054 & 0.8052 \\ 
57113.6697 & -4.0459 & 0.8306 & 0.4570 & 0.5831 & 0.9115 & 0.0076 & 0.7976 \\ 
57114.7286 & -5.8965 & 0.9335 & 0.4707 & 0.6021 & 0.8533 & 0.0102 & 0.7974 \\ 
57137.6987 & 0.4611 & 0.9086 & 0.4530 & 0.5858 & 0.8346 & 0.0055 & 0.8006 \\ 
57139.7151 & 4.5178 & 0.6327 & 0.4541 & 0.5843 & 0.8100 & 0.0056 & 0.8001 \\ 
57142.7166 & 7.4822 & 0.7658 & 0.4531 & 0.5882 & 0.8188 & 0.0055 & 0.8001 \\ 
57143.6682 & 3.3888 & 0.6213 & 0.4551 & 0.5875 & 0.8208 & 0.0054 & 0.8030 \\ 
57144.6615 & 3.1809 & 0.6798 & 0.4544 & 0.5884 & 0.8250 & 0.0070 & 0.8005 \\ 
57145.6867 & 3.1437 & 0.8123 & 0.4609 & 0.6055 & 0.8232 & 0.0077 & 0.8018 \\ 
57147.7393 & -1.2659 & 0.7483 & 0.4559 & 0.5908 & 0.8100 & 0.0056 & 0.7980 \\ 
57148.6849 & -5.9414 & 0.7604 & 0.4551 & 0.5869 & 0.8291 & 0.0054 & 0.8006 \\ 
57170.6373 & -6.4304 & 0.6514 & 0.4515 & 0.5838 & 0.8399 & 0.0051 & 0.7952 \\ 
57172.6821 & -2.8096 & 0.8633 & 0.4536 & 0.5842 & 0.8583 & 0.0061 & 0.7997 \\ 
57173.5731 & -1.7334 & 0.8738 & 0.4529 & 0.5811 & 0.8339 & 0.0073 & 0.7938 \\ 
57175.5927 & 4.2651 & 0.6846 & 0.4558 & 0.5851 & 0.7993 & 0.0071 & 0.7950 \\ 
57176.5824 & 3.0572 & 0.8997 & 0.4544 & 0.5902 & 0.8487 & 0.0083 & 0.7967 \\ 
57178.6893 & 2.8957 & 0.8069 & 0.4547 & 0.5901 & 0.8697 & 0.0067 & 0.7984 \\ 
57204.5383 & 0.2218 & 0.6601 & 0.4560 & 0.5836 & 0.9393 & 0.0060 & 0.8002 \\ 
57205.5096 & -3.3256 & 0.8361 & 0.4544 & 0.5910 & 0.8940 & 0.0060 & 0.8032 \\ 
57206.5318 & -3.4802 & 0.8711 & 0.4560 & 0.5843 & 0.9110 & 0.0061 & 0.7974 \\ 
57207.4762 & -6.2995 & 0.7452 & 0.4554 & 0.5858 & 0.8685 & 0.0059 & 0.7977 \\ 
57208.5256 & -7.9140 & 1.0426 & 0.4540 & 0.5879 & 0.8754 & 0.0079 & 0.7965 \\ 
57209.6188 & -5.2291 & 0.9096 & 0.4550 & 0.5843 & 0.8657 & 0.0075 & 0.7994 \\ 
57239.4941 & 3.8668 & 0.5425 & 0.4588 & 0.5871 & 0.8993 & 0.0053 & 0.8031 \\ 
57240.4922 & 4.5738 & 0.6486 & 0.4563 & 0.5931 & 0.9140 & 0.0056 & 0.8084 \\ 
57241.5110 & 0.2247 & 0.9067 & 0.4559 & 0.5849 & 0.8829 & 0.0067 & 0.8003 \\ 
57242.4737 & -2.8127 & 2.1974 & 0.4697 & 0.6022 & 1.4265 & 0.0286 & 0.8034 \\ 
57249.5255 & -5.6587 & 0.6795 & 0.4543 & 0.5997 & 0.8689 & 0.0068 & 0.8083 \\ 
57250.4923 & -2.2261 & 0.9313 & 0.4574 & 0.5977 & 0.8539 & 0.0081 & 0.8074 \\ 
57251.4762 & -4.7863 & 0.6921 & 0.4537 & 0.5836 & 0.8432 & 0.0047 & 0.8043 \\ 
57258.4289 & 3.9843 & 0.6944 & 0.4574 & 0.5842 & 0.8789 & 0.0057 & 0.8035 \\ 
57259.4298 & 2.5805 & 0.7458 & 0.4564 & 0.5818 & 0.8292 & 0.0054 & 0.8045 \\ 
57260.4874 & 2.0615 & 0.6685 & 0.4554 & 0.5866 & 0.9011 & 0.0047 & 0.8092 \\ 
57261.4934 & 1.1212 & 0.7153 & 0.4549 & 0.5852 & 0.9128 & 0.0050 & 0.8062 \\ 
57262.4796 & -0.0115 & 0.8523 & 0.4569 & 0.5874 & 0.9662 & 0.0070 & 0.8159 \\ 
57263.4798 & 1.6670 & 0.6632 & 0.4551 & 0.5940 & 0.9088 & 0.0058 & 0.8124 \\ 
57264.4790 & -1.3328 & 0.7019 & 0.4562 & 0.5945 & 0.9352 & 0.0068 & 0.8138 \\ 
57274.4658 & 0.7708 & 0.8334 & 0.4590 & 0.5897 & 0.9777 & 0.0070 & 0.8094 \\ 
57275.4619 & 3.2613 & 0.6845 & 0.4554 & 0.5897 & 0.9435 & 0.0061 & 0.8090 \\ 
57276.4603 & 4.8241 & 0.6564 & 0.4589 & 0.5960 & 0.9247 & 0.0059 & 0.8089 \\ 
57277.4582 & 4.7622 & 0.6351 & 0.4571 & 0.5861 & 0.8972 & 0.0049 & 0.8062 \\ 
57282.4744 & 4.7420 & 1.1354 & 0.4568 & 0.5829 & 0.8610 & 0.0052 & 0.8058 \\ 
57285.4754 & -5.3744 & 0.9515 & 0.4560 & 0.5917 & 0.8352 & 0.0085 & 0.8122 \\ 
57286.4849 & -5.3450 & 0.7508 & 0.4552 & 0.5878 & 0.8516 & 0.0065 & 0.8086 \\ 
57287.4630 & -5.0051 & 0.7946 & 0.4536 & 0.5888 & 0.8447 & 0.0063 & 0.8086 \\ 
57290.5193 & -1.8430 & 0.7608 & 0.4559 & 0.5891 & 0.8549 & 0.0078 & 0.8091 \\ 
57291.4861 & -2.2635 & 0.7752 & 0.4553 & 0.5888 & 0.8829 & 0.0074 & 0.8090 \\ 
57293.4607 & 0.0000 & 0.6706 & 0.4566 & 0.5860 & 0.8294 & 0.0072 & 0.8046 \\ 
57294.4756 & 4.9841 & 1.0672 & 0.4635 & 0.5879 & 0.9403 & 0.0106 & 0.8086 \\ 
57296.4301 & 9.1730 & 1.1774 & 0.4638 & 0.5932 & 0.9673 & 0.0102 & 0.8110 \\ 
57303.4081 & -0.2229 & 0.7379 & 0.4581 & 0.5862 & 0.9215 & 0.0057 & 0.8080 \\ 
57472.7269 & 4.6796 & 1.1044 & 0.4591 & 0.5887 & 0.8474 & 0.0100 & 0.8155 \\ 
57474.7285 & 6.5082 & 0.6286 & 0.4566 & 0.5919 & 0.8705 & 0.0054 & 0.8169 \\ 
57475.7086 & 7.2436 & 0.9217 & 0.4597 & 0.5860 & 0.8609 & 0.0074 & 0.8144 \\ 
57607.5019 & 3.7040 & 1.1875 & 0.4580 & 0.5956 & 0.8568 & 0.0088 & 0.8154 \\ 
57608.4638 & 2.4916 & 0.7224 & 0.4571 & 0.5901 & 0.8072 & 0.0053 & 0.8145 \\ 
57609.5581 & 4.0824 & 1.6268 & 0.4688 & 0.5973 & 0.8372 & 0.0141 & 0.8139 \\ 
57935.6549 & -0.6930 & 0.9038 & 0.4557 & 0.5903 & 0.9079 & 0.0075 & 0.8024 \\ 
57936.5276 & 1.5962 & 0.8964 & 0.4561 & 0.5887 & 0.9448 & 0.0088 & 0.8086 \\ 
57944.5151 & 2.8329 & 0.5128 & 0.4561 & 0.5854 & 0.9470 & 0.0057 & 0.8103 \\ 
57971.3846 & -2.5676 & 0.8980 & 0.4538 & 0.5881 & 0.9127 & 0.0064 & 0.8089 \\ 
57974.4582 & 1.8452 & 0.8170 & 0.4566 & 0.5925 & 0.8635 & 0.0068 & 0.8069 \\ 
57977.4881 & 4.2857 & 0.8850 & 0.4567 & 0.5961 & 0.9621 & 0.0075 & 0.8117 \\ 
57978.4393 & 5.6847 & 0.7480 & 0.4568 & 0.5924 & 0.9582 & 0.0056 & 0.8090 \\ 
57979.4464 & 8.4808 & 0.7172 & 0.4584 & 0.5886 & 0.9438 & 0.0049 & 0.8094 \\ 
57980.4294 & 5.6384 & 0.8731 & 0.4603 & 0.5888 & 0.9438 & 0.0061 & 0.8128 \\ 
57981.4270 & 7.7884 & 0.8546 & 0.4583 & 0.5875 & 0.9166 & 0.0066 & 0.8077 \\ 
57982.4171 & 4.5221 & 1.0451 & 0.4610 & 0.5958 & 1.0005 & 0.0082 & 0.8147 \\ 
57984.4080 & 1.2098 & 0.7717 & 0.4566 & 0.5879 & 0.9227 & 0.0053 & 0.8092 \\ 
57989.3714 & -3.6573 & 0.9132 & 0.4568 & 0.5953 & 0.8604 & 0.0068 & 0.8169 \\ 
57993.4588 & 0.5419 & 0.6713 & 0.4556 & 0.5913 & 0.9000 & 0.0068 & 0.8187 \\ 
57996.4369 & 4.6074 & 1.0328 & 0.4587 & 0.5883 & 0.8542 & 0.0078 & 0.8111 \\ 
58000.4093 & 6.8929 & 1.0264 & 0.4588 & 0.5911 & 0.8793 & 0.0079 & 0.8115 \\ 
58005.4765 & -5.8464 & 0.8280 & 0.4576 & 0.5943 & 0.9780 & 0.0066 & 0.8109 \\ 
58006.4841 & -3.9495 & 0.8971 & 0.4567 & 0.5937 & 0.9824 & 0.0070 & 0.8091 \\ 
58007.4650 & -7.5220 & 0.8376 & 0.4588 & 0.5879 & 0.9833 & 0.0068 & 0.8083 \\ 
58008.4648 & -3.2108 & 0.8609 & 0.4584 & 0.5904 & 0.9610 & 0.0051 & 0.8108 \\ 
58009.4755 & -4.3398 & 1.1500 & 0.4658 & 0.5906 & 1.0225 & 0.0099 & 0.8064 \\ 
58010.4846 & -4.2457 & 0.7485 & 0.4620 & 0.5900 & 1.0320 & 0.0077 & 0.8192 \\ 
58022.3695 & -0.2366 & 0.7895 & 0.4590 & 0.5898 & 0.9144 & 0.0056 & 0.8119 \\ 
58024.4562 & -1.6361 & 0.9246 & 0.4585 & 0.5894 & 0.9257 & 0.0061 & 0.8126 \\ 
58026.3661 & -4.9618 & 0.7807 & 0.4570 & 0.5847 & 0.8861 & 0.0051 & 0.8122 \\ 
58031.4482 & -4.1795 & 0.8973 & 0.4598 & 0.5856 & 0.8790 & 0.0077 & 0.8130 \\ 
58037.4262 & 1.8943 & 1.1604 & 0.4631 & 0.5910 & 0.9526 & 0.0096 & 0.8099 \\ 
58044.3890 & -4.5156 & 0.9186 & 0.4606 & 0.5879 & 0.9441 & 0.0062 & 0.8091 \\ 
58333.5067 & 2.5931 & 0.7935 & 0.4628 & 0.5866 & 1.0021 & 0.0059 & 0.8091 \\ 
58334.5021 & 1.1240 & 0.8291 & 0.4598 & 0.5887 & 0.9903 & 0.0056 & 0.8092 \\ 
58335.4729 & -0.0012 & 1.8726 & 0.4696 & 0.6122 & 1.0690 & 0.0157 & 0.8101 \\ 
58403.3701 & 4.3495 & 0.9156 & 0.4615 & 0.5880 & 1.0956 & 0.0069 & 0.8172 \\ 
58406.3686 & 7.5291 & 0.7414 & 0.4633 & 0.5913 & 1.0581 & 0.0065 & 0.8145 \\ 
58700.5049 & 8.2617 & 1.3557 & 0.4645 & 0.5901 & 1.1238 & 0.0127 & 0.8034 \\ 
58700.5160 & 6.8301 & 1.7573 & 0.4668 & 0.5946 & 1.1076 & 0.0163 & 0.8053 \\ 
58724.4723 & 2.4381 & 0.8577 & 0.4606 & 0.5944 & 1.1096 & 0.0064 & 0.8098 \\ 
58725.4676 & 1.6989 & 0.8991 & 0.4608 & 0.5916 & 1.0699 & 0.0071 & 0.8037 \\ 
58791.3321 & 6.0912 & 1.4268 & 0.4631 & 0.5969 & 1.1541 & 0.0095 & 0.8232 \\ 
58792.3266 & 6.2628 & 1.1013 & 0.4653 & 0.5944 & 1.0891 & 0.0081 & 0.8185 \\ 
59037.4711 & -1.3563 & 1.0557 & 0.4621 & 0.5970 & 1.0555 & 0.0073 & 0.8157 \\ 
59038.5999 & -1.8614 & 1.3793 & 0.4637 & 0.5986 & 1.1123 & 0.0096 & 0.8177 \\ 
59039.6133 & 2.4374 & 1.0650 & 0.4580 & 0.6015 & 1.1118 & 0.0065 & 0.8198 \\ 
59040.6150 & 2.8533 & 1.1285 & 0.4604 & 0.6013 & 1.1343 & 0.0093 & 0.8190 \\ 
59068.5029 & 7.2526 & 0.9858 & 0.4603 & 0.5966 & 1.1570 & 0.0077 & 0.8207 \\ 
59069.5561 & 2.5671 & 0.7651 & 0.4594 & 0.5999 & 1.1577 & 0.0068 & 0.8256 \\ 
59070.5927 & 4.8134 & 0.7815 & 0.4610 & 0.5930 & 1.1337 & 0.0066 & 0.8220 \\ 
59072.5652 & 9.1256 & 0.8643 & 0.4634 & 0.5926 & 1.2270 & 0.0074 & 0.8355 \\ 

\end{longtable}
\end{scriptsize}

}

%Priors S-index model
\begin{table*}[t]
\centering
\caption{Priors used for the S-index model with {\tt juliet}.}
\label{tab:gj720a_priors_details_sindex}
\begin{tabular}{l c c r}

\hline
\hline
\noalign{\smallskip}
Parameter & Prior & Unit & Description \\
\noalign{\smallskip}	
\hline	
\noalign{\smallskip}
\multicolumn{4}{c}{\textit{S-index parameters}} \\
\noalign{\smallskip}
$\gamma_0$ &  $\mathcal{U}$(-10, 10) & m$\rm s^{-1}$ & zero point for S-index\\
$\sigma$		 & $\mathcal{L} \mathcal{U}$(0.01, 10) & m$\rm s^{-1}$ & Extra jitter term for S-index \\
\noalign{\smallskip}							    
\multicolumn{4}{c}{\textit{GP parameters}} \\
\noalign{\smallskip}		
$\sigma_{\rm GP}$	 & $\mathcal{U}$(0, 20)   & m$\rm s^{-1}$ & Amplitude of the GP for the S-index\\
$\alpha_{\rm GP}$	&  $\mathcal{L} \mathcal{U}$ ($10^{-5}$, 1) & $\rm d^{-2}$    &  Inverse (squared) length-scale of the external parameter\\
$\Gamma_{\rm GP}$	& $\mathcal{L} \mathcal{U}$(0.01, 100) & ... & Amplitude of the sine-part of the kernel\\
$P_{\rm rot, GP}$ & $\mathcal{U}$(1, 500) & d & Period of the GP quasi-periodic component for the S-index\\ 		
\noalign{\smallskip}	
\hline	
\noalign{\smallskip}
\end{tabular}
\tablefoot{The prior labels of $\mathcal{U}$ and $\mathcal{L} \mathcal{U}$ represent uniform and loguniform distribution, respectively.}
\end{table*}

%PriorsBM+GP+1pl juliet
\begin{table*}[t]
%\begin{small}
\centering
\caption{Priors used for GJ~720~Ab fitting the RV+GP model with {\tt juliet}. The model statistically prefered and used to determine the final planetary orbital parameters.} 
\label{tab:gj720a_priors_details}
\begin{tabular}{l c c r}

\hline
\hline
\noalign{\smallskip}
Parameter & Prior & Unit & Description \\
\noalign{\smallskip}	
\hline	
\noalign{\smallskip}
\multicolumn{4}{c}{\textit{RV parameters}} \\
\noalign{\smallskip}
$\gamma_0$ &  $\mathcal{U}$(-10, 10) & m$\rm s^{-1}$ & RV zero point for HARPS-N\\
$\sigma$		 & $\mathcal{U}$(0.01, 10) & m$\rm s^{-1}$ & Extra jitter term for HARPS-N \\
\noalign{\smallskip}							    
\multicolumn{4}{c}{\textit{GP parameters}} \\
\noalign{\smallskip}		
$\sigma_{\rm GP,RV}$	 & $\mathcal{U}$(0, 10)   & m$\rm s^{-1}$ & Amplitude of the GP for the RVs\\
$\alpha_{\rm GP,RV}$	&  $\mathcal{J}$ ($10^{-11}$, $10^{-6}$) & $\rm d^{-2}$    &  Inverse (squared) length-scale of the external parameter\\
$\Gamma_{\rm GP,RV}$	& $\mathcal{J}$(0.01, 100) & ... & Amplitude of the sine-part of the kernel\\
$P_{\rm rot, GP,RV}$ & $\mathcal{U}$(30, 50) & d & Period of the GP quasi-periodic component for the RVs\\
\noalign{\smallskip}							    
\multicolumn{4}{c}{\textit{Planet parameters} }\\	
\noalign{\smallskip}										    
$P$ 			& $\mathcal{N}$ (19.5, 0.5) 	& d	& Period of planet b \\
$T_0$ (BJD-2,456,400) & $\mathcal{U}$ (0,15)		& d &  Time of periastron passage\\
$e$ 				& $\mathcal{U}$ (0, 0.8)		& ... & Orbital eccentricity of planet b\\
$\omega$ 	&	$\mathcal{U}$ (0, 360)	& deg	& Periastron angle of planet b\\
$K$		&  $\mathcal{U}$ (0, 10)  	&  m$\rm s^{-1}$      & RV semi-amplitude of planet b   \\    				
\noalign{\smallskip}	
\hline	
\noalign{\smallskip}
\end{tabular}
\tablefoot{The prior labels of $\mathcal{N}$, $\mathcal{U}$ and $\mathcal{J}$ represent normal, uniform, and jeffrey distribution, respectively. The reference time for $T_0$ is BJD-2,456,400. }
%\end{small}
\end{table*}

%Summary of different models
\begin{table*}[t]
\centering
\caption{Summary of all hyperparameters of the different models applied to GJ~720~A RVs implemented with {\tt juliet} with their corresponding priors and uncertainties.}
\label{tab:summary_prios}
\begin{tiny}
\begin{tabular}{l c c c c c c c c c}
\hline
\hline
\noalign{\smallskip}
 				 & $\sigma_{\rm GP}$ &  $\alpha_{\rm GP}$	& $\Gamma_{\rm GP}$ & $P_{\rm rot, GP}$ & $P$ & $T_0$ & $e$ & $\omega$ & $K$ \\
  				 & (m$\rm s^{-1}$)  &  ($\rm d^{-2}$) 	& ... & (d) & (d) & (BJD) & ... & (deg) & (m$\rm s^{-1}$)  \\				 
\noalign{\smallskip}	
\hline	
\noalign{\smallskip}
BM+GP & $\mathcal{U}$(0, 15) & $\mathcal{J}$ ($10^{-20}$, $10^{-5}$) & $\mathcal{J}$(0.01, 100) & $\mathcal{U}$(1, 1000) & ... & ... & ... & ... & ... \\
BM+GP+LT$^{*}$ & $\mathcal{U}$(0, 15) & $\mathcal{J}$ ($10^{-20}$, $10^{-3}$) & $\mathcal{J}$(0.01, 15) & $\mathcal{U}$(1, 1000) & ... & ... & ... & ... & ... \\

\noalign{\smallskip}	
\hline	
\noalign{\smallskip}

BM+1pl & ... & ... & ... & ... & $\mathcal{U}$ (1, 50) & $\mathcal{U}$ (0,50) & $\mathcal{U}$ (0, 0.8) & $\mathcal{U}$ (0, 360) & $\mathcal{U}$ (0, 10)  \\

\noalign{\smallskip}	
\hline	
\noalign{\smallskip}

BM+GP+1pl$^{(1)}$& $\mathcal{U}$(0, 10) & $\mathcal{J}$ ($10^{-11}$, $10^{-3}$) & $\mathcal{J}$(0.01, 100) & $\mathcal{U}$(1, 1000) & $\mathcal{N}$ (19.5, 0.5) & $\mathcal{U}$ (0,15) & $\mathcal{U}$ (0, 0.8) & $\mathcal{U}$ (0, 360) & $\mathcal{U}$ (0, 10) \\
BM+GP+1pl$^{(2)}$ & $\mathcal{U}$(0, 10) & $\mathcal{J}$ ($10^{-11}$, $10^{-6}$) & $\mathcal{J}$(0.01, 100) & $\mathcal{U}$(30, 50) & $\mathcal{N}$ (19.5, 0.5) & $\mathcal{U}$ (0,15) & $\mathcal{U}$ (0, 0.8) & $\mathcal{U}$ (0, 360) & $\mathcal{U}$ (0, 10)  \\
BM+GP+1pl+LT$^{*}$ & $\mathcal{U}$(0, 15) & $\mathcal{J}$ ($10^{-20}$, $10^{-3}$) & $\mathcal{J}$(0.01, 100) & $\mathcal{U}$(1, 1000) & $\mathcal{N}$ (19.5, 0.5) & $\mathcal{U}$ (0,15) & $\mathcal{U}$ (0, 0.8) & $\mathcal{U}$ (0, 360) & $\mathcal{U}$ (0, 10) \\

\noalign{\smallskip}	
\hline	
\noalign{\smallskip}

BM+2pl & ... & ... & ... & ... & $\mathcal{N}$ (19.5, 0.5) & $\mathcal{U}$ (0,30) & $\mathcal{U}$ (0, 0.8) & $\mathcal{U}$ (0, 360) & $\mathcal{U}$ (0, 10) \\
			   & ... & ... & ... & ... & $\mathcal{N}$ (42.1, 0.1) & $\mathcal{U}$ (0,40) & $\mathcal{U}$ (0, 0.8) & $\mathcal{U}$ (0, 360) & $\mathcal{U}$ (0, 10) \\
BM+GP+2pl & $\mathcal{U}$(0, 15) & $\mathcal{J}$ ($10^{-11}$, $10^{-3}$) & $\mathcal{J}$(0.01, 100) & $\mathcal{U}$(1, 1000) &  $\mathcal{N}$ (19.5, 0.5) & $\mathcal{U}$ (0,15) & $\mathcal{U}$ (0, 0.8) & $\mathcal{U}$ (0, 360) & $\mathcal{U}$ (0, 10) \\
 			& ... & ... & ... & ... &  $\mathcal{N}$ (42.1, 0.5) & $\mathcal{U}$ (0,50) & ... & ... & $\mathcal{U}$ (0, 10) \\

\noalign{\smallskip}	
\hline	
\noalign{\smallskip}
\end{tabular}
\end{tiny} 
\tablefoot{$^{*}$ The LT hyperparameters are $\rm rv_{interc}$ and $\rm rv_{slope}$ indicating the intercept and slope parameters, respectively. Their priors were set with a uniform distribution, $\mathcal{U}$(-100, 100). $^{(1), (2)}$: The difference between these two models is the prior of the GP $\rm P_{rot}$ parameter. For the first model the $\rm P_{rot}$ was set with a wide prior value (1--1000)\,d (including all possible RV signals) while the second one was set around the value of the possible rotation period of the star (30--50)\,d. The prior labels of $\mathcal{N}$, $\mathcal{U}$ and $\mathcal{J}$ represent normal, uniform, and jeffrey distribution, respectively. The reference time for $T_0$ is BJD-2,456,400. All the models include the BM model that includes the RV zero point and an extra jitter term for the HARPS-N RVs, their hyperparameters and priors value were set as $\mathcal{U}$(-10, 10) and $\mathcal{U}$(0.01, 10), respectively.}
\end{table*}

%Results from different juliet models
\begin{table*}[!tb]
\begin{small}
\centering
\caption{Summary of the final parameter values of GJ~720~A for the different models that we followed using {\tt juliet}.}
\label{tab:gj720a_final_values_diff_models_juliet}
\begin{tabular}{l c c c c c c c}
\hline
\hline
\noalign{\smallskip}
Parameter & 	S-index (BM+GP) & BM+GP 					& BM+1pl 						& BM+GP+1pl$^{(1)}$ 						& BM+GP+1pl$^{(2)}$ 					& BM+2pl 			& BM+GP+2pl \\
\noalign{\smallskip}	
\hline	
\noalign{\smallskip}
\multicolumn{8}{c}{\textit{Base model (BM)} } \\
\noalign{\smallskip}	

%Parameter 							S-index (BM+GP)				& 	  BM+GP 					& BM+1pl 						& BM+GP+1pl 	(1--1000d)				& BM+GP+1pl 					& BM+2pl 					& BM+GP+2pl \\

$\gamma_0$ (m/s)						&    $0.93^{+0.05}_{-0.04}$ 		&    $-0.89^{+2.11}_{-2.71}$ 		&  $-0.09^{+0.22}_{-0.22}$ 		& $-0.29^{+1.82}_{-2.31}$ 			& $-0.53^{+2.29}_{-2.55}$ 		&  $-0.17^{+0.21}_{-0.21}$ 	&  $-0.22^{+0.97}_{-1.32}$ \\
$\sigma$	(m/s)							&    $0.03^{+0.003}_{-0.002}$		&    $1.62^{+0.16}_{-0.14}$		&  $3.25^{+0.19}_{-0.13}$ 		& $1.47^{+0.14}_{-0.15}$ 				& $0.14^{+0.14}_{-0.13}$ 			&  $2.25^{+0.18}_{-0.16}$ 	&  $1.59^{+0.18}_{-0.20}$ \\

\noalign{\smallskip}	
\hline
\noalign{\smallskip}						    
\multicolumn{8}{c}{\textit{GP parameters}} \\
\noalign{\smallskip}	

$\sigma_{\rm GP}$ (m/s)					&$0.13^{+0.04}_{-0.03}$  			&    	$5.41^{+2.04}_{-1.24}$  		& 	...						& $4.02^{+2.32}_{-1.19}$  			& $4.44^{+2.32}_{-1.36}$  		&	...					&  $2.63^{+1.57}_{-0.63}$  	 \\
$\alpha_{\rm GP}$ ($\rm 10^{-6}d^{-2}$)		&$50.1^{+114.5}_{-20.5}$			&      $0.74^{+0.54}_{-0.30}$		& 	...						& $1.62^{+3.38}_{-0.78}$ 				& $1.41^{+1.27}_{-0.64}$ 			&	...					&  $4.31^{+15.05}_{-2.57}$  \\
$\lambda$ (d) $^{(*)}$					&$141.28^{+93.5}_{-220.9}$		&	$1162.48^{+1360.83}_{-1825.74}$	&	...					& $785.67^{+543.99}_{-1132.28}$		& $842.15^{+887.36}_{-1250.0}$	&	...					&  $481.68^{+257.77}_{-623.78}$ \\		
$\Gamma_{\rm GP}$						&$0.13^{+0.13}_{-0.06}$			&  	$2.84^{+1.65}_{-1.04}$		&  	...						& $1.09^{+1.11}_{-0.54}$ 				& $0.99^{+0.84}_{-0.47}$ 			& 	...					&  $0.37^{+7.62}_{-0.33}$ \\
$P_{\rm rot, GP}$ (d)					&$36.05^{+1.39}_{-1.44}$			&  	$38.93^{+0.03}_{-0.03}$		&  	...						& $35.24^{+0.13}_{-0.12}$ 			& $35.23^{+0.10}_{-0.11}$ 		& 	...					&  $113.89^{+381.51}_{-0.43}$ 	 \\

\noalign{\smallskip}
\hline	
\noalign{\smallskip}							    
\multicolumn{8}{c}{\textit{Planet 1 parameters} }\\	
\noalign{\smallskip}	
	
$P$  (d) 								& 	...						& 	...						&  $19.484^{+0.007}_{-0.006}$ 		&  $19.467^{+0.006}_{-0.006}$ 			& $19.466^{+0.005}_{-0.005}$ 		& $19.486^{+0.006}_{-0.007}$ 		& $19.475^{+0.006}_{-0.006}$ 	 \\
$T_0$ (d)					 			& 	...						& 	...						&  $26.57^{+0.53}_{-0.43} $ 		&  $6.80^{+0.45}_{-0.45} $ 			& $6.81^{+0.43}_{-0.42} $ 		& $24.38^{+1.32}_{-18.67} $ 		& $6.42^{+0.44}_{-0.42} $ 	 \\
$e$ 									&  	...						&  	...						&  $0.16^{+0.04}_{-0.04} $		&  $0.12^{+0.05}_{-0.06} $			& $0.12^{+0.05}_{-0.06} $			& $0.09^{+0.07}_{-0.06} $			& $0.12^{+0.06}_{-0.06} $		  \\
$\omega$ (deg) 						&  	...						&  	...						&  $169.69^{+57.78}_{-33.43}$		&  $110.71^{+23.12}_{-24.01}$ 			& $110.22^{+23.97}_{-24.28}$ 		& $114.60^{+57.25}_{-52.62}$ 		& $101..23^{+24.91}_{-26.31}$ 	  \\
$K$ (m/s)								& 	...						& 	...						&  $3.97^{+0.27}_{-0.37}$			&  $4.70^{+0.28}_{-0.30}$ 			& $4.72^{+0.27}_{-0.27}$ 			& $3.81^{+0.29}_{-0.31}$ 			& $4.10^{+0.28}_{-0.27}$ 		  \\
%$\gamma_0$ (m/s)$^{(2)}$ 				& 	...						& 	...						&  							& 								& $-0.53^{+2.29}_{-2.54}$ 		& 						&\\

\noalign{\smallskip}
\hline
\noalign{\smallskip}							    
\multicolumn{8}{c}{\textit{Planet 2 parameters} }\\	
\noalign{\smallskip}	

$P$  (d) 								& 	...						& 	...						&  	...						&  	...							& 	...						&  $42.11^{+0.04}_{-0.03}$	& $42.10^{+0.06}_{-0.06}$ 	 \\
$T_0$ (d)								& 	...						& 	...						&  	...						&  	...							& 	...						&  $37.02^{+1.84}_{-2.08} $	& $7.55^{+39.88}_{-3.28} $ 	 \\
$e$ 									&  	...						&  	...						&   	...						&   	...							&  	...						&  ...						& ...						  \\
$\omega$ (deg) 						&  	...						&  	...						&   	...						&   	...							&  	...						&  ...						& ...					 	  \\
$K$ (m/s)								& 	...						& 	...						&  	...						&  	...							& 	...						&  $2.16^{+0.32}_{-0.31}$ 	& $1.51^{+0.31}_{-0.29}$ 		  \\
\noalign{\smallskip}	
\hline	
\noalign{\smallskip}
\end{tabular}
\tablefoot{$^{(1), (2)}$: The difference between these two models is the prior of the GP $\rm P_{rot}$ parameter. For the first model the $\rm P_{rot}$ was set with a wide prior value (1--1000)\,d (including all possible RV signals) while the second one was set around the value of the possible rotation period of the star (30--50)\,d. $^{(*)}$: The $\lambda$ parameter corresponds with the length-scale of the GP kernel expressed in day units, it was derived from $\alpha_{\rm GP}$ parameter. The reference time for $T_0$ is BJD-2,456,400.}
\end{small}
\end{table*}

%emcee GP analysis
\begin{table*}[t]
\centering
\caption{Same as for Tab \ref{tab:gj720a_priors_details} but for {\tt emcee} instead of {\tt juliet}. Parameters $e$ and $\omega$ are derived from the explored parameters $\sqrt{e} sin(\omega)$ and $\sqrt{e} cos(\omega)$}
\label{tab:nino_t1}
\begin{tabular}{l c c r}
\hline
\hline
\noalign{\smallskip}
Parameter & Prior & Unit & Description \\
\noalign{\smallskip}	
\hline	
\noalign{\smallskip}
\multicolumn{4}{c}{\textit{RV parameters}} \\
\noalign{\smallskip}
$\gamma_0$ &  $\mathcal{U}$(-10, 10) & m$\rm s^{-1}$ & RV zero point for HARPS-N\\
$\sigma$		 & $\mathcal{U}$(0.01, 10) & m$\rm s^{-1}$ & Extra jitter term for HARPS-N \\
\noalign{\smallskip}							    
\multicolumn{4}{c}{\textit{GP parameters}} \\
\noalign{\smallskip}		
$B_{\rm GP,RV}$	 & $\mathcal{L} \mathcal{U}$($10^{-2}$, 100)   & m$\rm s^{-1}$ & Amplitude of the GP for the RVs\\
$C_{\rm GP,RV}$	&  $\mathcal{L} \mathcal{U}$($10^{-2}$, 100) & ...    &  Additive factor impacting on the amplitude of the GP for the RVs\\
$L_{\rm GP,RV}$	& $\mathcal{L} \mathcal{U}$(300, $10^5$) & d & Length-scale of exponential part of the GP for the RVs\\
$P_{\rm rot, GP,RV}$ & $\mathcal{L} \mathcal{U}$(30, 50) & d & Period of the GP quasi-periodic component for the RVs\\
\noalign{\smallskip}							    
\multicolumn{4}{c}{\textit{Planet parameters} }\\	
\noalign{\smallskip}										    
$P$ 			& $\mathcal{U}$ (5, 25) 	& d	& Period of planet b \\
$T_0$ (BJD-2,456,400) & $\mathcal{U}$ (0,15)		& d &  Time of periastron passage\\
$e$ 				& $\mathcal{U}$ (0, 0.8 )		& ... & Orbital eccentricity of planet b\\
$\omega$ 	&	$\mathcal{U}$ (0, 360)	& deg	& Periastron angle of planet b\\
$K$		&  $\mathcal{U}$ (0, 10 )  	&  m$\rm s^{-1}$      & RV semi-amplitude of planet b   \\    				
\noalign{\smallskip}	
\hline	
\noalign{\smallskip}
\end{tabular}
\tablefoot{The prior labels of $\mathcal{U}$ and $\mathcal{L}\mathcal{U}$ represent uniform and loguniform distribution, respectively. The reference time for $T_0$ is BJD-2,456,400. }
\end{table*}

\end{appendix}

\end{document}